\documentclass[10pt,aps,prd,reprint,nofootinbib,floats,floatfix,amsfonts,amssymb,amsmath,preprintnumbers,notitlepage,superscriptaddress,twocolumn]{revtex4-1}

\usepackage{url}
\usepackage{graphicx,color} 
\usepackage{hyphenat} 
\usepackage{amsmath, amssymb, amsfonts,amsbsy,amsthm} 
\usepackage{float}
\usepackage{subcaption}
\usepackage{soul,xcolor} 
\setstcolor{red}

\newcommand{\orcid}[1]{\href{https://orcid.org/#1}{\includegraphics[scale=0.15]{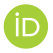}}}

\usepackage[normalem]{ulem}

\begin{document}

\title{Correlated 0.01Hz - 40 Hz seismic and Newtonian noise and its impact on future gravitational-wave detectors  }

\author{Kamiel Janssens\orcid{0000-0001-8760-4429}}
\affiliation{Universiteit Antwerpen, Prinsstraat 13, 2000 Antwerpen, Belgium}
\affiliation{Universit\'e C$\hat{o}$te d’Azur, Observatoire de la C$\hat{o}$te d’Azur, CNRS, Artemis, 06304 Nice, France}
\author{Guillaume Boileau \orcid{0000-0002-3576-6968}}
\affiliation{Universit\'e C$\hat{o}$te d’Azur, Observatoire de la C$\hat{o}$te d’Azur, CNRS, Laboratoire Lagrange, 06304 Nice, France}
\author{Nelson Christensen\orcid{0000-0002-6870-4202}}
\affiliation{Universit\'e C$\hat{o}$te d’Azur, Observatoire de la C$\hat{o}$te d’Azur, CNRS, Artemis, 06304 Nice, France}
\author{Nick van Remortel \orcid{0000-0003-4180-8199}}
\affiliation{Universiteit Antwerpen, Prinsstraat 13, 2000 Antwerpen, Belgium}
\author{Francesca Badaracco \orcid{0000-0001-8553-7904}}
\affiliation{Università degli studi di Genova, via Dodecaneso 33, 16146, Italy}
\affiliation{INFN, Sez. di Genova, via Dodecaneso 33, 16146, Italy}
\author{Benjamin Canuel \orcid{0000-0002-1378-2334}}
\affiliation{LP2N, Laboratoire Photonique, Num{\'e}rique et Nanosciences, Universit{\'e} Bordeaux--IOGS--CNRS:UMR 5298, rue F. Mitterrand, F--33400 Talence, France}
\author{Alessandro Cardini}
\affiliation{INFN, sezione di Cagliari, I-09042, Monserrato (Cagliari), Italy}
\author{Andrea Contu}
\affiliation{INFN, sezione di Cagliari, I-09042, Monserrato (Cagliari), Italy}
\author{Michael W. Coughlin\orcid{0000-0002-8262-2924}}
\affiliation{School of Physics and Astronomy, University of Minnesota, Minneapolis, Minnesota 55455, USA}
\author{Jean-Baptiste Decitre}
\affiliation{Laboratoire Souterrain \`{a} Bas Bruit (LSBB), CNRS: UAR3538, Avignon University, Rustrel F-84400,France}
\author{Rosario De Rosa}
\affiliation{Universitá Federico II Napoli, 80126 Napoli, Italy}
\affiliation{INFN - sezione di Napoli, 80126 Napoli, Italy}
\author{Matteo Di Giovanni \orcid{0000-0003-4049-8336}}
\affiliation{La Sapienza Università di Roma, I-00185 Roma, Italy}
\affiliation{INFN, Sezione di Roma, I-00185 Roma, Italy}
\author{Domenico D'Urso \orcid{0000-0002-8215-4542}}
\affiliation{Department of Chemistry, Physics, Mathematics and Natural Science, Università degli Studi di Sassari, I-07100, Sassari, Italy}
\affiliation{INFN, sezione di Cagliari, I-09042, Monserrato (Cagliari), Italy}
\author{Stéphane Gaffet\orcid{0000-0003-4726-2764}}
\affiliation{Laboratoire Souterrain \`{a} Bas Bruit (LSBB), CNRS: UAR3538, Avignon University, Rustrel F-84400,France}
\author{Carlo Giunchi \orcid{0000-0002-0174-324X}}
\affiliation{Istituto Nazionale di Geofisica e Vulcanologia, Sezione di Pisa, Italy}
\author{Jan Harms \orcid{0000-0002-7332-9806}}
\affiliation{Gran Sasso Science Institute (GSSI), I-67100 L'Aquila, Italy}
\affiliation{INFN, Laboratori Nazionali del Gran Sasso, I-67100 Assergi, Italy}
\author{Soumen Koley \orcid{0000-0002-5793-6665}}
\affiliation{Gran Sasso Science Institute (GSSI), I-67100 L'Aquila, Italy}
\affiliation{INFN, Laboratori Nazionali del Gran Sasso, I-67100 Assergi, Italy}
\author{Valentina Mangano}
\affiliation{INFN, Sezione di Roma, I-00185 Roma, Italy}
\author{Luca Naticchioni \orcid{0000-0003-2918-0730}}
\affiliation{INFN, Sezione di Roma, I-00185 Roma, Italy}
\author{Marco Olivieri}
\affiliation{Sezione di Bologna, Istituto Nazionale di Geofisica e Vulcanologia, Bologna, Italy}
\author{Federico Paoletti}
\affiliation{INFN - sezione di Pisa, 56127 Pisa, Italy}
\author{Davide Rozza \orcid{0000-0002-7378-6353}}
\affiliation{Department of Chemistry, Physics, Mathematics and Natural Science, Università degli Studi di Sassari, I-07100, Sassari, Italy}
\affiliation{INFN, sezione di Cagliari, I-09042, Monserrato (Cagliari), Italy}
\author{Dylan O. Sabulsky\orcid{0000-0001-7421-2821}}
\affiliation{Laboratoire Souterrain \`{a} Bas Bruit (LSBB), CNRS: UAR3538, Avignon University, Rustrel F-84400,France}
\author{Shahar Shani-Kadmiel\orcid{0000-0003-2215-6164}}
\affiliation{R\&D Department of Seismology and Acoustics, KNMI, De Bilt, The Netherlands}
\author{Lucia Trozzo}
\affiliation{INFN - sezione di Napoli, 80126 Napoli, Italy}

\date{\today}

\begin{abstract}
We report correlations in underground seismic measurements with horizontal separations of several hundreds of meters to a few kilometers in the frequency range 0.01\,Hz to 40\,Hz. These seismic correlations could threaten science goals of planned interferometric gravitational-wave detectors such as the Einstein Telescope as well as atom interferometers such as MIGA and ELGAR. We use seismic measurements from four different sites, i.e.  the former Homestake mine (USA) as well as two candidate sites for the Einstein Telescope, Sos Enattos (IT) and Euregio Maas-Rhein (NL-BE-DE) and the site housing the MIGA detector, LSBB (FR). At all sites, we observe significant coherence for at least 50\% of the time in the majority of the frequency region of interest.
Based on the observed correlations in the seismic fields, we predict levels of correlated Newtonian noise from body waves.
We project the effect of correlated Newtonian noise from body waves on the capabilities of the triangular design of the Einstein Telescope's to observe an isotropic gravitational-wave background (GWB) and find that, even in case of the most quiet site, its sensitivity will be affected up to $\sim$20\,Hz. 
The resolvable amplitude of a GWB signal with a negatively sloped power-law behaviour would be reduced by several orders of magnitude. However, the resolvability of a power-law signal with a slope of e.g. $\alpha=0$ ($\alpha=2/3$) would be more moderately affected by a factor $\sim$ 6-9 ($\sim$3-4) in case of a low noise environment.
Furthermore, we bolster confidence in our results by showing that transient noise features have a limited impact on the presented results. 

\end{abstract}

\maketitle


\section{Introduction}
\label{sec:Introduction}

Searches for unmodeled and/or long duration gravitational-wave (GW) signals, such as the isotropic GW background (GWB)~\cite{Christensen_2018}, are more susceptible to be biased by correlated noise. One such example are correlations in magnetic field fluctuations over Earth-scale distances, such as the Schumann resonances~\cite{Schumann1,Schumann2}. Their potential effect on GWB searches with Earth-based GW interferometers -- LIGO \cite{2015}, Virgo \cite{VIRGO:2014yos} and KAGRA \cite{PhysRevD.88.043007} -- has been extensively investigated \cite{Thrane:2013npa,Thrane:2014yza,Coughlin:2016vor,Himemoto:2017gnw,Coughlin:2018tjc,Himemoto:2019iwd,Meyers:2020qrb,PhysRevD.107.022004}. Moreover, the effect of correlated lightning glitches on searches for GW bursts, such as core collapse supernova, was studied \cite{Kowalska-Leszczynska:2016low,PhysRevD.107.022004}. Furthermore, the effect of Schumann resonances on the Einstein Telescope (ET) was investigated~\cite{PhysRevD.104.122006} and shown to be a limiting noise source for the search for a GWB below $\sim$ 30Hz, in case ET has a similar magnetic coupling as LIGO/Virgo.

ET is the European proposal for a third-generation, Earth-based interferometric GW detector \cite{Punturo:2010zz}. 
Its baseline proposal exists out of three nested detectors. Each detector is composed of two interferometers, one optimized for low frequency observations and the other devoted to high frequency observations. This configuration is often referred to as "xylophone". The detectors are arranged in an equilateral triangle with opening angle of $\pi/3$ and arm lengths of 10 km. In this paper, we ignore the details of the xylophone configuration~\cite{Hild:2010id} and we treat ET as consisting of three interferometers. 
This assumption has no effect on our studies. 
In a proposed alternative design, ET is composed of two separate L-shaped interferometers in xylophone configuration, located in two far sites. Given the large separation of these two separated detector sites, it is deemed that the effect of correlations in seismic noise is negligible in the frequency band of interest for the ET. Therefore the noise projections in this paper are only relevant in case the triangular design is chosen. However, the investigation of correlated noise over a distance of 10km is relevant for both designs as it provides information on potential correlated noise coupling between input and end masses of a singular interferometer.

\begin{figure}
    \centering
    \includegraphics[width=\linewidth]{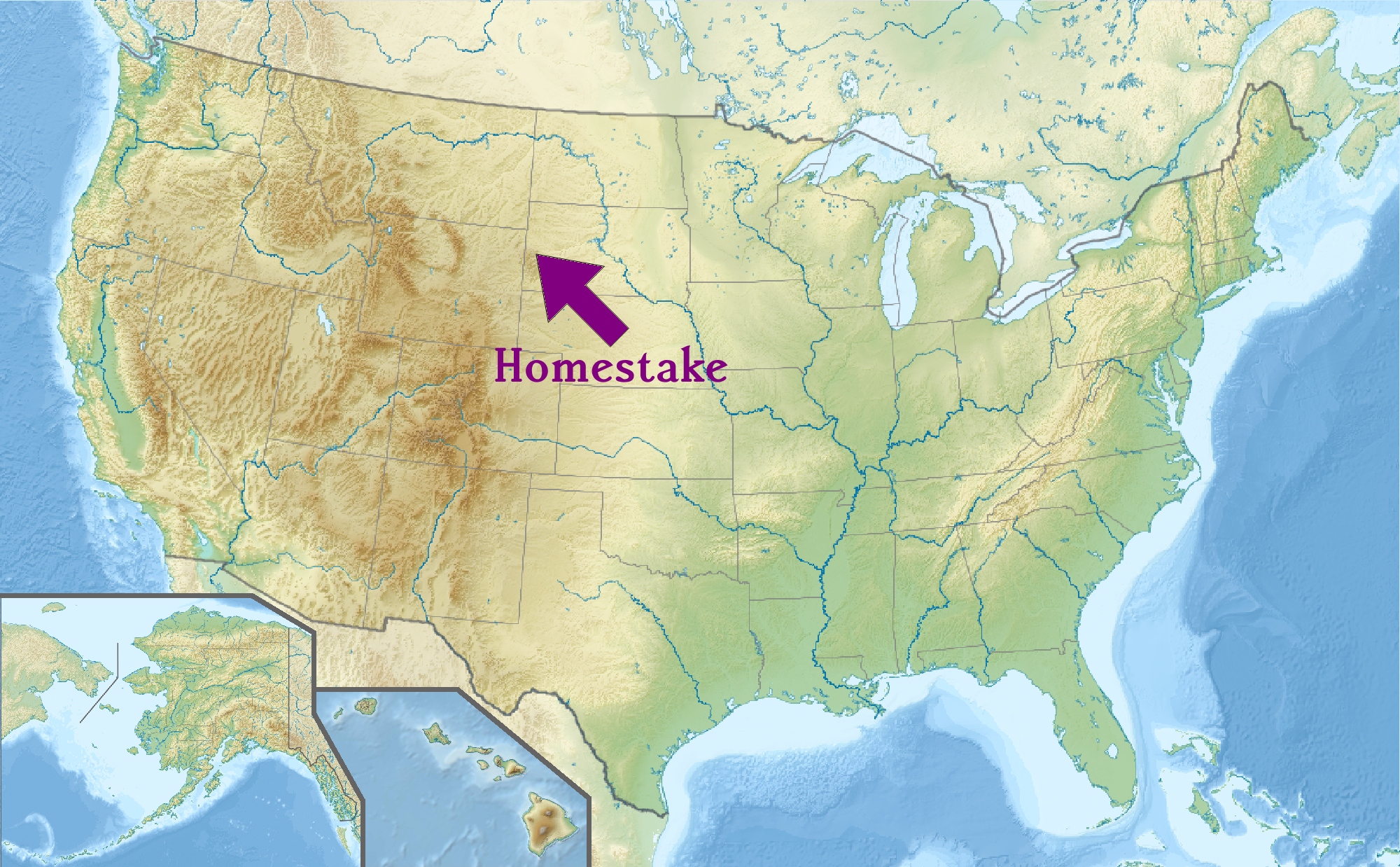}
    \includegraphics[width=\linewidth]{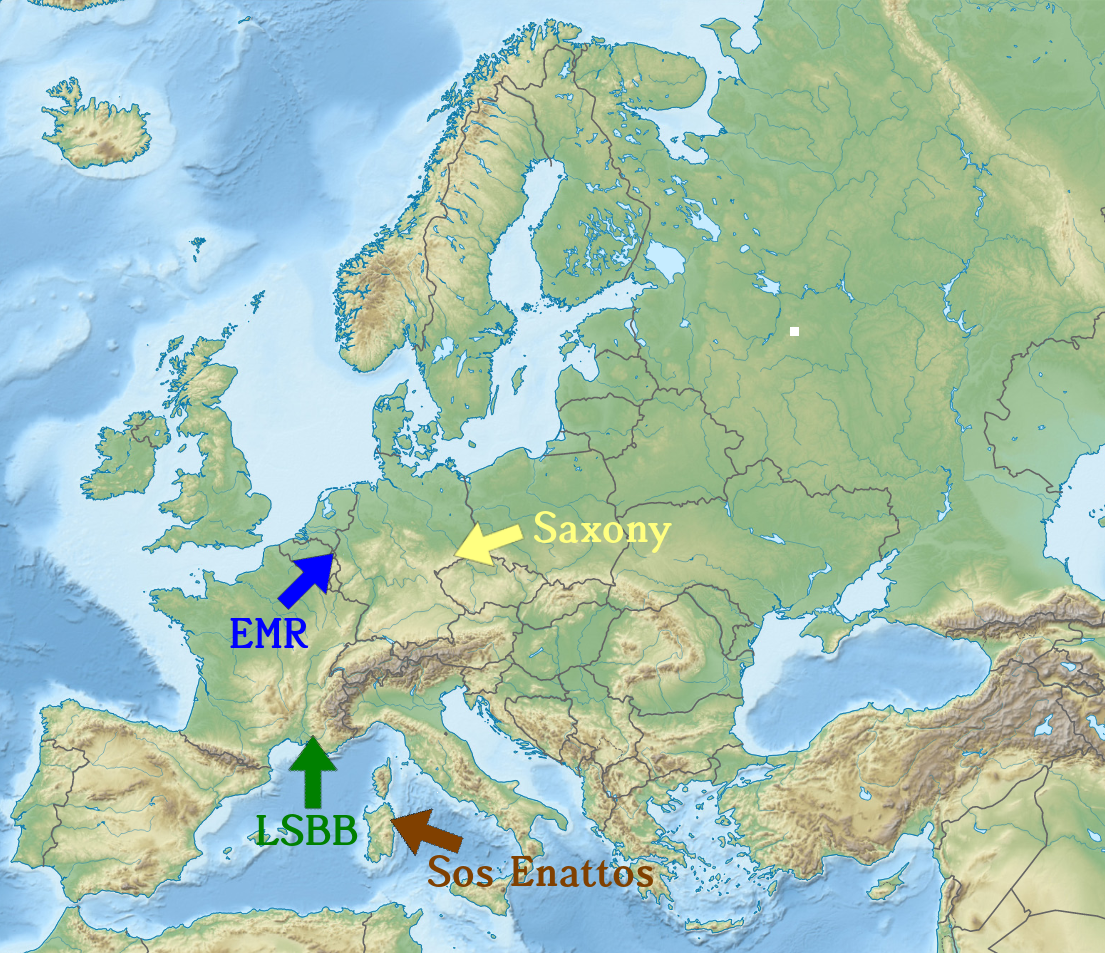}
    \caption{Top: map of the USA highlighting the location of the former Homestake mine \cite{10.1785/0220170228}. Bottom: map of Europe with the locations of LSBB (green) and of the two ET candidate sites, namely Sos Enattos (brown) and Euregio Maas-Rhein (EMR) (blue). We also highlight the third ET candidate site in Saxony (yellow) that was not considered for the analyses of this paper. The maps are taken from \cite{mapUSA} and \cite{mapEU} and modified according to the creative commons license 3.0.}
    \label{fig:maps}
\end{figure}

Official candidate sites for the ET are the area in the vicinity of the Sos Enattos mine in Sardinia\cite{Naticchioni_2014,Naticchioni_2020,10.1785/0220200186,Allocca:2021opl,digiovanni23,Saccorotti_2023}, Italy, and the Euregio Maas-Rhein (EMR) at the intersection of the Belgian, Dutch, and German borders \cite{Amann:2020jgo,Koley_2022,Bader_2022}. The region of Saxony in Germany has been recently proposed as another possible candidate. The different sites are indicated in the bottom panel of Fig. \ref{fig:maps}.
Moreover, we highlight the USA's proposal for a third generation interferometric GW detector: Cosmic Explorer (CE)\cite{Reitze2019Cosmic}. However, we do not consider CE in our study as it is planned to have two widely separated detectors.

\begin{figure}
    \centering
    \includegraphics[width=\linewidth]{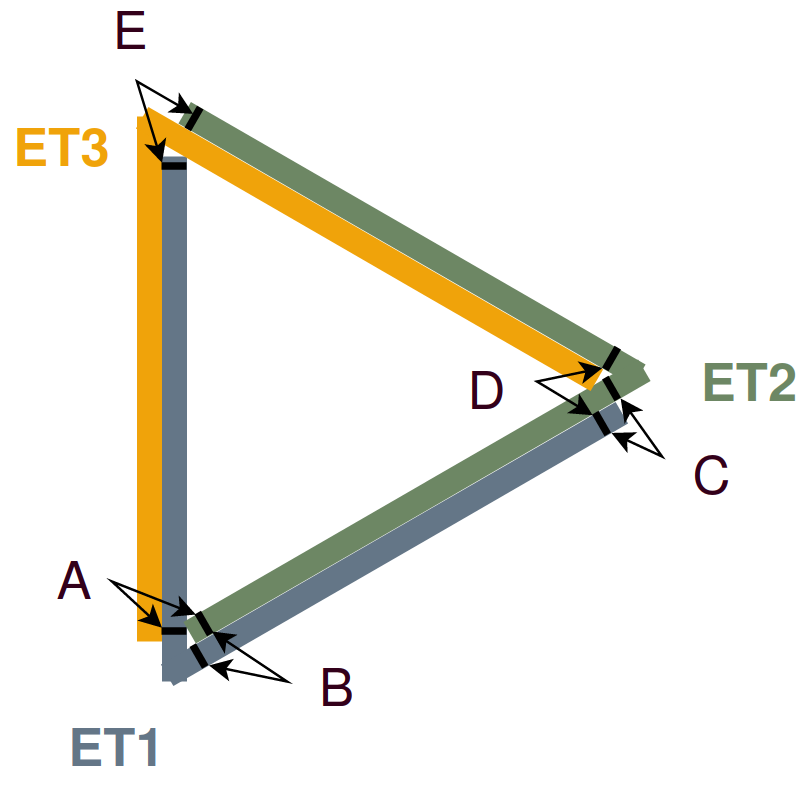}
    \caption{Scheme of the proposed ET triangular configuration (the low- and the high-frequency detectors are not showed). Considering the ET1-ET2-baseline, we can identify 5 possible coupling locations where seismic and Newtonian noise can correlate on distances between 300m and 600m: A to E \cite{PhysRevD.106.042008}. Additionally there are multiple possible coupling locations for correlated noise on distances of about 10.5km (not indicated).}
    \label{fig:ETDiagram}
\end{figure}

Due to the nested design of three almost co-located interferometers in a triangular configuration, there are several potential coupling locations for correlated noise to enter on distance scales of 300\,m to 600\,m into the different ET interferometers, as illustrated in Fig. \ref{fig:ETDiagram} \cite{PhysRevD.106.042008}. 
Even though ambient seismic fields rapidly lose coherence over large distances for frequencies higher than 1 Hz \cite{Coughlin_2014_seismic,Coughlin2019_seismic}, in \cite{PhysRevD.106.042008} the authors have shown that on distance scales of several hundreds of meters, significant correlations in seismic noise are present at least 50\% of the time up to 40\,Hz. Please note that the results in \cite{PhysRevD.106.042008} are based on surface measurements at EMR and underground measurements at Homestake.
The correlations in seismic noise result in correlations in Newtonian noise (NN) \cite{Harms2019,PhysRevD.30.732,PhysRevD.58.122002}, which is a force exerted on GW test-masses caused by density fluctuations in the surrounding medium. Correlations in NN from body waves could seriously affect the search for a GWB with the ET by orders of magnitude \cite{PhysRevD.106.042008}. These results were also considered in a recent study comparing the scientific benefits of a triangular detector configuration versus a configuration with two non co-located, L-shaped detectors \cite{Branchesi:2023mws}.

In previous works, potential coupling locations on distance scales of $\sim$ 10.5\,km were neglected. However, as can be seen on Fig. \ref{fig:ETDiagram} (not indicated), there are multiple possible coupling locations between the input and end test masses of the different detectors.
For these distances, the seismic noise is expected to have lost coherence in the frequency band of relevance for the ET.

\begin{figure}
    \centering
    \includegraphics[width=\linewidth]{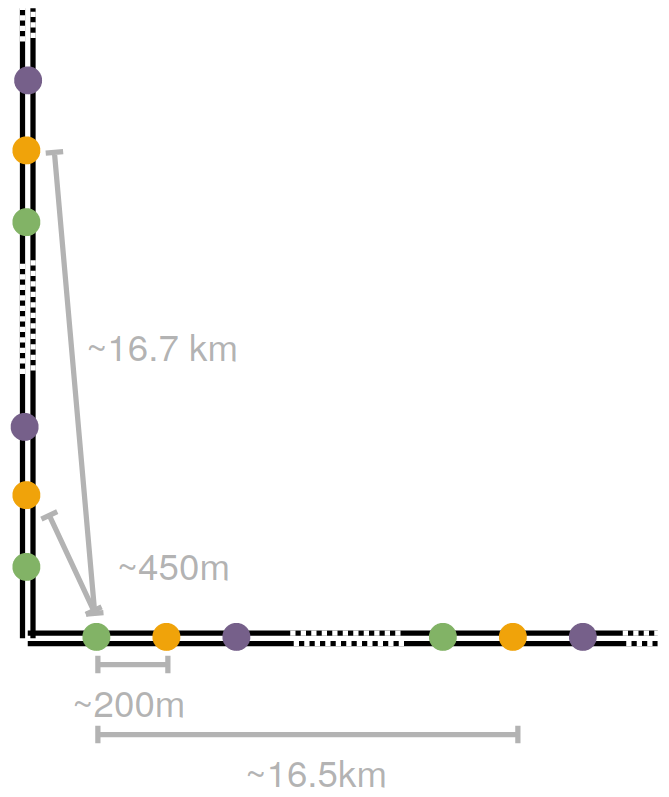}
    \caption{Scheme of the ELGAR configuration, where the dots represent the atom interferometers. Each color corresponds to a single 16,3 km gradiometer of the 2D array. We give examples of different correlation lengths that must be taken into account to calculate the impact of NN \cite{Chaibi2016}. In this illustration, we have assumed the distance between the beamsplitter and the first atom interferometer to be 200 m.}
    \label{fig:ELGARDiagram}
\end{figure}

In this work, we aim to investigate seismic correlations on distance scales from several hundreds of meters up to tens of kilometers with the goal to provide further insights in potential coupling to the different ET detectors. However, these correlation studies are also of particular interest for GW searches using atom interferometry. Earth-based atom interferometers aim to be sensitive to GWs in the frequency range 0.1\,Hz-10\,Hz with their peak sensitivity typically around 1Hz-2Hz~\cite{Canuel_2018,Canuel:2019abg}. At those frequencies, the seismic waves have longer correlation lengths up to several kilometers. In Fig.~\ref{fig:ELGARDiagram}, we illustrate the set-up of a proposal for a future atom interferometer, ELGAR\footnote{European Laboratory for Gravitation and Atom-interferometric Research} \cite{Canuel:2019abg}. This L-shaped detector consists in a 2D array of atom gradiometers.
Each arm is composed of N= 80 single gradiometers of baseline L=16,3 km placed every 200 m. The GW signal is obtained by the difference of the averaged gradiometric signal in each arm. This signal extraction method implies that all NN correlations over distances from 200 m up to tens of kilometers are relevant for the detector sensitivity \cite{Chaibi2016}.

In this work, we aim to further understand the effect of correlations in seismic and NN and how they could impact searches for GWs. First of all, we use underground seismic data from four different sites and a total of nine different sensor pairs with horizontal separations between $\sim$230\,m and $\sim$10\,km.
This reduces the effect of any site dependence that might be present in the earlier results of \cite{PhysRevD.106.042008}, where underground seismic measurements of a single site were used. Furthermore, we probe correlations on longer distances as in \cite{PhysRevD.106.042008}, where the largest separation between seismic sensors was 810m.
Finally, we do not only focus on the frequency band of interest for the ET (1Hz-40Hz)\footnote{In the ET community this frequency band is often referred to as the low-frequency (LF) band.}, but also on lower frequencies of interest for atom interferometry (0.01Hz-1Hz).
In this work we do not make a projection of the correlated seismic and NN on present or future atom interferometers. However, we do discuss seismic correlations over the entire frequency range 0.01Hz-40Hz, which can be used to serve for the calculation of noise projections for atom interferometers, similar to those in earlier work~\cite{PhysRevD.107.022004,Junca_2019}.

For GW interferometric detectors we will not discuss the effect of seismic and NN on their instantaneous sensitivity, nor methods to perform NN subtraction as these have been extensively studied in the literature \cite{PhysRevD.86.102001,PhysRevD.92.022001,Coughlin_2014_seismic,Coughlin_2016,PhysRevLett.121.221104,Tringali_2019,Badaracco_2020,Badaracco_2019,NN_Sardinia2020,10.1785/0220200186,Bader_2022,Koley_2022, digiovanni23,vanBeveren_2023}. However, we focus on the impact of correlated NN on the search for a GWB, which is (one of) the most sensitive search(es) to correlated noise.

In Sec.~\ref{sec:Scope}, we highlight the different analyses that are performed. In Sec.~\ref{sec:Homestake}-\ref{sec:Terziet}, we present the seismic results for each site. Afterwards in Sec.~\ref{sec:SiteComparison}, we discuss the results of the different sites, by using reference measurement per site. In Sec.~\ref{sec:CorrNN}, we use the observed seismic correlations to make a projection for the levels of correlated NN. In Sec.~\ref{sec:Glicth}, we investigate the effect of transient seismic noise and its impact on the results presented in Sec.~\ref{sec:Homestake}-\ref{sec:CorrNN}. Finally, in Sec.~\ref{sec:Conclusion}, we conclude our results and highlight possibilities for future work.

\section{Scope of performed analysis}
\label{sec:Scope}

In this study we analyze underground seismic data from four different geographical locations. We only focus on horizontal seismic waves and the subsequent NN from body waves. We focus on these measurements as an earlier study \cite{PhysRevD.106.042008} has shown that the effect of correlated NN from Rayleigh waves on the ET and its search for a GWB is modest compared to the effect from NN from body waves. The key factor is that ET will be built underground, drastically reducing the effect from surfaces waves above 1\,Hz.
In this earlier study \cite{PhysRevD.106.042008}, they analysed both vertical and horizontal seismic noise from underground data at the Homestake mine in the USA. Moreover, they used four different sensor pairs with distances between 255\,m and 810\,m. Because we use slightly different parameters when analysing our data, we re-analyse the data from Homestake to get a one-to-one comparison with this earlier results and have a better comparison with the new results presented in this paper.

As shown on the maps in Fig.~\ref{fig:maps}, the geographical locations from which data are used are: the former Homestake mine in the USA\cite{10.1785/0220170228}, the MIGA\footnote{Matter-wave laser Interferometric Gravitation Antenna}~\cite{Canuel_2018} site at the `Laboratoire Souterrain \`a Bas Bruit' (LSBB) in France and the two candidate sites for the ET, the former Sos Enattos mine in Italy and the EMR region in the border region of the Netherlands, Belgium and Germany. No data from the third possible candidate site located in the region of Saxony in Germany are used, as at the time of this paper no long term ($>$ 1 month) underground measurements were performed with a horizontal separation of  at least several hundreds of meters between two seismometers. It will be interesting to perform a similar analysis when such data becomes available.

The different sensor pairs and the distance between the sensors, as well as their depth with respect to the surface, are listed in Tab.~\ref{tab:sensors}. For all pairs we only analyse the correlations in seismic noise between the NS seismic measurement of the two sensors. Only for the GAS-RAM pair at LSBB we analyse the three other dirictionality pairs: EW-EW, NS-EW and EW-NS, to compare these results from LSBB, with a similar analysis for Homestake \cite{PhysRevD.106.042008}.
Furthermore, in Tab. \ref{tab:sensors} we list the range over which the sensors are sensitive. In the discussion of the results and figures we only show those frequency regions and exclude the frequency bands where the measurements are dominated by e.g. sensor self-noise.

\begin{table*}[]
\begin{tabular}{|l|l|l|l|l|l|l|l|l|l|}
\hline
Location & Name 1 & Name 2 & Model 1 & Model 2 &  Horizontal distance & Vertical distance & Depth 1 & Depth 2 & Frequency range \\ \hline
Homestake (US) & D2000 & E2000 & STS-2 & STS-2 &  $\sim 405$ m & $\sim 0$ m & 610 m & 610m  & 0.01Hz-40Hz\\ \hline
LSBB (FR) & GAS & RAM & STS-2 & STS-2 &  $\sim 600$ m & $\sim 14$ m & 260 m & 500 m & 0.01Hz-40Hz \\ \hline
 & MGS & RAM & STS-2 & STS-2 &  $\sim 750$ m & $\sim 4$ m & 240 m & 500 m & 0.01Hz-40Hz  \\ \hline
 & MGS & GAS & STS-2 & STS-2 &  $\sim 850$ m & $\sim 11$ m & 240 m & 260 m & 0.01Hz-40Hz \\ \hline
Sos Enattos (IT) & SOE1 & SOE2 & T120H & T360 &  $\sim 230$ m & $\sim 27$ m & 84 m & 111 m & 0.01Hz-10Hz \\ \hline
 &SOE1 & SOE3 & T120H & T240 &  $\sim 380$ m & $\sim 76$ m & 84 m & 160 m & 0.01Hz-10Hz \\ \hline
 & SOE2 & SOE3 & T360 & T240 &  $\sim 370$ m & $\sim 49$ m & 111 m & 160 m  & 0.01Hz-10Hz \\ \hline
 & P2 & P3 & T120Q & T120Q &  $\sim 10000$ m & $\sim 50$ m & 264 m & 252 m & 0.01Hz-10Hz \\ \hline
EMR (NL) & TERZ & CTSN & STS-5A & LE3DBH   & $\sim$ 2417 m & $\sim$ 9.4 m & 250 m & 250 m & 0.2Hz-18Hz  \\ \hline
\end{tabular}
\caption{Table summarizing the sensor pairs that are used in the correlation analysis in this paper. Please note that the depth is with respect to the surface and not with respect to the sea level. As an example: GAS, RAM and MGS at LSBB are all located at approximately the same sea level height, but have significantly different depths.}
\label{tab:sensors}
\end{table*}

For Homestake and LSBB, we analyse a full year of data and compare the results from different months to establish seasonal variations in the level of seismic noise. For Homestake we use data from 2016\footnote{We use the entire year of data, excluding Dec 2016, since no high quality data was available for this month.} and for LSBB from 2018\footnote{The sensors were not operational during the month of May therefore this month can not be include in the analysis.}. The data periods always start and end at 00:00:00 UTC, e.g. for LSBB 1 Jan 00:00:00 UTC 2018 - 1 Jan 00:00:00 UTC 2019.
At the time of analysis, the other sites did not always have a continuous data taking period of one year, so only a sub-set of a few months is analyzed. Throughout the paper, we use the months of Jan. and Aug. to present the results and make comparisons between the different sites, where we consider these results to be representative for winter and summer, respectively.
For Sos Enattos, we analyzed data from Aug 2021, as well as the month of January from the years 2022 and 2023 for two of the pairs P2-P3 and SOE2-SOE3, respectively. For EMR, we use data recorded during the month of January 2023 as at the time of the analysis no long term high quality data was available for the month of August. The sensors are still actively acquiring more data and future studies could look to include more results. 

The data is analyzed with two different sets of parameters optimized for the frequency regions of interest for atom interferometers (0.01Hz - 1Hz) and ET ($>$ 1Hz). We use a frequency resolution of 0.005Hz and average data in chunks of 6h for the low frequency studies. This enables us to resolve coherence up to $\sim 9.3 \times 10^{-3}$, while still having $\mathcal{O}$(100) chunks per month to get a good sense of the variability of the data during every month. To be able to resolve smaller coherence up to $\sim 3.8 \times 10^{-4}$, for the high frequency region we use a frequency resolution of 0.1Hz and average data over chunks of 4h, ensuring $\mathcal{O}$(200) chunks per month.
In the interest of being concise, the figures in this work show the results using these two different parameter sets in one unified plot. More specifically, between 0.01Hz and 1Hz the results with 5\,mHz resolution are shown, whereas for frequencies above 1Hz the results of the analyses with a frequency resolution of 0.1Hz are presented. This is clearly indicated in the relevant figures. Finally the parameters used for the different analysis are also summarized in Tab. \ref{tab:params}.

\begin{table*}[]
\begin{tabular}{|l|l|l|l|}
\hline
 & Freq. res. (FFT length) & Chunk length & Resolvable coherence  \\ \hline
 0.01Hz-1Hz & 5mHz (200s) & 6h &  $\sim 9.3 \times 10^{-3}$   \\ \hline
 1Hz - 40Hz & 0.1Hz (10s) & 4h &  $\sim 3.8 \times 10^{-4}$   \\ \hline  
 Glitch study & 0.1Hz (10s) & 1min &  N.A.   \\ \hline  
\end{tabular}
\caption{Summary of the used analysis parameters. Individual fft segments are averaged together to a chunk of data, which forms the starting point of the results described in this work.}
\label{tab:params}
\end{table*}

Finally, for the study of transient effects on our results, which is presented in Sec. \ref{sec:Glicth}, we analyzed the data with a frequency resolution of 0.1Hz and averaged over 1 min chunks. This duration was inspired based on an earlier study of seismic glitchiness at Sos Enattos \cite{Allocca:2021opl}, where they highlight 1 min is a realistic estimate for a possible signal from a coalescing Intermediate Mass Black Hole (IMBH).

\section{Homestake}
\label{sec:Homestake}

In \cite{PhysRevD.106.042008}, the Homestake data of the sensor pair D2000-E2000 was analysed, as well as several other sensor pairs with different horizontal separations. However, in these earlier results, a frequency resolution of 0.01\,Hz was used and data was averaged over 24h segments with a resolvable coherence of about $\sim 10^{-3}$. They used data from Mar 2015 to Dec 2016. In this section we study the seasonal effect, for which we use different analysis parameters, as described in previous section. The parameters are optimized to ensure that we can uncover a sufficiently low coherence, while at the same time have enough data chuncks to show the variation within each month in a percentile plot.

\begin{figure*}[t]
    \centering
    \includegraphics[width=0.49\textwidth]{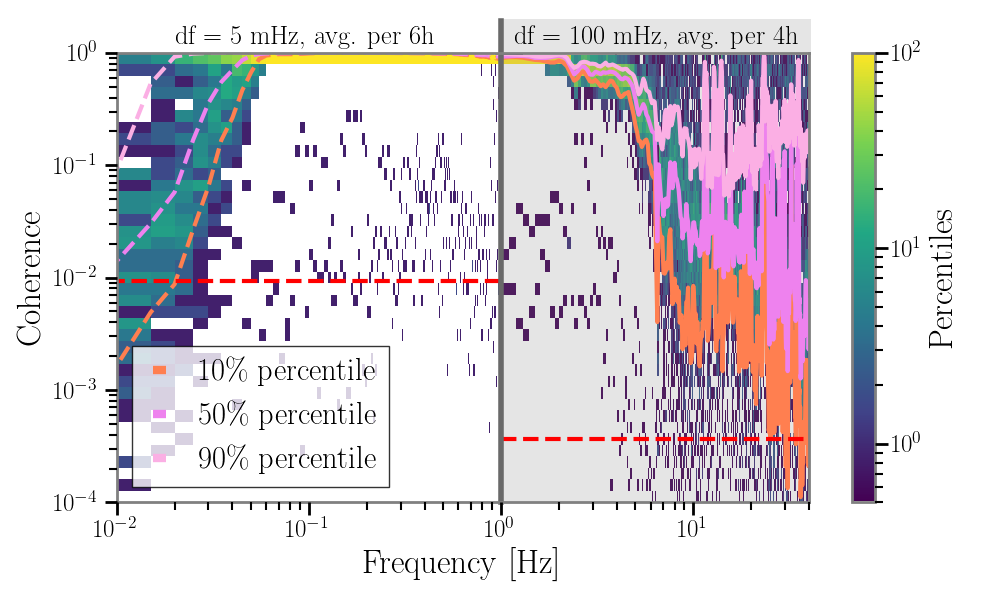}
    \includegraphics[width=0.49\textwidth]{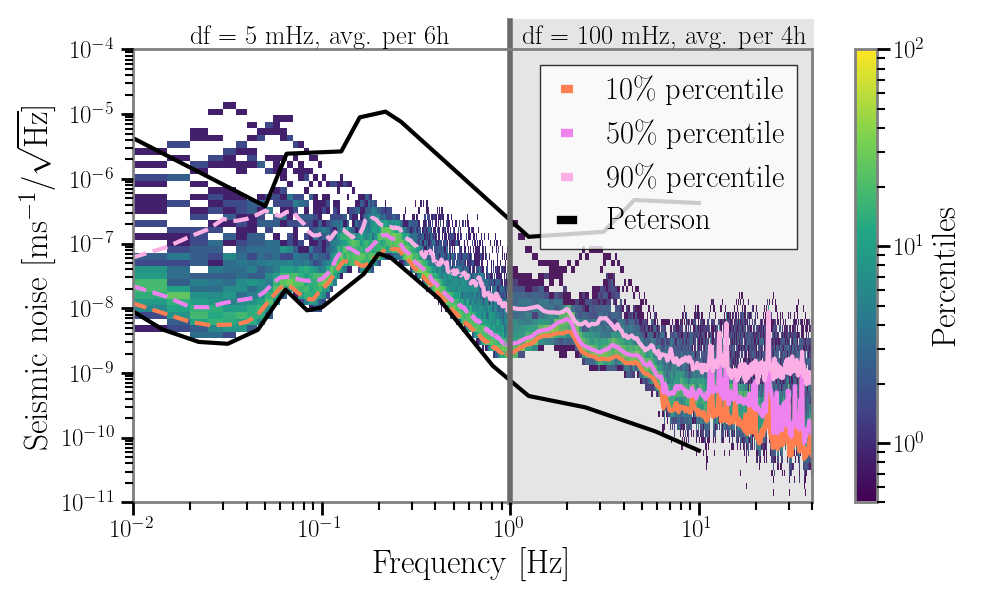}
    \caption{ The coherence (left panel) and CSD (right panel) between the underground seismometers (NS components) D2000 and E2000 (Homestake, USA) during the month of August.
    The 10$^{\text{th}}$, 50$^{\text{th}}$ and 90$^{\text{th}}$ percentiles are shown in respectively light pink, dark pink and light orange, where dashed (full) lines are used for the analyses with the different parameters $<$ 1Hz ($\geq$ 1Hz). The red dashed line (left panel) represents the level of coherence expected from Gaussian data. The black curves (right panel) represent the low and high noise models by Peterson \cite{Pet1993}.}
    \label{fig:D2000-E2000_CohCSD_5mHz}
\end{figure*}

Focusing on frequencies below 1\,Hz, we see in the left panel of Fig.~\ref{fig:D2000-E2000_CohCSD_5mHz} that during the month of August 2016 the seismic noise is almost fully coherent around the microseismic peak, which is coming from sea activity~\cite{Pet1993}.
To determine whether the observed coherence is significant or not, we compare it to the coherence expected from Gaussian data which goes approximately as 1/N, where N is the number of time segments over which was averaged.
At the lowest frequencies the coherence decreases with decreasing frequency. However at 0.01Hz (0.02Hz), at least 50\% (90\%) of the time there is significant coherence to the level of $\sim 9.3 \times 10^{-3}$. Here we want to point out that the decrease in coherence below 0.04Hz has most likely a non-physical origin and arises through data processing leakage.
Above 1\,Hz, we observe significant coherence 90\% of the time up to $\sim$ 40Hz. This result shows that the month of August has higher coherence compared to the results presented for the period Mar 2015 - Dec 2016 in earlier work \cite{PhysRevD.106.042008}. The authors of \cite{PhysRevD.106.042008} stated that the lower coherence observed, for the 10\% percentile below 1\,Hz, in their data was likely due to higher levels of anthropogenic noise leading to degraded coherence. Since this is not observed in any of the months analysed in this paper, these events seem to be limited to the 2015 data.

The right panel of Fig. \ref{fig:GlitchRemoval-D2000-E2000_CohCSD_100mHz} shows the accompanying cross-spectral density (CSD) of the seismic noise. These correlations in the seismic noise are in agreement with the earlier results \cite{PhysRevD.106.042008} if we take into account we do not expect a perfect match as here we only present a small subset of the same data.

\begin{figure*}
    \centering
    \includegraphics[width=0.49\textwidth]{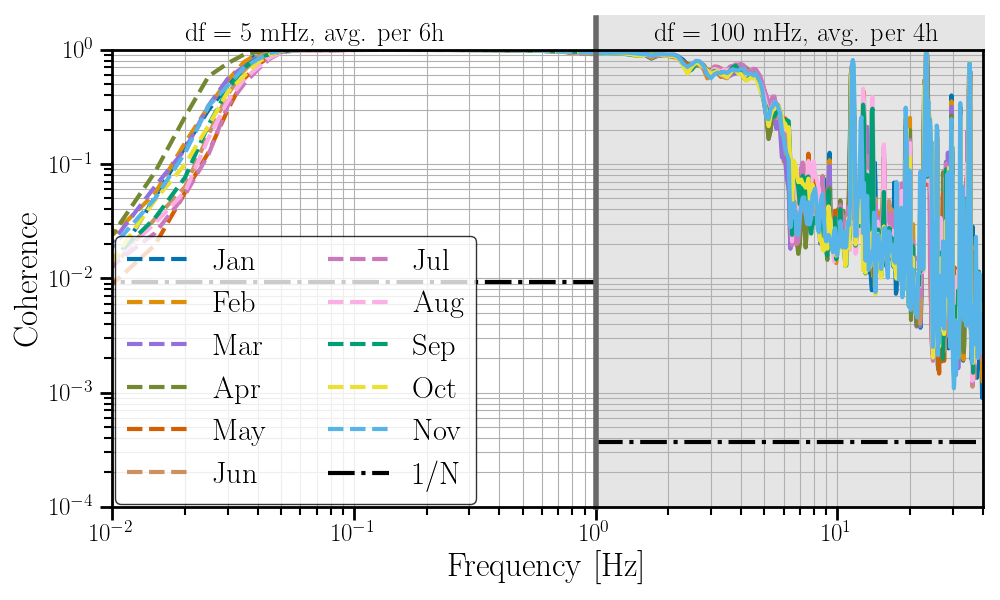}
    \includegraphics[width=0.49\textwidth]{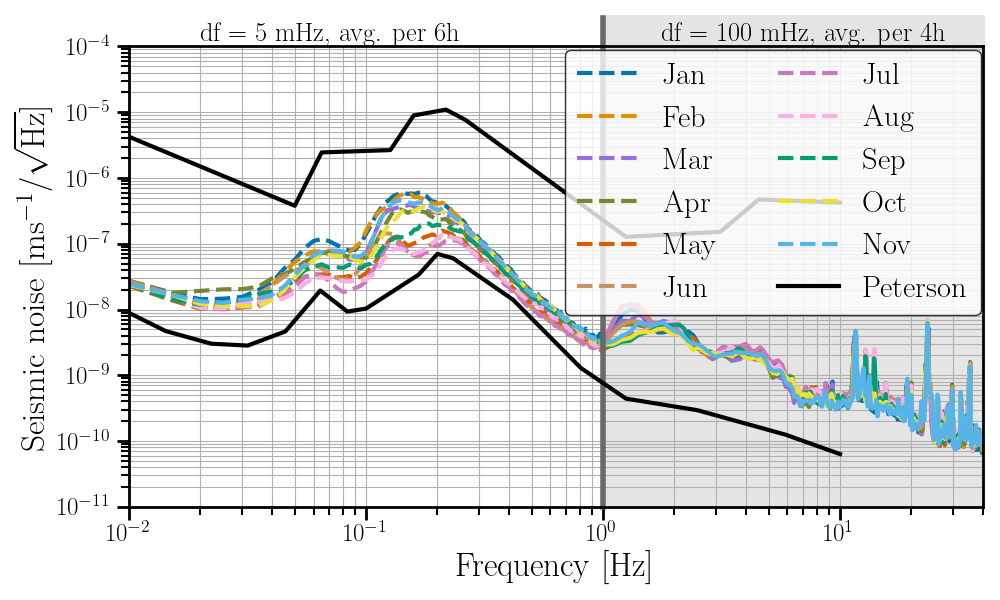}    
    \caption{The median coherence (left panel) and CSD (right panel) of the underground seismometers at Homestake ($\Delta x\approx$405m and depth = 610m) as a function of the month of the year 2016. 
    The black dot-dashed line (left panel) represents the level of coherence expected from Gaussian data. In the right panel, the Peterson low and high noise models are shown in black..}
    \label{fig:Homestake_monthly_CohCSD_5mHz}
\end{figure*}

Fig.~\ref{fig:Homestake_monthly_CohCSD_5mHz} compares median observed coherence (left panel) and CSD (right panel) for the different months of the year. Notice that, as mentioned earlier, the month of Dec is missing due to the absence of good quality data during this period. The only significant difference in coherence across different months is observed below 0.04Hz, which is linked to non-physical effects. However around the first and secondary microseism peaks between 0.04\,Hz and 0.3\,Hz the observed seismic noise during winter is larger, almost up to an order of magnitude at the secondary microseism peak~\cite{Stutzmann2009,Junca_2019,digiovanni23}. At higher frequencies, no clear seasonal pattern is observed.

\section{LSBB}
\label{sec:LSBB}

In this section, we use the shortest distance pair (GAS-RAM, $\Delta$x $\approx$ 600m) as the reference pair for the LSBB site. For this pair of sensors, we present the coherence and seismic noise percentiles for the NS-NS direction, investigate the seasonal fluctuations as well as the effect of using aligned and perpendicular measurement directions. At the end, we compare the median coherence and median seismic noise for the three pairs at this site.

\begin{figure*}
    \centering
    \includegraphics[width=0.49\textwidth]{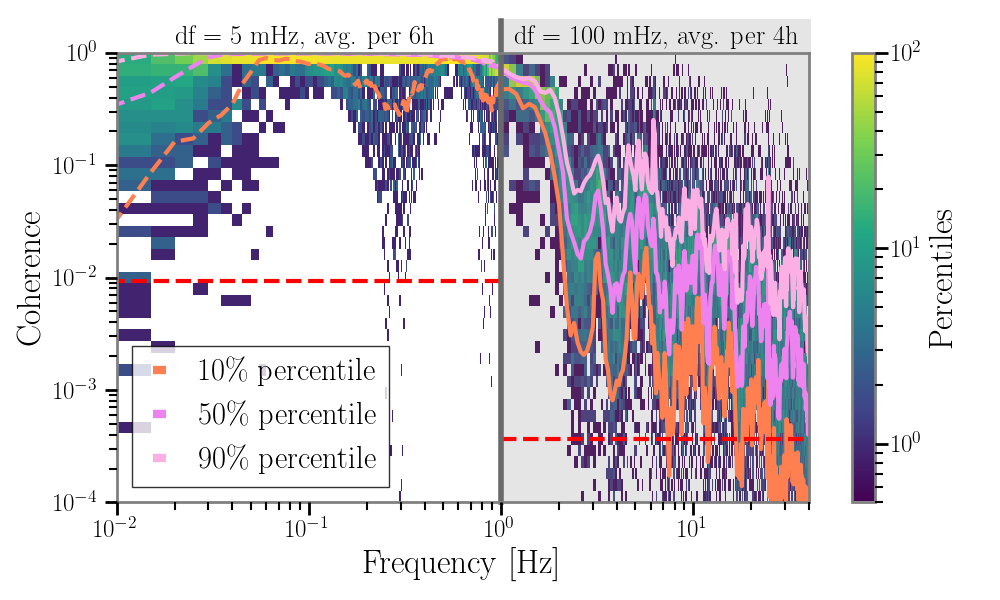}
    \includegraphics[width=0.49\textwidth]{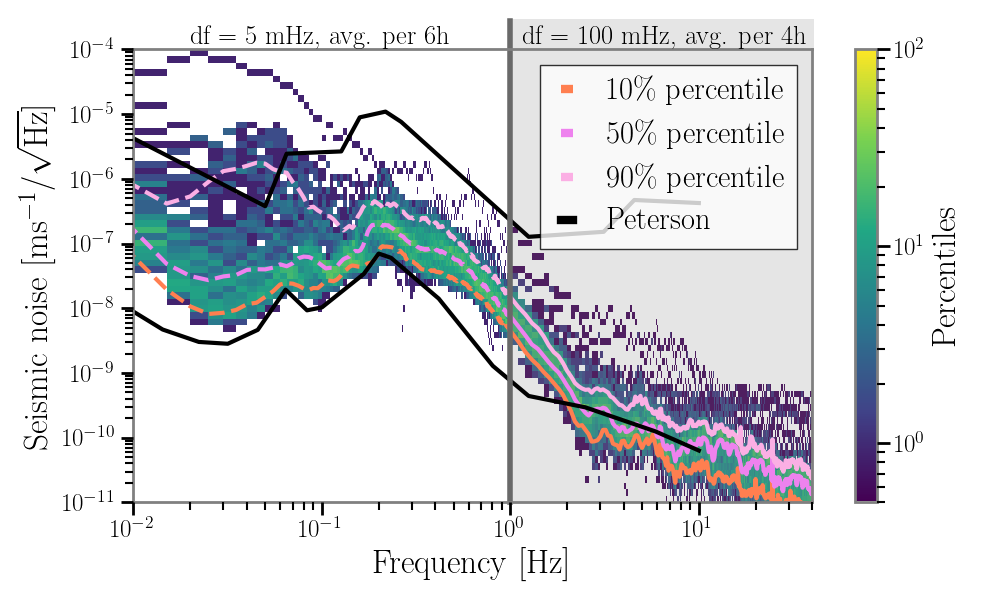}
    \caption{The coherence (left panel) and CSD (right panel) between the underground seismometers (NS components) GAS and RAM (LSBB, FR) during the month of August. 
    The 10$^{\text{th}}$, 50$^{\text{th}}$ and 90$^{\text{th}}$ percentiles are shown in respectively light pink, dark pink and light orange, where dashed (full) lines are used for the analyses with the different parameters $<$ 1Hz ($\geq$ 1Hz). The red dashed line (left panel) represents the level of coherence expected from Gaussian data. The black curves (right panel) represent the low and high noise models by Peterson \cite{Pet1993}.}
    \label{fig:GAS-RAM_CohCSD_5mHz}
\end{figure*}

The left panel of Fig.~\ref{fig:GAS-RAM_CohCSD_5mHz} shows the observed coherence between GAS and RAM. Significant coherence is observed more than 90\% of the time in the entire frequency range of 0.01\,Hz - 20\,Hz. Furthermore, above 20\,Hz at least 50\% of the time significant coherence is observed. A decrease in coherence is observed around 0.3\,Hz for the 10\% percentile. Even though we do not have a clear explanation for this feature, it seems likely there is a site specific noise source at this frequency. The magnitude of this effect depends from month to month with some months being almost unaffected. The accompanying CSD is shown in the right panel of Fig.~\ref{fig:GAS-RAM_CohCSD_5mHz}. We notice that the LSBB site is seismically very quiet, where its CSD sometimes is even lower than Peterson's low noise model. However, note that Peterson's low noise model is derived for power spectral density (PSD) values and not for CSDs. The PSDs of GAS and RAM (not shown here) are quiet above $\sim 2$Hz with the 10\% percentile about a factor 2 or less above Peterson's low noise limit.

\begin{figure*}
    \centering
    \includegraphics[width=0.49\textwidth]{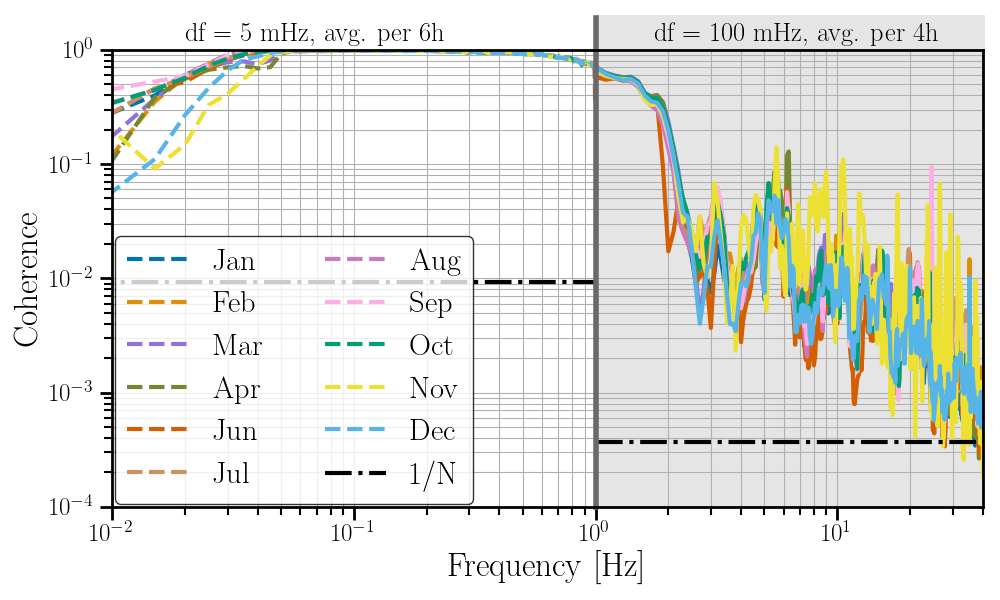}
    \includegraphics[width=0.49\textwidth]{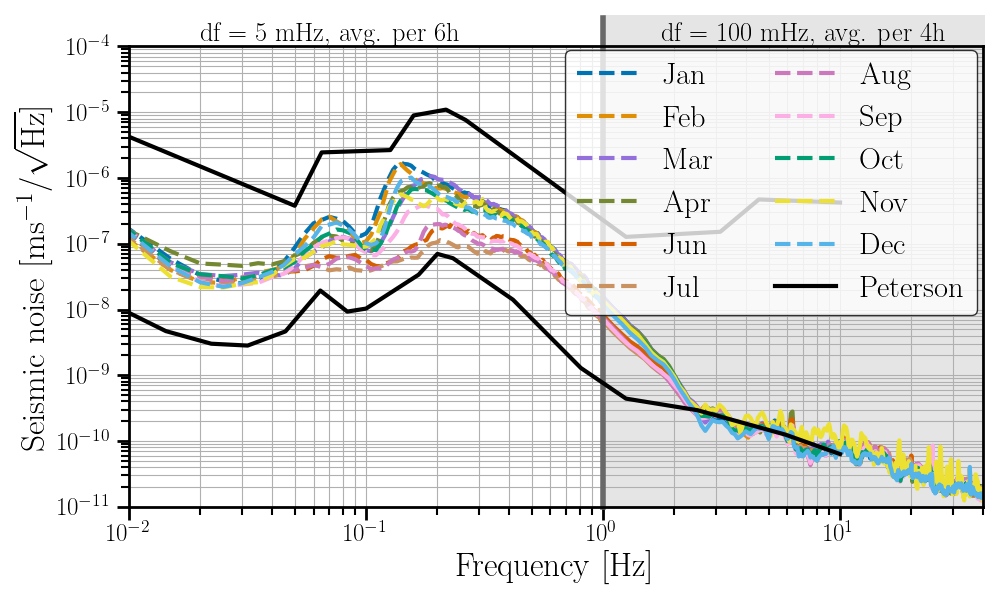}
    \includegraphics[width=0.49\textwidth]{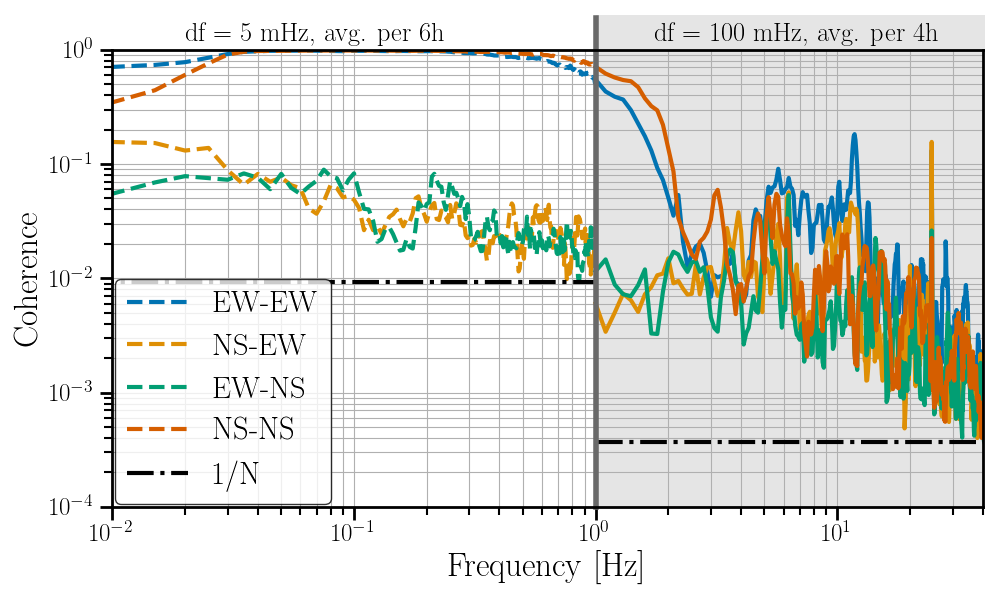}
    \includegraphics[width=0.49\textwidth]{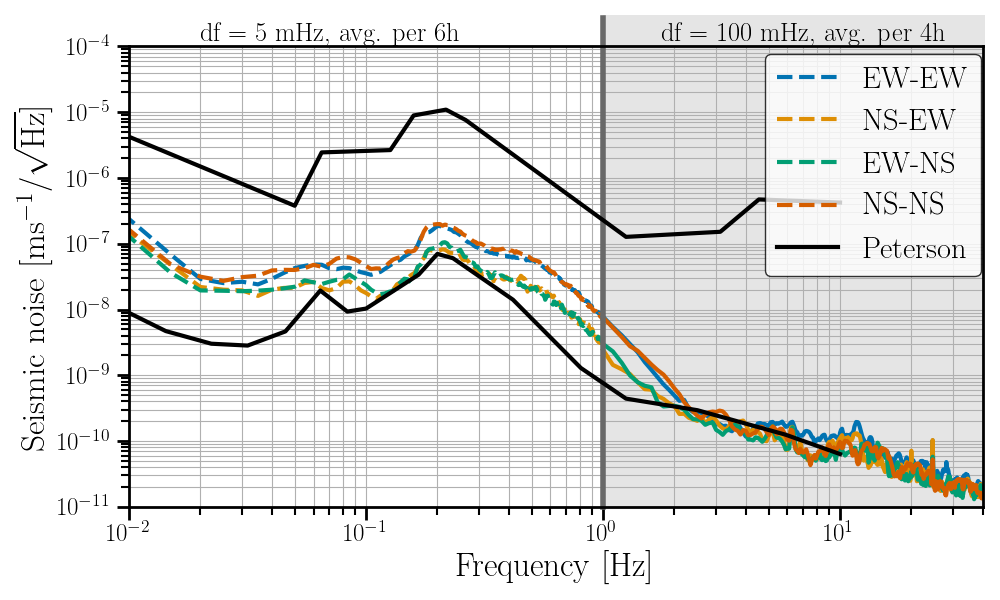}
    \includegraphics[width=0.49\textwidth]{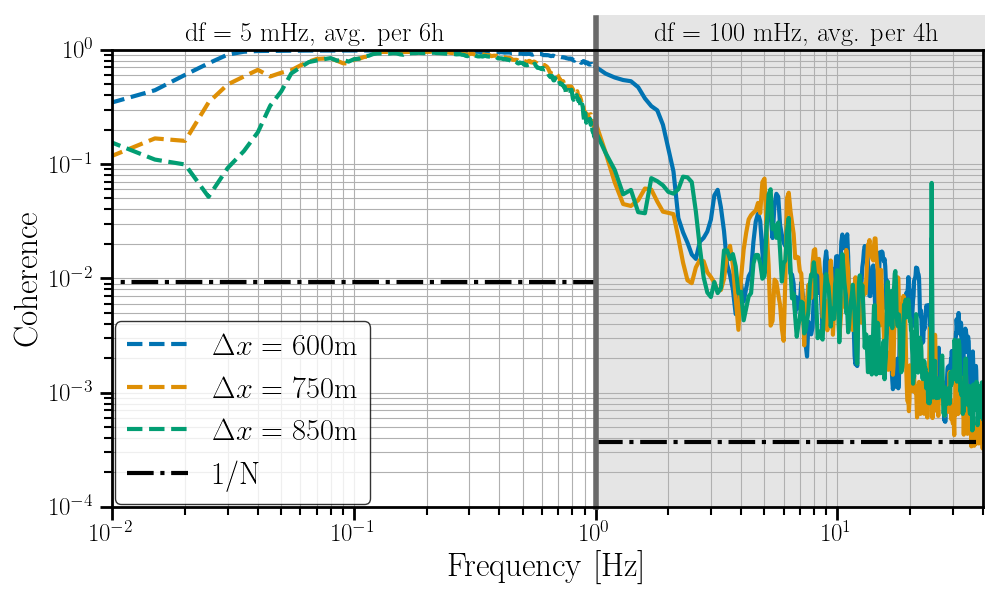}
    \includegraphics[width=0.49\textwidth]{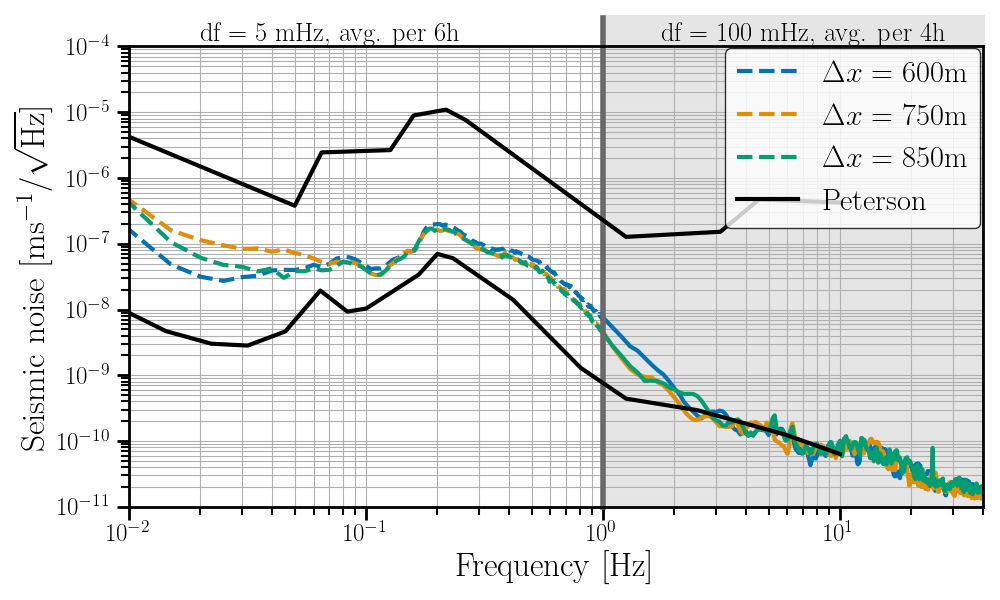}
    \caption{The median coherence (left panel) and CSD (right panel) of the underground seismometers at LSBB as a function of the month of the year (top panels), the sensors orientations (middle panels) and the horizontal separation (bottom panels). For the top four panels the GAS and RAM sensors were used. For the bottom four panels data from the month of August was used. 
    The black dot-dashed line (left panels) represents the level of coherence expected from Gaussian data. In the right panels, the Peterson low and high noise models are shown in black.}
    \label{fig:LSBB_CohCSD_5mHz}
\end{figure*}

The top two panels of Fig.~\ref{fig:LSBB_CohCSD_5mHz} compare median observed coherence (top left panel) and CSD (top right panel) for the different months of the year. Similar to the Homestake analysis, the fluctuation in observed coherence below $\sim$0.04Hz is linked to data processing effects. 
Similar to Homestake, and as in the literature, higher levels of seismic noise is observed during winter months at the microseism peaks (0.05Hz - 0.5Hz). In the case of LSBB this excess seismic noise during winter extends up to $\sim$ 1\,Hz-2\,Hz. This could possibly be explained due to LSBBs close proximity to the Mediterranean sea ($<$ 100 km) whereas Homestake closest ocean is located more than 1500\,km away.
Furthermore, we also would like to note that LSBB is near to underground natural water masses.

In the two middle panels of Fig. \ref{fig:LSBB_CohCSD_5mHz} we compare the coherence (middle left panel) and CSD (middle right panel) for the four different combinations one can make between the seismic wave measurements in the horizontal plane: NS-NS, NS-EW, EW-NS and EW-EW. Similar to the results from Homestake presented in earlier work \cite{PhysRevD.106.042008} \footnote{Note that in \cite{PhysRevD.106.042008} they only presented similar results for frequencies above 0.05\,Hz.}, there is no difference observed in either coherence or CSD for frequencies above $\sim$2\,Hz. For lower frequencies, the coherence is lower when correlating perpendicular observing directions of the two different sensors. However, the observed coherence is still significant for at least 50\% of the time and the CSD is at most a factor two smaller for the perpendicular orientations.

As can be seen in the bottom left panel of Fig. \ref{fig:LSBB_CohCSD_5mHz}, we find a decreased seismic coherence between 0.4\,Hz and 2\,Hz for more distant sensors.
This is similar to what was found for underground seismic coherence at Homestake in earlier work\cite{PhysRevD.106.042008}. 
The CSD presented in the bottom right panel of Fig. \ref{fig:LSBB_CohCSD_5mHz} experiences a minimal, to negligible effect with respect to distance in the frequency band 0.4\,Hz and 2\,Hz. Above 2\,Hz, no effect is observed.
Additionally below 0.1\,Hz sensors separated by a larger distance seem to observe lower coherence. The seismic correlated noise in this frequency region is different for the different pairs, however their is no clear pattern with respect to the horizontal separation.

\section{Sos Enattos}
\label{sec:SosEnattos}

\begin{figure*}
    \centering
    \includegraphics[width=0.49\textwidth]{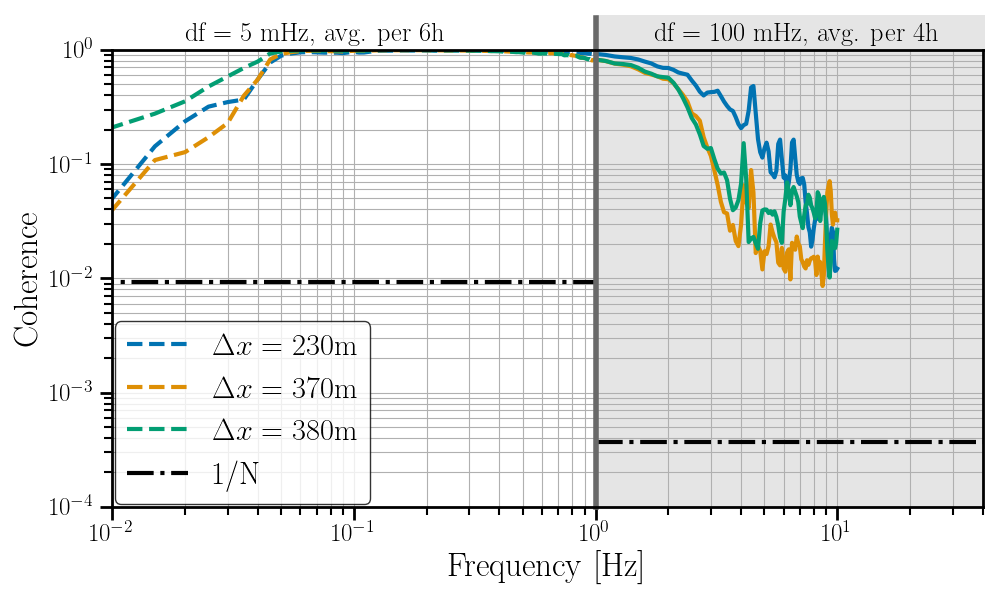}
    \includegraphics[width=0.49\textwidth]{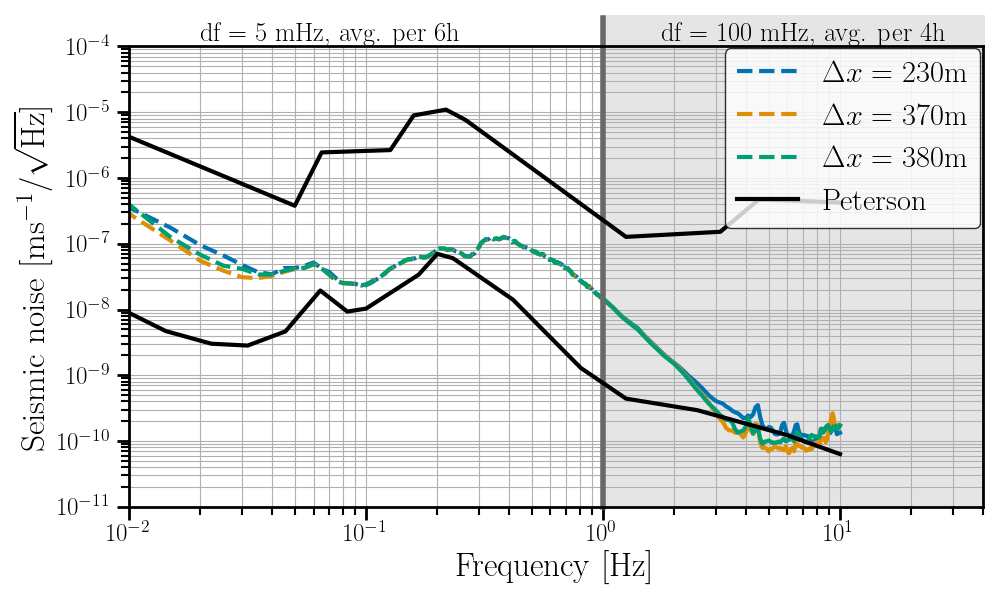}
    \caption{The median coherence (left panel) and CSD (right panel) of the underground seismometers at Sos Enattos as a function of distance for Aug 2021. 
    The black dot-dashed line (left panel) represents the level of coherence expected from Gaussian data. The Peterson low and high noise models are shown in black (right panel).}
    \label{fig:SosEnattos_distance_CohCSD_5Hz}
\end{figure*}

The coherence between the different sensors with a horizontal separation of several hundreds of meters is shown in the left panel of Fig.~\ref{fig:SosEnattos_distance_CohCSD_5Hz}. Even though the observed coherence below $\sim$0.05\,Hz is different, there seems to be no distance dependant relationship. As discussed earlier, this is most likely a non-physical effect.
Between 1\,Hz and $\sim$ 7\,Hz, the observed coherence for the shortest distance pair is higher compared to the two other pairs. The correlations in the seismic noise, see right panel of Fig. \ref{fig:SosEnattos_distance_CohCSD_5Hz}, for this shortest distance pair of sensors are (marginally) larger in the frequency range 3\,Hz-7\,Hz, compared to the other sensor pairs.

Apart from these three pairs of sensors with a horizontal separation of several hundreds of meters, we also analysed two sensors (P2 and P3) which are located approximately 10\,km from each other. This distance is both relevant for the ET as well as atom interferometers. It is namely the distance between two ET end stations in the triangular baseline and the approximate distance scale on which multiple atom gradiometers are deployed in the ELGAR detector, respectively. Furthermore, for both the triangular and two L baseline for the ET, this is the distance between the input and output optics of one single interferometer.

\begin{figure*}
    \centering
    \includegraphics[width=0.49\textwidth]{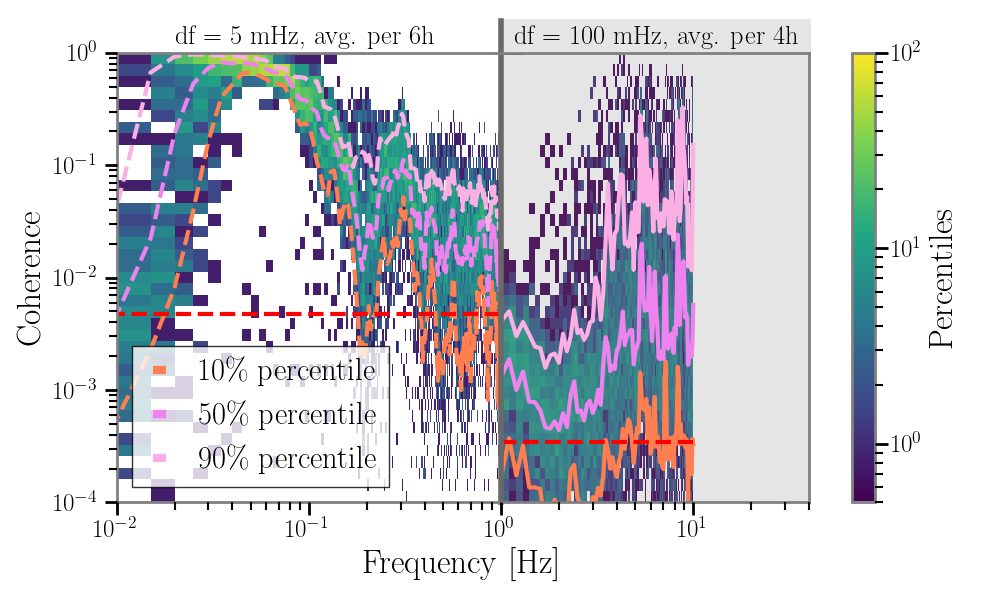}
    \includegraphics[width=0.49\textwidth]{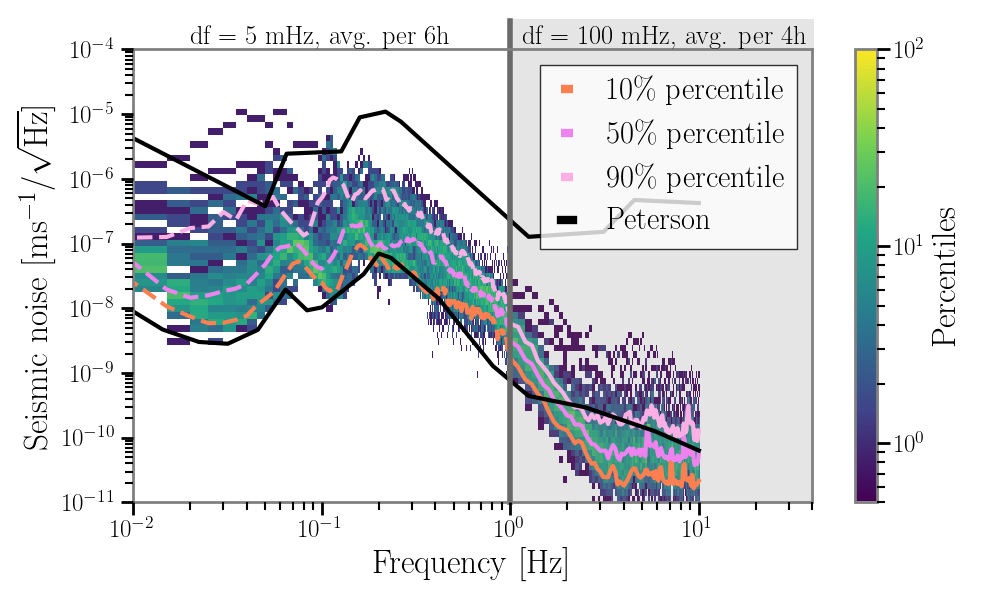}
    \caption{The coherence (left panel) and CSD (right panel) between the underground seismometers (NS components) P2 and P3 (Sos Enattos, IT) with an approximate horizontal separation of 10km and depth of $\geq$ 250m. 
    The 10$^{\text{th}}$, 50$^{\text{th}}$ and 90$^{\text{th}}$ percentiles are shown in respectively light pink, dark pink and light orange, where dashed (full) lines are used for the analyses with the different parameters $<$ 1Hz ($\geq$ 1Hz). The red dashed line (left panel) represents the level of coherence expected from Gaussian data. The black curves (right panel) represent the low and high noise models by Peterson \cite{Pet1993}.}
    \label{fig:P2-P3_cohCSD}
\end{figure*}

For this long distance pair of underground seismometers, we find 90\% (50\%) of the time significant coherence in the frequency range 0.02\,Hz-0.3\,Hz (0.01\,Hz-1\,Hz) as shown in the left panel of Fig.~\ref{fig:P2-P3_cohCSD}. Between 2\,Hz and 10\,Hz, there are a number of spectral features which lead to significant coherence over a broader frequency range. Some of these frequencies are most likely caused by electromagnetic interference affecting the digitizer and/or cabling. Such an example is the line at 8.3\,Hz which is the modulation frequency of the Italian GSM network, i.e. GHz signal packets are transmitted with a frequency of 8.3\,Hz~\cite{10.3389/fphy.2023.1094921}. It is likely the other features are from non-seismic origin as well. However, additional research should further investigate this excess coherence and exclude any coherence being from seismological origin.

The right panel of Fig.~\ref{fig:P2-P3_cohCSD} shows the accompanying levels of correlated seismic noise.

\section{Euregio Maas-Rhein}
\label{sec:Terziet}

For EMR, we only present results for frequencies above 0.2\,Hz as one of the sensors (CTSN) is dominated by sensor self-noise for lower frequencies. Furthermore the sensors at EMR are only sampled at 40\,Hz and the data has a rapidly decreasing sensitivity for frequencies above 18Hz.

\begin{figure*}
    \centering
    \includegraphics[width=0.49\textwidth]{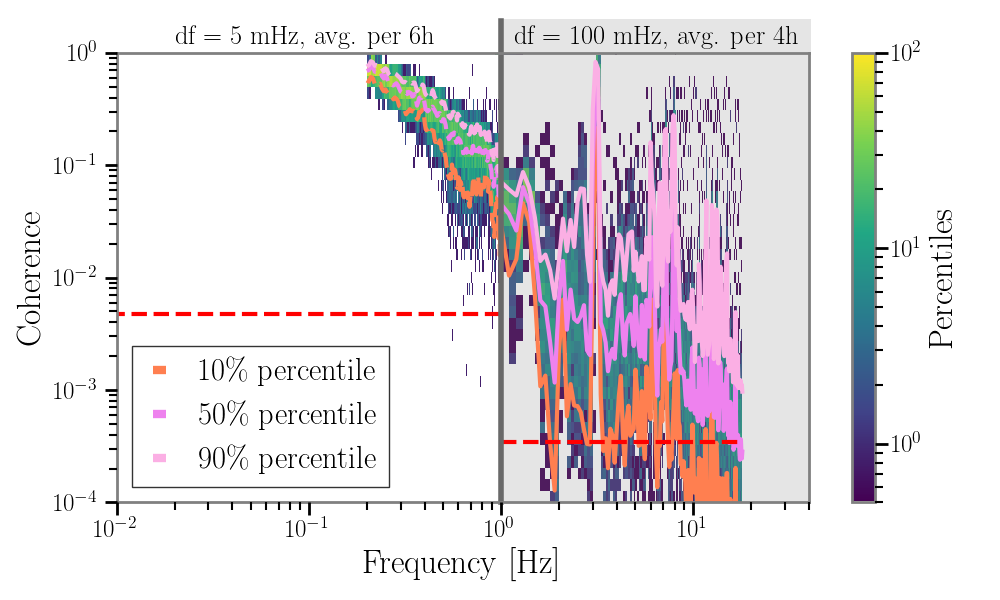}
    \includegraphics[width=0.49\textwidth]{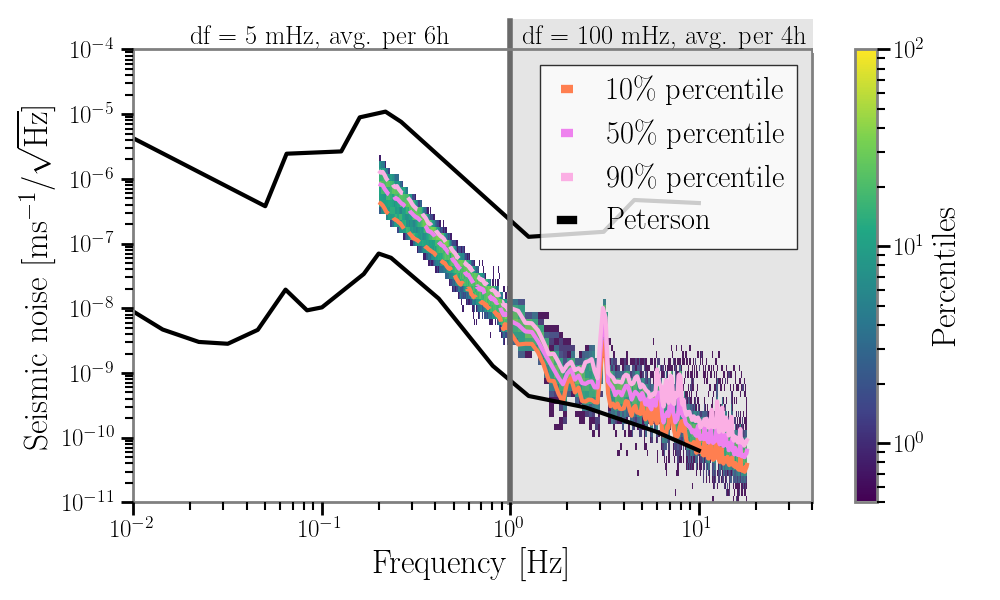}
    \caption{The coherence (left panel) and CSD (right panel) between the underground seismometers (NS components) TERZ and CTSN (EMR, NL) with an approximate horizontal separation of $\sim$2.4km and depth of $\sim$250m. (Jan 2023) 
    The 10$^{\text{th}}$, 50$^{\text{th}}$ and 90$^{\text{th}}$ percentiles are shown in respectively light pink, dark pink and light orange, where dashed (full) lines are used for the analyses with the different parameters $<$ 1Hz ($\geq$ 1Hz). The red dashed line (left panel) represents the level of coherence expected from Gaussian data. The black curves (right panel)represent the low and high noise models by Peterson \cite{Pet1993}.}
    \label{fig:TERZ-CTSN_CohCSD_100mHz}
\end{figure*}

As can be expected due to the larger horizontal separation between the sensors, the observed coherence is lower compared to the other sites. However, as shown in the left panel of Fig. \ref{fig:TERZ-CTSN_CohCSD_100mHz}, 90\% of the time there is still significant coherence up to $\sim$ 2Hz, as well as at several highly coherent frequencies above 2\,Hz. Furthermore, 50\% of the time there is significant coherence up to $\sim$ 16Hz. The accompanying CSD is shown in the right panel of Fig.~\ref{fig:TERZ-CTSN_CohCSD_100mHz}.

\section{Discussion}
\label{sec:SiteComparison}

In Sec. \ref{sec:Homestake}-\ref{sec:Terziet} we introduced the data for the four different sites considered in this study. Here we compare the data from the different sites during the months of January. Similar results for August are provided in the Appendix. 
Making this comparison is very challenging due to the large variety of parameters which are not the same between the different measurements at different sites. Often, more than one of the following relevant parameters are different for each site: location, geographical topology, sensor separation, sensor type, sensor depth, sampling frequency, different year, etc.
Based on the available data we try to make some general conclusions by comparing and combining all the seismic data from the different sites. However to really further understand the effect of each individual parameter such as sensor depth or separation etc, a systematic study should be performed where these parameters are carefully controlled for. Currently, such a study is under development at LSBB.

Given all these different parameters the results below should not be considered as a site comparison but rather as a demonstration of possible ranges of the figures of merit involved. We use the following sensor pairs for the different sites: D2000-E2000 (Homestake), GAS-RAM (LSBB), SOE2-SOE3(Sos Enattos) and TERZ-CTSN (EMR).

\begin{figure*}
    \centering
    \includegraphics[width=0.49\textwidth]{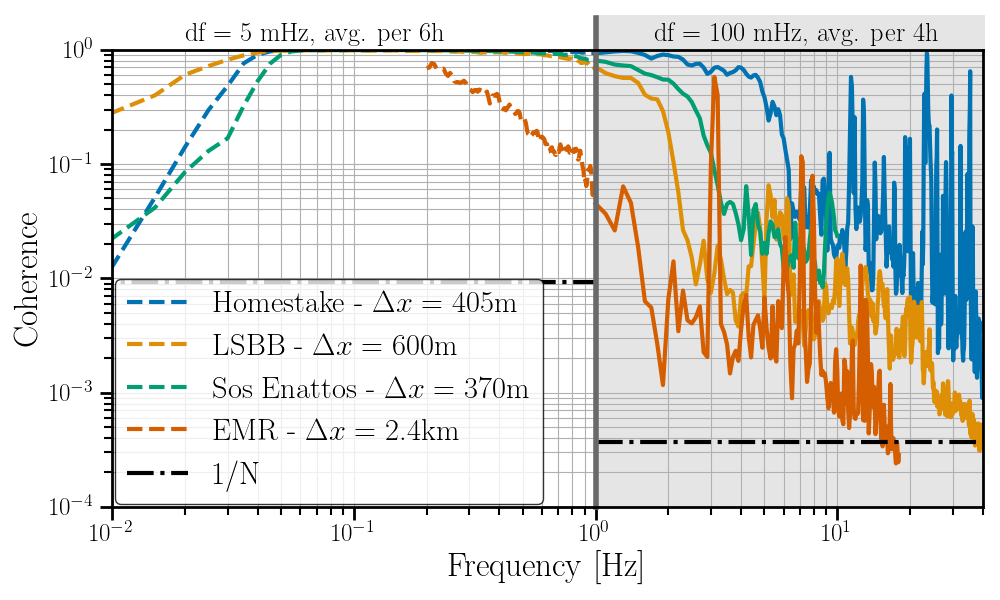}
    \includegraphics[width=0.49\textwidth]{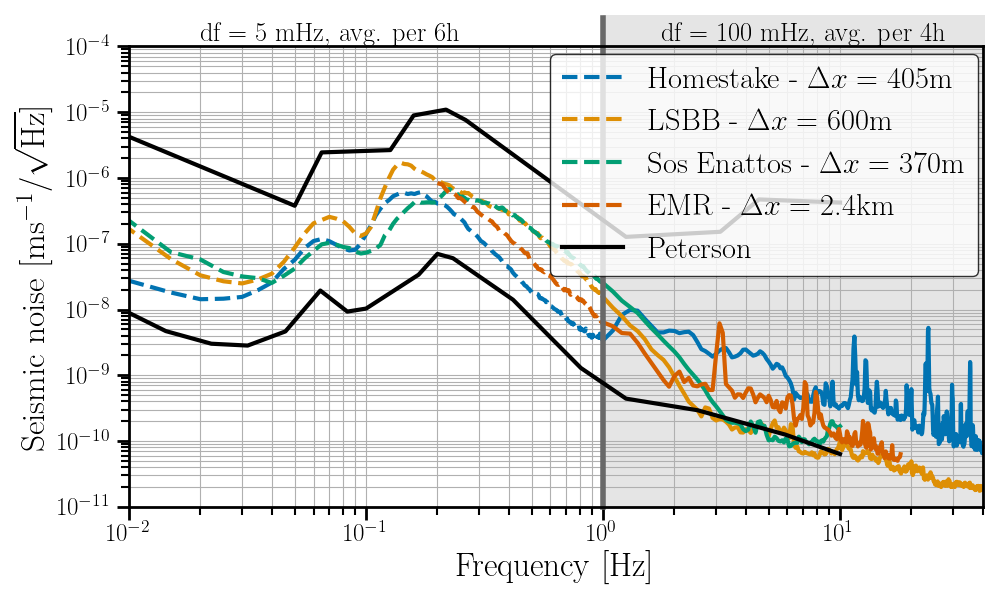}
    \caption{The median coherence (left panel) and CSD (right panel) of the underground seismometers for the different geographical locations studied in this paper for the month of January. 
    Fore more details on the sensors specifications, see Tab. \ref{tab:sensors}.
    Note: the data is not from the same year. The data $<$1Hz (dashed curves) are analysed using 200 second long segments which are averaged per 6\,h-window. Above 1Hz the data (full curves) are analysed using 10 second long segments which are averaged per 4\,h-window. 
    The black dot-dashed line (left panel) represents the level of coherence expected from Gaussian data. The Peterson low and high noise models are shown in black (right panel.}
    \label{fig:Comparison_Jan_distance_CohCSD_100mHz}
\end{figure*}

Based on the coherence of the different sites represented in the left panel of Fig.~\ref{fig:Comparison_Jan_distance_CohCSD_100mHz}, we conclude that we observe significant coherence at least 50\% of the time for frequencies between 0.01\,Hz and 40\,Hz for underground seismometers with a sub kilometer separation. Even in the case of a separation of $\sim$2.4km, as is the case for TERZ-CTSN, we observe significant coherence 50\% of the time up to frequencies of about 16\,Hz. Furthermore, we would like to point out that the coherence of EMR is high around the microseism peak. In Sec.~\ref{sec:SosEnattos}, we showed that even on distance scales of $\sim$ 10\,km, the seismic noise between 0.01\,Hz and 1\,Hz is coherent at least 50\% of the time.
Future investigations might be needed to probe such kilometer-long distance scales in the low frequency region in more detail. However, this is already a first order demonstration that future kilometer-long baseline atom interferometers, such as ELGAR~\cite{Canuel:2019abg}, could potentially be impacted by correlated seismic and NN and should investigate this in more detail.

When looking at the observed correlated seismic noise as shown in the right panel of Fig. \ref{fig:Comparison_Jan_distance_CohCSD_100mHz}, we find that the levels of correlated seismic noise for all the different sites typically does not differ by more than one order of magnitude.
Below 1Hz Homestake has the lowest levels of seismic noise, apart for the frequency range 0.07\,Hz-0.3\,Hz when the seismic noise at Sos-Enattos is even lower. 
This is in line with the expectation that the seismic noise from ocean waves is lower at Homestake due to the large distance to the closest ocean, whereas the Sos-Enattos and LSBB sites are in (very) close proximity to the Mediterranean sea. The reason why Sos Enattos has the lowest CSD between 0.07\,Hz and 0.3\,Hz, might be linked to the fact that the secondary microseism peak seems to reach its maximum at a slightly higher frequency. This might be due to many different factors which vary for the diverse different sites.

At higher frequencies, Homestake becomes the noisiest site.
At the other end, LSBB is an extremely quiet site in the high frequency region and Sos Enattos has low levels of correlated seismic noise up to $\sim$8Hz. However, for the latter, the levels of correlated noise increases between 8Hz and 10Hz, even though this correlation may be due to a non-seismic origin as stated in section \ref{sec:SosEnattos}. Although the seismometers at the EMR site are located at a much larger separation from each other, the observed levels of correlated seismic noise is somewhere in between the different sites.

\section{Correlated Newtonian noise}
\label{sec:CorrNN}

We calculate the levels of NN from body waves in an identical way to earlier work \cite{PhysRevD.106.042008}, i.e. we use Eq.~\ref{eq:BWNN}. Similar to this previous work we assume the bulk density $\rho_{0,\text{Bulk}}$ to be $2800 \text{kg\,m}^{-3}$. Additionally, we assume $p=1/3$, which accounts for the different mixing ratio of P- and S-waves. Choosing the correct value of p is non-trivial, as it strongly depends on the seismic sources (far or close) as well as on the geology. It could be argued that p should be chosen by relying on the equipartitioning of the energy in the assumption of a diffuse field. For a diffuse field the value of p depends on P- and S-wave velocities \cite{f24a9b6cf61740668d35dac9c3c317ab}. However, the presence of close seismic sources makes it difficult to make any assumptions by exploiting the equipartition of the
energy. Therefore we use $p=1/3$ as in this way the correlated NN is at most off by a factor 2 for
both of the most extreme values of p=0 and p=1 (see Eq. \ref{eq:BWNN}).
Note, the effectiveness with which NN can be subtracted depends on the actual value of p \cite{Badaracco_2024}. We purposefully do not use more accurate values, which depends on the site, as the goal of this paper is to probe a site independent order of magnitude rather than providing accurate site specific results. $S_{\xi_x}$ represents the PSD, or in our case the CSD, of the displacement caused by the body-waves along the arm direction and L is the length of the interferometer.

\begin{equation}\label{eq:BWNN}
    S_{\text{Body-wave}}(f) = \left(\frac{4\pi}{3}G\rho_{0,\text{Bulk}} \right)^2 (3p+1) \frac{1}{L^2(2\pi f)^4}S_{\xi_x}(f)
\end{equation}

\subsection{The search for a GWB}

The search for an (isotropic) GWB is very sensitive, if not the most, to correlated noise sources. Therefore, we now describe what one tries to measure when looking for an isotropic GWB and how we can project the effect of correlated noise on this figure of merit. As stated earlier, this projection is only considered to be relevant for the triangular design of the ET. The seismic noise is deemed to be uncorrelated in the relevant frequency band for the two L design, due to the large separation between the separate L-shaped interferometers.

When searching for an isotropic GWB one typically tries to measure its energy density, $\text{d}\rho_{\rm GW}$, contained in a logarithmic frequency interval, $\text{d}\ln f$. Furthermore one divides by the critical energy density $\rho_{\rm c} = 3H_0^2c^2/(8\pi G)$ for a flat Universe to construct a dimensionless figure of merit $\Omega_{\rm GW}(f)$~\cite{Christensen_2018,PhysRevD.46.5250,PhysRevD.59.102001,LivingRevRelativ20}:

\begin{equation}
    \label{eq:omgeaGWB}
    \Omega_{\rm GW}(f) = \frac{1}{\rho_{\rm c}}\frac{\text{d}\rho_{\rm GW}}{\text{d}\ln f}~,
\end{equation} 
where $H_0$ is the Hubble-Lemaître constant, $c$ is the speed of light and $G$ is Newton's constant. We use the 15-year Planck value of 67.9 km s$^{-1}$ Mpc$^{-1}$ for $H_0$~\cite{Planck:2015fie}. 

When searching for an isotropic, Gaussian, stationary and unpolarized GWB, one can construct the cross-correlation statistic $\hat{C}_{IJ}(f)$,
\begin{equation}
    \label{eq:cross-correlationstatistic}
\hat{C}_{IJ}(f) = \frac{2}{T_{\textrm{obs}}} \frac{{\rm{Re}}[\Tilde{s}^*_I(f)\Tilde{s}_J(f)]}{\gamma_{IJ}(f)S_0(f)}~,
\end{equation}
which is an unbiased estimator of $\Omega_{\rm GW}(f)$ in the absence of correlated noise~\cite{PhysRevD.59.102001,LivingRevRelativ20}. $I$ and $J$ represent the two interferometers and $\Tilde{s}_I(f)$ is the Fourier transform of the time domain strain data $s_I(t)$ measured by interferometer $I$. $\gamma_{IJ}$ is the normalized overlap reduction function which encodes the baseline's geometry~\cite{PhysRevD.46.5250,Romano:2016dpx}. $S_0(f)$ is a normalisation factor given by $S_0(f)=(9H_0^2)/(40\pi^2f^3)$ and $T_{\textrm{obs}}$ is the total observation time of the data-collecting period\footnote{The normalisation factor $S_0(f)$ for ET differs from that one of e.g. LIGO-Virgo-KAGRA by a factor of 3/4, due to the different opening angle between the interferometers' arms ($\pi$/2 for LIGO-Virgo-KAGRA and $\pi$/3 for ET) \cite{Romano:2016dpx}.}.

In line with earlier studies on the impact of correlated noise on the ET \cite{PhysRevD.104.122006,PhysRevD.106.042008} we refer to the three different ET interferometers as ${\rm ET}_1, {\rm ET}_2, {\rm ET}_3$, which we assume to have identical sensitivity. Furthermore we neglect the difference in $\gamma_{IJ}$ between the baseline pairs $IJ =  {\rm ET}_1{\rm ET}_2; {\rm ET}_1{\rm ET}_3; {\rm ET}_2{\rm ET}_3$, since the relative difference between the overlap reduction functions of the different arms is smaller than $5\times 10^{-7}$ for frequencies under 1 kHz ~\cite{PhysRevD.104.122006}. In the remainder of the paper, we use the ${\rm ET}_1{\rm ET}_2$-baseline as our default observing baseline.

Similar to earlier work~\cite{PhysRevD.106.042008}, we can construct equivalent cross-correlation statistics for the correlated NN:

\begin{equation}
    \label{eq:C_Seis_NN}
    \begin{aligned}
        \hat{C}_{{\rm NN},{\rm ET}_1{\rm ET}_2}(f) \frac{{\rm S}_{\rm Body-wave}}{\gamma_{{\rm ET}_1{\rm ET}_2}(f)S_0(f)},\\
    \end{aligned}
\end{equation}
where ${\rm S}_{\rm Body-wave}$ was introduced in Eq \ref{eq:BWNN}.

The sensitivity of a search for an isotropic GWB can be related to the instantaneous sensitivity of the ET interferometer, referred to as the one-sided amplitude spectral density (ASD) $P_{\rm ET}(f)$, as follows~\cite{PhysRevD.46.5250,PhysRevD.59.102001,LivingRevRelativ20}:

\begin{equation}
    \label{eq:sigmaGWB}
    \sigma_{{\rm ET}_1{\rm ET}_2}(f) \approx \sqrt{\frac{1}{2T_{\rm obs}\Delta f}\frac{P_{\rm ET}^2(f)}{\gamma_{{\rm ET}_1{\rm ET}_2}^2(f)S_0^2(f)}}~,
\end{equation}
with $\Delta f$ the frequency resolution. Here we have assumed identical sensitivity in the different ET interferometers ${\rm ET}_1, {\rm ET}_2, {\rm ET}_3$. $\sigma_{{\rm ET}_1{\rm ET}_2}(f)$ is the standard deviation on the cross-correlation statistic defined in Eq.~\ref{eq:cross-correlationstatistic}, in the small signal-to-noise ratio (SNR) limit and absence of correlated noise. Because the GWB one tries to observe is very weak, the former is a realistic assumption. The effect of the latter is the focus of this section.

As many of the expected signals for an isotropic GWB behave as a power-law, a more appropriate sensitivity to such a signal than $\sigma_{{\rm ET}_1{\rm ET}_2}(f)$ would be one that takes into account this broadband character of the expected signal. Such a broadband sensitivity is given by the so called power-law integrated (PI) curve: $\Omega^{\rm PI}_{{\rm ET}}(f)$. $\Omega^{\rm PI}_{{\rm ET}}(f)$ is constructed using $\sigma_{{\rm ET}_1{\rm ET}_2}(f)$ such that at any frequency a power-law signal $\Omega_{\rm GW}(f)$ with an SNR of 1 for the ${\rm ET}_1{\rm ET}_2$ baseline is tangent to this PI-sensitivity curve~\cite{Thrane:2013oya}. This makes $\Omega^{\rm PI}_{{\rm ET}}(f)$ the relevant figure of merit to identify correlated broadband noise sources that could impact the search for an isotopic GWB.

\subsection{Impact of correlated NN on the search for a GWB}

\begin{figure}
    \centering
    \includegraphics[width=\linewidth]{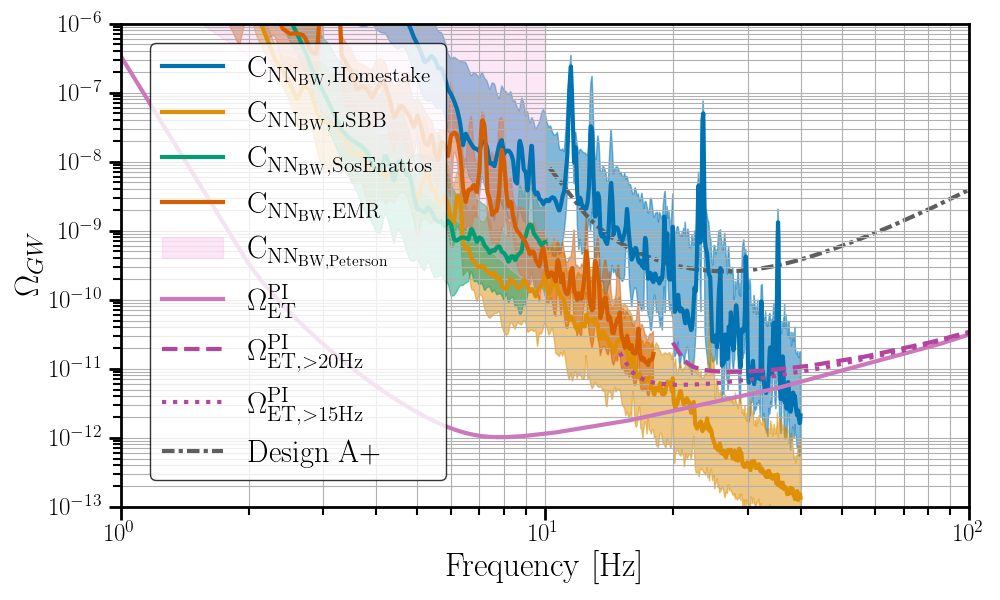}
    \caption{The projected impact from correlated NN from body-waves, as calculated in this section, for the seismic data from the month of January for the different locations, see the text for details on the used sensors, their distances and depths. 
    As a comparison we make the same projection using the Peterson low noise and high noise models. For the broadband ($\Omega^{\rm PI}_{{\rm ET}}$) sensitivity to a GWB we assumed one year of observation time (100\% duty cycle). The one year PI curve of the A+ design for the LIGO Hanford LIGO Livingston and Virgo detectors is represented by the dot-dashed curve. This curve was obtained using the open data provided by the LVK collaborations \cite{O3IsotropicDataset} and was first presented in \cite{KAGRA:2021kbb}. Please note: in this paper we present the 1$\sigma$ PI-curve, whereas in \cite{KAGRA:2021kbb} the 2$\sigma$ PI-curve is shown.}
    \label{fig:StochBudget_Jan_NN}
\end{figure}

 With Fig.~\ref{fig:StochBudget_Jan_NN} we present how correlated NN caused by the observed seismic correlations presented earlier in this paper affects the search for an isotropic GWB with the ET. Fig.~\ref{fig:StochBudget_Jan_NN} shows the noise budget using the seismic data from the month of January.
Since in the region of interest for the ET (i.e. $>$1Hz) there is little difference in the amplitude of seismic noise across seasons, we do not provide the results for August in the main text. However, for completeness, you can find these results in Fig. \ref{fig:StochBudget_Aug_NN} in the Appendix. 
Furthermore, we highlight that these budgets are calculated identically to the Homestake results in \cite{PhysRevD.106.042008} as well as the results in \cite{Branchesi:2023mws}.
The data used for the NN budgets in Fig. \ref{fig:StochBudget_Jan_NN} is the same as the data used in Fig. \ref{fig:Comparison_Jan_distance_CohCSD_100mHz}.

In line with the results presented in \cite{PhysRevD.106.042008}, correlated NN from body-waves, assuming a seismic environment as observed at Homestake, would dominate the power-law integrated sensitivity curve for broadband GWB signals up to $\sim$ 40Hz about 50\% of the time. This would limit ETs sensitivity to isotropic GWB signals up to 20\,Hz-30\,Hz to levels similar as planned to be achieved by LIGO and Virgo during their fifth observing run (Design A+). Even in the most optimistic scenario as discussed by the projection of the correlated NN based on observed correlations at LSBB, the search for an isotropic GWB would be limited by correlated NN up to $\sim$ 20\,Hz for at least 50\% of the time. 
In \cite{Branchesi:2023mws} a couple of assumptions were used to get an estimate of the lowest possible levels of correlated NN affecting the search for a GWB. For this, the authors of \cite{Branchesi:2023mws} multiplied Peterson's low noise limit with the observed coherence at Homestake as shown in \cite{PhysRevD.106.042008}. This lead to a minimal impact of correlated NN above $\sim$ 10Hz. However, based on the results described in this paper, it seems that this assumption is overly optimistic. Neither of the two candidate sites, Sos Enattos and EMR, has correlated seismic noise that low. Even the seismically quiet site LSBB has considerably higher levels of correlated NN which are at least one order of magnitude larger at 10\,Hz. 

To understand the impact of this noise source on the analysis for an isotropic GWB it is important to understand what the PI curve reflects. As stated earlier, this integrates over $\sigma_{{\rm ET}_1{\rm ET}_2}(f)$ as a function of frequency. 
Consider a broadband power-law signal: $\Omega_{\mathrm{GW}}(f) = \Omega_{\mathrm{ref}} \left(\frac{f}{f_{\mathrm{ ref}}} \right)^{\alpha}$. For all negatively sloped (i.e. $\alpha<0$) signals, the correlated NN noise presents a significant problem seriously limiting the science potential. However, many of the expected signals have positive power-law slopes. Some examples are the GWB from unresolved compact binary coalesence (CBC) events ($\alpha=2/3$) \cite{MM_PartI, PhysRevLett.116.131102, Regimbau_2011} and the GWB from core collapse supernovae ($\alpha=3$) \cite{10.1046/j.1365-8711.1999.02194.x}. Multiple cosmological GWBs, e.g. cosmic strings \cite{LIGOScientific:2021nrg}, predict a flat GWB ,i.e. $\alpha\approx 0$.
For such signals the impact of the correlated noise might be less dramatic.
Not using the data below 20Hz (15Hz) out of concern of noise contamination results in the dark purple dashed (dotted) PI curves in Fig.~\ref{fig:StochBudget_Jan_NN}. The $>$20Hz ($>$15Hz) PI is not affect by the LSBB correlated NN levels up to the 90\% (50\%) percentiles. This illustrates that in certain circumstances it might be beneficial to disregard the low frequency data in the case of correlated noise contamination. 
In searching for power-law signals with large positive slopes such as $\alpha=3$, the low frequency regime (and therefore potential contamination from correlated NN noise) is irrelevant as the dominant contribution to the detectabilty comes from frequencies above 100Hz. For $\alpha=0$ ($\alpha=2/3$) one loses a factor $\sim$6-9 ($\sim$3-4) in starting the analysis at 15Hz-20Hz rather than at 1Hz to avoid contamination from correlated noise. This also implies that for certain searches (e.g. $\alpha=0$) it might be more beneficial to place aggressive data quality vetoes on times with large ambient seismic noise and remove up to 50\% of the data to lower the contamination of correlated noise. Namely, the gain from adding more data scales as $\sqrt{t}$, with $t$ the total observation time and might be outweighed by the gain of reduced noise contamination. See TABLE~\ref{tab:PIs} for a summary.

\begin{table*}[]
\begin{tabular}{|l|l|l|l|}
\hline
 & $\Omega^{\rm PI}_{{\rm ET}\geq{\rm 1Hz}}$ & $\Omega^{\rm PI}_{{\rm ET}\geq{\rm 15Hz}}$ & $\Omega^{\rm PI}_{{\rm ET}\geq{\rm 20Hz}}$  \\ \hline
 $\alpha = 0$ & 1e-12 & 6e-12 & 9e-12  \\ \hline
 $\alpha = 2/3$ & 2.1e-12 & 6e-12 &  7.8e-12   \\ \hline  
 $\alpha = 3$ & 2e-13 & 2e-13 &  2e-13   \\ \hline  
\end{tabular}
\caption{The approximate amplitude $\Omega_{\mathrm{GW}}(f)$ of the power-law signal tangent to the 1$\sigma$ PI-curves (as shown in Fig. \ref{fig:StochBudget_Jan_NN}) with different starting frequencies of the analysis. We use $f_{\rm ref}=$25Hz.}
\label{tab:PIs}
\end{table*}

Note, here we we have not assumed any level of noise subtraction. However, for the noise subtraction as discussed in earlier work \cite{PhysRevD.86.102001,PhysRevD.92.022001,Coughlin_2014_seismic,Coughlin_2016,PhysRevLett.121.221104,Tringali_2019,Badaracco_2020,Badaracco_2019,NN_Sardinia2020,10.1785/0220200186,Bader_2022,Koley_2022, digiovanni23}, one typically assumes to construct a Wiener filter. For frequencies above several Hz, the NN of body-waves is below the detector sensitivity on the typical timescales over which the Wiener filter is calculated. The noise sources only becomes problematic for the search for a GWB as in this case one correlates data over very long timescales of $\mathcal{O}$(1 yr) over which correlated noise sources can accumulate significance. Future research should further investigate the efficiency of these noise subtraction techniques if the NN noise is subthreshold compared to the detector sensitivity when determining the Wiener filter.

\section{Glitch study}
\label{sec:Glicth}

After the presentation of the earlier study investigating the impact of correlated NN based on seismic observations at Homestake \cite{PhysRevD.106.042008}, some concerns were raised to which extent the multi-hour long averages were dominated by a limited number of short but (very) loud time periods. To address these concerns in this paper, we perform a study of transient seismic noise in this section. For this study, we were inspired by an earlier study of the seismic glitchiness at Sos Enattos \cite{Allocca:2021opl}. There, they investigate the impact of seismic glitches on the inspiral signal of an intermediate mass black hole binary within a segment of 1 min\footnote{As described in \cite{Allocca:2021opl} the exact signal length depends on many parameters of the system and can range from ten seconds to many tens of seconds. The window of 1 min was chosen as a compromise between the shorter and longer signal duration and only serves as an indicative figure of merit.}.
Therefore, in this paper we are looking at the seismic glitchiness on one-minute time segments. For this study, we use data from the month of August for Homestake, LSBB and Sos Enattos and data from the month of January for EMR. Please note that this implies that the EMR results has a higher noise level around the microseism peaks than during summertime at the same location, which is the season considered at the other sites.

We want to state that the study performed in this section and the analysis of \cite{Allocca:2021opl} serve a different purpose and therefore should not be compared one-to-one. The goal of the authors of \cite{Allocca:2021opl} was to establish the effect of seismic glitchiness on the ET. On the other hand, the goal of this section is to understand if a small subset of short times lead to a significant bias of our estimate of ambient seismic noise over several hour long time segments. That is, are the percentiles presented in e.g. Figs \ref{fig:Comparison_Jan_distance_CohCSD_100mHz} and \ref{fig:StochBudget_Jan_NN} a good measure of the ambient seismic noise over several hours, or are they rather dominated by a small amount of large seismic transients.
Even though we get some information on the seismic glitchiness as a byproduct of our analysis, more data (e.g. 1yr as in \cite{Allocca:2021opl}) should be analysed to make clear statements on each site'sglitchiness.

\begin{figure*}
    \centering
    \includegraphics[width=0.49\textwidth]{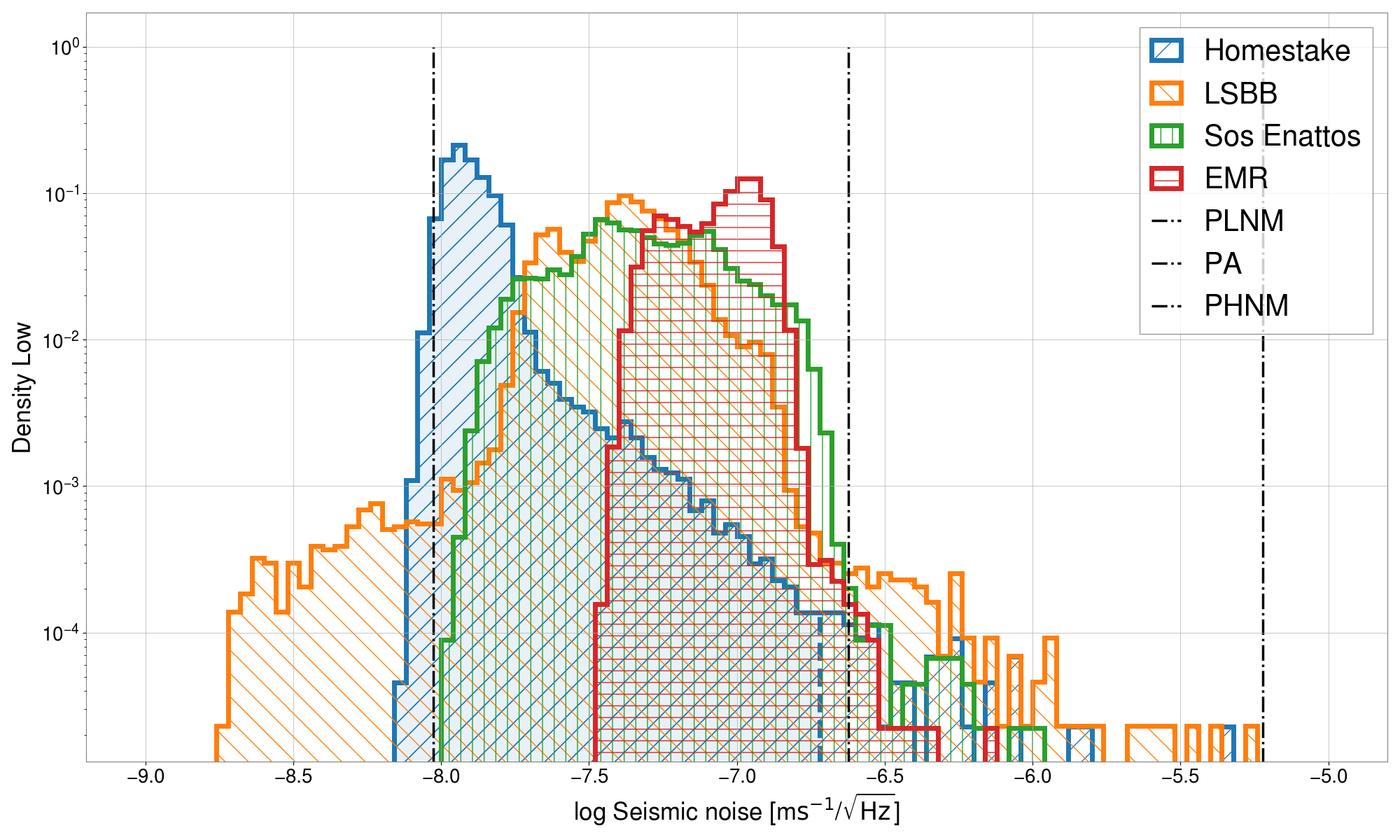}
    \includegraphics[width=0.49\textwidth]{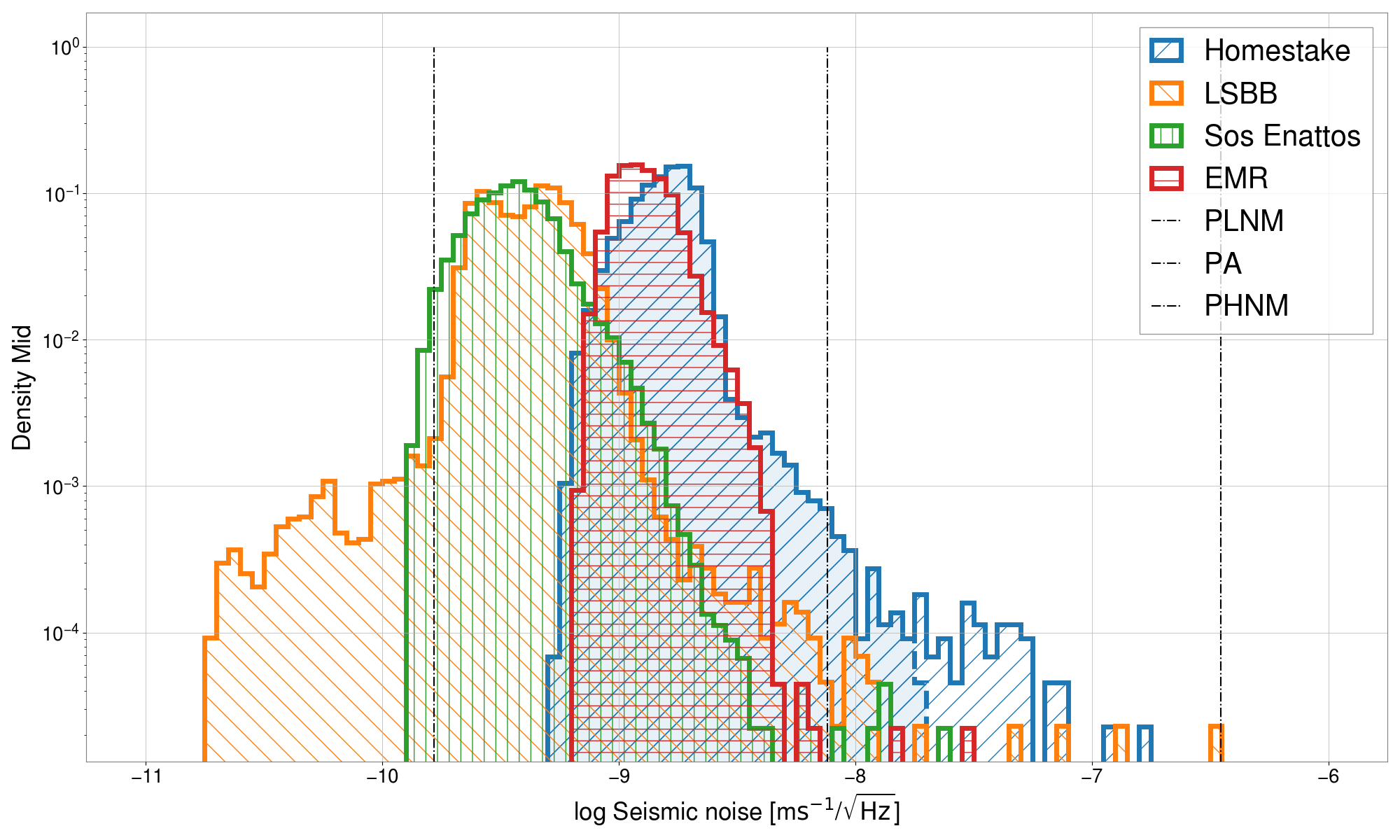}
    \includegraphics[width=0.49\textwidth]{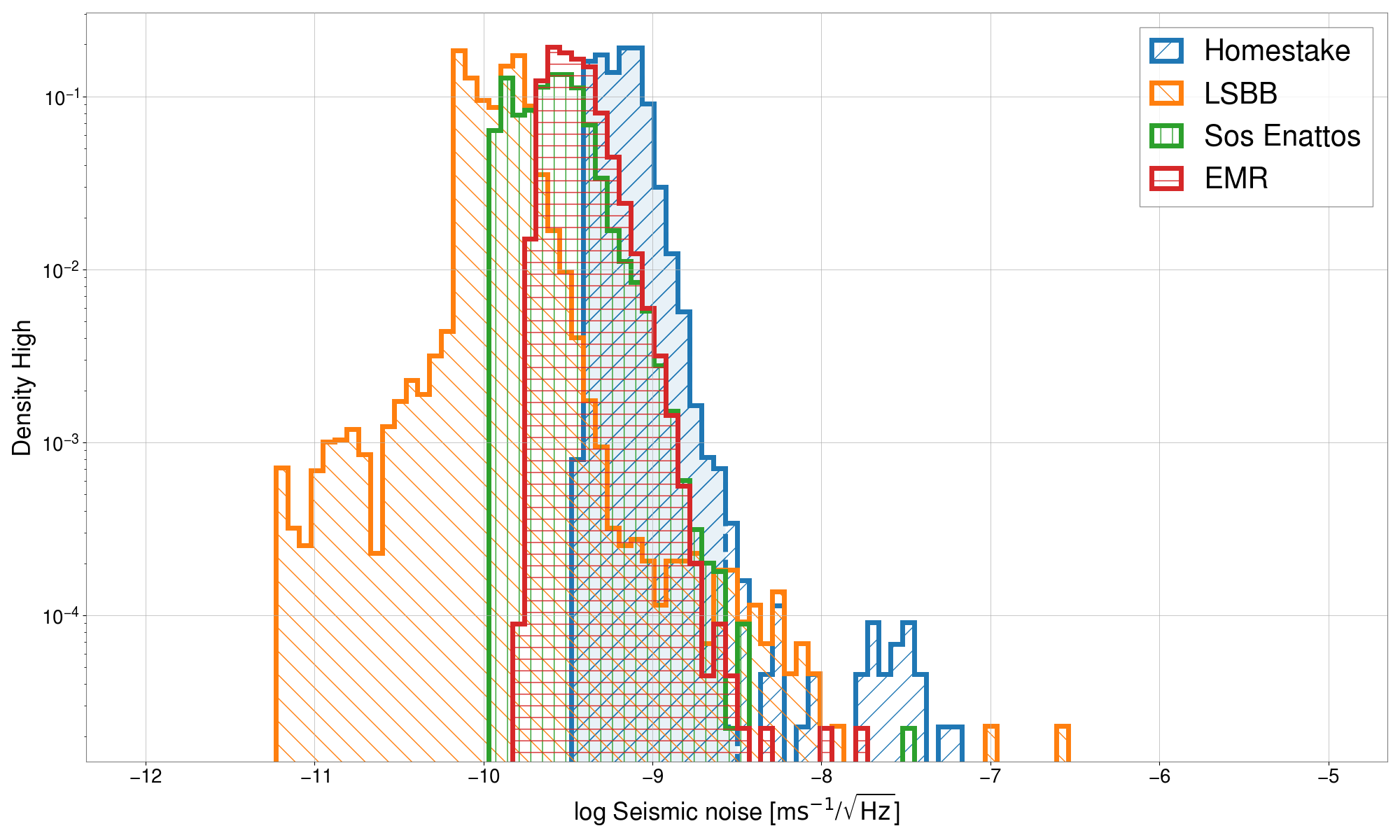}
    \caption{The logarithmic average seismic noise in the from Homestake (blue), LSBB (orange), Sos Enattos (green) and the EMR (red) for the low frequency band 0.1Hz-1Hz (top left), mid frequency band 1Hz-10Hz (top right) and high frequency band 10Hz-40Hz (bottom).
    The black curves (top panels) represent the logarithmic average of low and high noise models by Peterson \cite{Pet1993}, as well as the mean of the Peterson models. Data are analysed using 60-second segments.
    The blue dashed line corresponds to the cut, we sorted and deleted the $50^{\text th}$ highest PSDs.  }
    \label{fig:Hist_PSD1}
\end{figure*}

The sensor pairs we use are respectively D2000-E2000, GAS-RAM, SOE2-SOE3 and TERZ-CTSN where we only considered the correlation between seismic noise observed in the North-South component of each sensor. In \cite{Allocca:2021opl} the authors construct a type of SNR indicating the seismic glitchiness compared to ETs sensitivity. However, we make statements on the potential bias on our estimates of the ambient seismic noise caused by the glitchiness of the seismic data of the site itself. To this extent we consider three different frequency regions: 0.1Hz-1Hz, 1Hz-10Hz and 10Hz-40Hz (10Hz-18Hz in the case of the EMR sensor). We use the logarithmic average of the seismic noise in each of these frequency regions as an indicator to study the effect of glitchiness in each one-minute time segment on our estimates of the ambient seismic noise.
For the PSD of the first sensor\footnote{That is respectively D2000, GAS, SOE2 and TERZ for the different sites.} of each site we present their distributions in Fig. \ref{fig:Hist_PSD1}. Since the histograms for the PSD of the second sensor have similar behavior, we do not include these figures and focus our discussion on Fig. \ref{fig:Hist_PSD1}. In case of the low- and mid- frequency region we show the value expected from Peterson's low noise (PLNM), logarithmic average (PA) and high noise models (PHNM).

Based on these histograms a first comment we can make is that neither of the sites seems strongly dominated by a subset of loud outliers\footnote{In this context we consider an outlier to be a data point which is significantly disconnected from the bulk of the distribution. As an example the handfull of orange data points near -5.5 for the top left panel and near -7 for the top right and bottom panel are considered outliers.}. 
We carefully examined the time-frequency maps of each associated sensor, allowing us to analyse the behaviour of PSDs, CSDs, and coherence over the duration of a month. These analyses reveal recurring patterns in the PSDs. Specifically, we observed a day-night effect, with nights being significantly quieter than days. This can be attributed to human activity. Additionally, there are a limited number of moments when the PSD and CSD exhibit higher amplitude values, indicating the presence of micro-seismic phenomena that affect ambient measurements. The histograms shown in Fig. \ref{fig:Hist_PSD1} highlight the implicated PSD values, which result in an elongation of the distribution tail due to their higher amplitudes.

As a proof of concept, we remove a subset of the loudest one-minute segments and investigate the impact on the observed coherence and CSD. We demonstrate this proof of concept on one site, namely Homestake. We have chosen Homestake as it has a relatively long tail of more noisy segments in the mid- and high frequency range, but compared to LSBB for which this is also the case, Homestake is the site with the loudest predicted levels of correlated seismic noise in these frequency regions. For the PSD of both sensors, we identified in every frequency region the 50 loudest one-minute segments. These 300 noisy segments identified can be described by only 115 unique segments due to large overlap in the noisy segments. This equals to about 0.25\% of the total data. This choice of cut-off value is semi-arbitrary. However the distributions of the transient seismic noise are strongly site dependent and therefore it is hard to define an absolute value of which segment should be considered an outlier and which not. At the same time, we do not want to remove a large amount of data as this is expected to impact the observed percentiles just by the mere fact of removing the data. In Fig.~\ref{fig:Hist_PSD1}, we have indicated the 50 loudest bins in that frequency region by the blue dashed line for the histogram of the Homestake data. Please note that in total more bins from the histogram are removed based on the removal criteria in the different frequency regions as well as based on the PSD of the second sensor.

\begin{figure*}[t]
    \centering
    \includegraphics[width=0.49\textwidth]{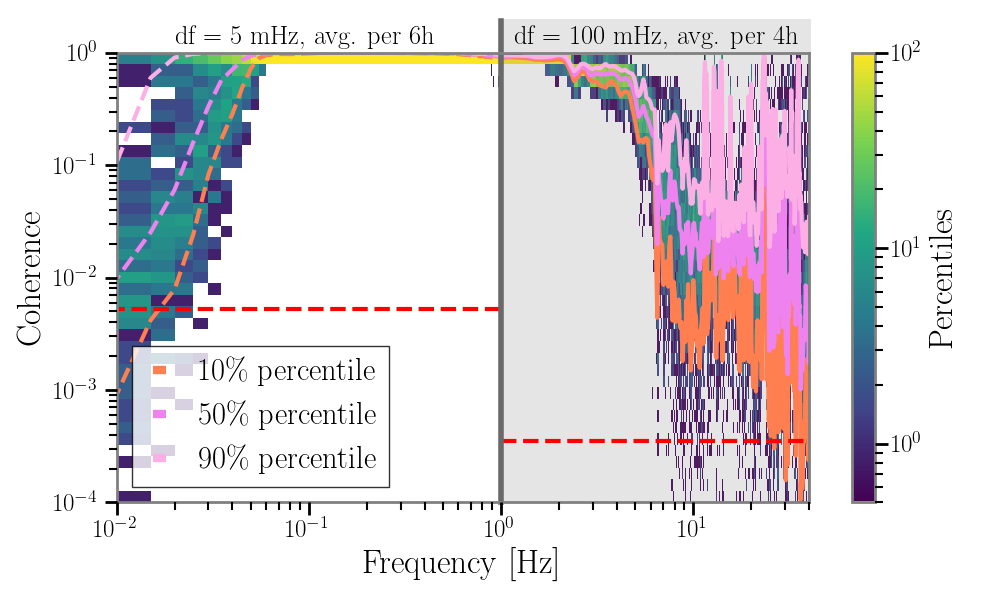}
    \includegraphics[width=0.49\textwidth]{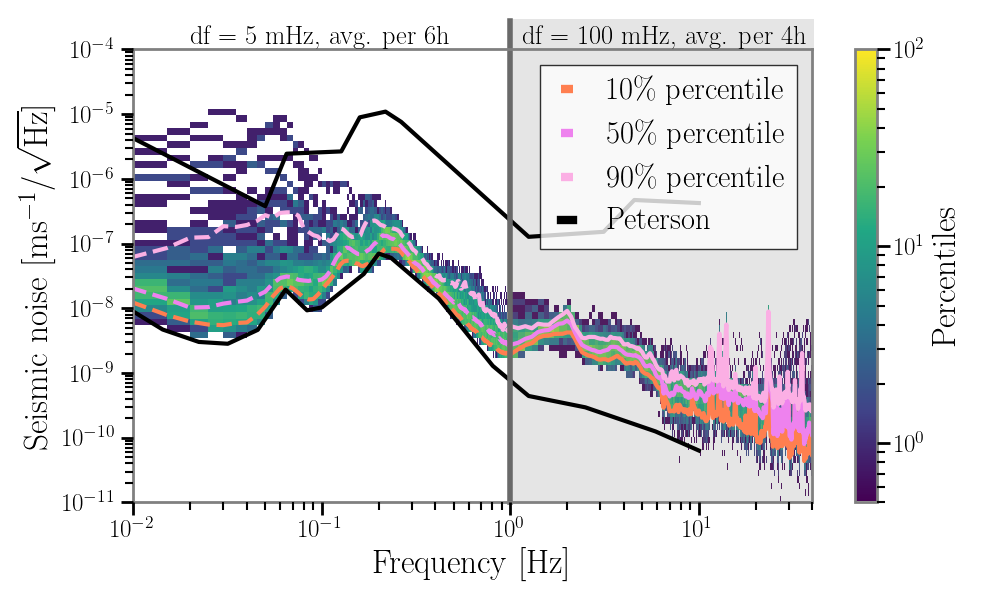}    
    \caption{The coherence (left panel) and CSD (right panel) between the underground seismometers (NS components) D2000 and E2000 at Homestake ($\Delta x\approx$405m and depth = 610m) where 115 noisy 1 min-segments were removed from the analysis as described in the text. The data $<$1Hz (Aug 2016) are analysed using 200 second long segments which are averaged per 6\,h-window. Above 1Hz the data are analysed using 10 second long segments which are averaged per 4\,h-window. The 10$^{\text{th}}$, 50$^{\text{th}}$ and 90$^{\text{th}}$ percentiles are shown in respectively light pink, dark pink and light orange, where dashed (full) lines are used for the analyses with the different parameters $<$ 1Hz ($\geq$ 1Hz). The red dashed line represents the level of coherence expected from Gaussian data which goes approximately as 1/N, where N is the number of time segments over which was averaged.}
    \label{fig:GlitchRemoval-D2000-E2000_CohCSD_100mHz}
\end{figure*}

After identifying the loudest segments, we run the analysis for the seismic noise of the month of August at Homestake again with the identified segements removed from the analysis. The coherence and CSD after removal, shown in Fig. \ref{fig:GlitchRemoval-D2000-E2000_CohCSD_100mHz}, should be compared to Fig. \ref{fig:D2000-E2000_CohCSD_5mHz}. The only difference between the data used for these figures is the removal of the loudest segments in Fig. \ref{fig:GlitchRemoval-D2000-E2000_CohCSD_100mHz}. 

When comparing the 10\% and 50\% percentiles for both coherence and the CSD before and after the removal of the loudest transients, almost no effect is observed. The largest effect is observed $>$ 10Hz for which at all times the seismic CSD after glitch removal is at most 20\% smaller compared to the CSD before glitch removal.  However, for the 90\% percentile a significant difference is observed. The coherence after glitch removal is about two to ten times lower for frequencies above 10Hz. The CSD after glitch removal is up to a factor two lower between 1Hz and 10Hz and up to a factor of four between 10Hz and 40Hz. On average the 90\% percentile of the CSD is two times lower after glitch removal. The correlated seismic noise is visibly cleaner after removing the loudest one-minute segments as compared to before.

Furthermore, we want to highlight that we tested the impact of removing 115 one-minute segments by removing the same number of segments arbitrarily chosen. This yields quasi-identical results to the analysis where no data was excluded, shown in Fig. \ref{fig:D2000-E2000_CohCSD_5mHz}. This proves the effect discussed above is indeed due to removing the loudest segments and not due to the act of removing 115 one-minute segments.

Based on the analysis above we can state that whereas the 90\% percentile of CSD (and coherence) are significantly impacted by a small number of loud transient time segments, the 10\% and 50\% percentiles are not. Both in earlier work \cite{PhysRevD.106.042008} as well as in this paper, all key conclusions are based on the 50\% percentiles. Therefore, we state that all key conclusions of this (and earlier \cite{PhysRevD.106.042008}) work were not disproportionately dominated by a small number of noisy segments, but the results form a good representation of the correlated noise of the seismic ambient environment.

\section{Conclusion}
\label{sec:Conclusion}
Next generation Earth-based GW atom interferometers, such as ELGAR, as well as interferometric detectors, such as the ET and CE, promise to be powerful instruments to observe GWs in the next decades. With the sensitivity range of atom interferometers to GWs, mainly between 0.1Hz-1Hz, and ET's unprecedented low frequency sensitivity in the range 1\,Hz-10\,Hz, they could open new windows into the GW universe.

However, earlier work \cite{PhysRevD.106.042008} demonstrated that correlated seismic noise and, particularly, correlated NN could seriously limit the search for an isotropic GWB with co-located ET detectors. In this work, we build further on these earlier results and improve them in multiple ways. Rather than using underground seismic data from one site, we use four geologically different sites, one of which houses the atom interferometer MIGA and two are candidate sites to house the ET. Additionally, we probe a wider range of horizontal separations between the seismic sensors ranging from 230\,m to 10\,km. The depth of the sensors varies between 84\,m and 610\,m. 
Furthermore, we probed the low frequency region 0.01\,Hz-0.1\,Hz, which is of interest for atom interferometers. Finally we performed a study of the seismic glitches of the sites and prove that seismic transients do not significantly affect our analyses and conclusions.

We analysed data from underground seismometers at the former Homestake mine (USA), the MIGA site at the `Laboratoire Souterrain \`a Bas Bruit' (FR) and two of the three candidate sites for the ET, the former Sos Enattos mine (IT) and the Euregio Maas-Rhein (NL-BE-DE).
The used sensor pairs have a wide variety of analysis parameters such as horizontal separation, depth, sensors model, geological environment, located in a borehole or a cavern, etc. Currently a systematic study is being performed at LSBB.
However, despite this wide variety, across all these we find (for sensors with a horizontal separation up to 2.4km) significant coherence at least 50\% of the time in frequency ranges for both atom interferometers as well as the ET. The levels of correlated seismic noise across these wide variety of sensors is at most about one order of magnitude when comparing the 50\% percentiles.
More concretely for sensors with a separation less than one kilometer we find at least 50\% of the time significant coherence in the entire frequency band 0.01Hz-40Hz. For a sensor pair in the EMR region with a separation of 2.4km we find significant coherence up to $\sim$16Hz for the 50\% percentile. Finally, a pair of seismometers separated by 10km at the Sos Enattos site demonstrate that even on this multi kilometer distance we find significant seismic coherence between 0.01Hz and 1Hz for the 50\% percentile.
For frequencies larger than 1Hz, these results are (almost) independent of the time of the year.

These seismic correlations are important to take into account for next generation atom interferometers and interferometric GW detectors. For the former, multiple atomic gradiometers are planned to be placed at distances of several hundred of meters to tens of kilometers. For the ET the baseline configuration consists of an equilateral triangle made up of three nested GW interferometers. In this configuration the end and central station of two different detectors are planned to be separated by a distance of several hundred of meters in the current design \cite{ETdesignRep}. Additionally, 10km is a relevant distance scale for the ET as this is the proposed length-scale for the detector's arms.

One of the goals from this and earlier \cite{PhysRevD.106.042008} research was to provide insight on whether the separation between the corner and end stations of the nested ET  detectors would matter for the correlated seismic noise. Based on the results presented in this work it seems that any displacement $<$1km is insufficient to decrease seismic coherence. A distance of $\sim$ 2.5km does provide significant reduction in seismic coherence. However even in such a case we find significant coherence up to $\sim$ 16Hz 50\% of the time or more. To truly eliminate all seismic coherence (to the level of $\sim 3.8 \times 10^{-4}$) above 1\,Hz one should have a horizontal separation between the corner and end stations of different ET detectors between 2.5\,km and 10\,km. However, such a large separation is rather unlikely.

The effect of the observed correlations of seismic noise, and subsequent NN, on atom interferometers should be projected in future work. However, the results presented in this work could form an ideal starting point to make such projections. Additionally our results demonstrate that one should be cautious and investigate these effects in more detail as we observe seismic correlations over the entire frequency band of interest (0.1\,Hz-10\,Hz) as well as over the entire range of distances expected for a realistic set-up ranging from hundreds of meters to $\mathcal{O}$(10)\,km. 

In the context of the ET, we make a projection of the expected contributions of NN from body waves on the search for an isotropic GWB. This search is the most sensitive to correlated noise sources as one tries to observe a weak background of correlated GW signals for which one typically has to integrate over $\mathcal{O}$(1 year) of data. Previous investigations have shown that ET's instantaneous sensitivity would probably be affected by NN by a factor 3-5, which is considered to be realistically removable by noise subtraction methods such as Wiener filtering \cite{PhysRevD.86.102001,PhysRevD.92.022001,Coughlin_2014_seismic,Coughlin_2016,PhysRevLett.121.221104,Tringali_2019,Badaracco_2020,Badaracco_2019,NN_Sardinia2020,10.1785/0220200186,Bader_2022,Koley_2022, digiovanni23}.
However, as shown in \cite{PhysRevD.106.042008}, the search for an isotropic GWB is subject to very small coherent noise signals.
In this work, we have shown that the projections based on underground seismic measurements at Homestake, of which similar results were presented in \cite{PhysRevD.106.042008}, are probably more pessimistic compared to a realistic situation at one of the ET candidate sites. However, at the same time we demonstrated that the assumptions made in \cite{Branchesi:2023mws} are probably too optimistic as all measurements, even those at extremely low noise seismic sites above 10Hz, predict stronger effects.
More concretely, these results predict that during at least 50\% of the time the search for an isotropic GWB would be dominated by NN from body-waves up to 20Hz. This assumes our most quiet seismic observations at LSBB. For the Homestake site the effect extents up to 40Hz.

The correlated noise contamination in the low frequency regime has a different impact depending on which type of broadband isotropic GWB one is looking for. The resolvable amplitude for a negatively sloped power-law signal will be reduced by several orders of magnitude. However, most of the signals currently searched for in this frequency band have a flat or positive power-law slope, e.g. $\alpha=0,2/3,3$ \cite{KAGRA:2021kbb}. We demonstrate that by only analysing data above 15Hz-20Hz one would loose up to a factor $\sim$ 6-9 ($\sim$3-4) for a power-law signal with a slope of $\alpha=0$ ($\alpha=2/3$). Such analysis could be free of correlated noise in case of low noise levels such as at the LSBB site. The detectability of a signal with $\alpha=3$ is unaffected as it is dominated by the data at higher frequencies.

Finally, the measurements at Sos Enattos seem to suggest significant seismic correlations on length-scales of 10km are situated below 1Hz. Therefore, any correlation effects between test masses located 10km-far are to
be excluded in the frequency band of interest for ET. However, at current detectors increased noise levels are observed above several Hz during times of loud seismic activity at lower frequencies. Such an example are slow scattered light glitches caused by increased microseism activity between 0.1Hz and 0.3Hz \cite{Davis_2021}. These more complex coupling mechanisms are likely to be detector dependant in which case they are less likely to enter coherently in different interferometers. However, more work might be needed in the future to entirely exclude this additional pathway for correlated noise to enter into GWB searches with co-located ET detectors. Furthermore, the effect of the correlated seismic noise for frequencies below 1Hz on the angular control of the instruments is to be further understood.

\acknowledgements

K.J. was supported by FWO-Vlaanderen via grant number 11C5720N during part of this work.
M.W.C acknowledges support from the National Science Foundation with grant numbers PHY-2308862 and PHY-2117997.
The authors’ grateful acknowledgment also goes to the contribution of the Fondi di Ateneo per la ricerca 2019, Fondi di Ateneo per la ricerca 2020 of the University of Sassari and the support of the Italian Ministry of Education, University and Research within the PRIN 2017 Research Program Framework, no. 2017SYRTCN and University of Sassari.

\bibliography{references}

\begin{thebibliography}{71}%
\makeatletter
\providecommand \@ifxundefined [1]{%
 \@ifx{#1\undefined}
}%
\providecommand \@ifnum [1]{%
 \ifnum #1\expandafter \@firstoftwo
 \else \expandafter \@secondoftwo
 \fi
}%
\providecommand \@ifx [1]{%
 \ifx #1\expandafter \@firstoftwo
 \else \expandafter \@secondoftwo
 \fi
}%
\providecommand \natexlab [1]{#1}%
\providecommand \enquote  [1]{``#1''}%
\providecommand \bibnamefont  [1]{#1}%
\providecommand \bibfnamefont [1]{#1}%
\providecommand \citenamefont [1]{#1}%
\providecommand \href@noop [0]{\@secondoftwo}%
\providecommand \href [0]{\begingroup \@sanitize@url \@href}%
\providecommand \@href[1]{\@@startlink{#1}\@@href}%
\providecommand \@@href[1]{\endgroup#1\@@endlink}%
\providecommand \@sanitize@url [0]{\catcode `\\12\catcode `\$12\catcode
  `\&12\catcode `\#12\catcode `\^12\catcode `\_12\catcode `\%12\relax}%
\providecommand \@@startlink[1]{}%
\providecommand \@@endlink[0]{}%
\providecommand \url  [0]{\begingroup\@sanitize@url \@url }%
\providecommand \@url [1]{\endgroup\@href {#1}{\urlprefix }}%
\providecommand \urlprefix  [0]{URL }%
\providecommand \Eprint [0]{\href }%
\providecommand \doibase [0]{http://dx.doi.org/}%
\providecommand \selectlanguage [0]{\@gobble}%
\providecommand \bibinfo  [0]{\@secondoftwo}%
\providecommand \bibfield  [0]{\@secondoftwo}%
\providecommand \translation [1]{[#1]}%
\providecommand \BibitemOpen [0]{}%
\providecommand \bibitemStop [0]{}%
\providecommand \bibitemNoStop [0]{.\EOS\space}%
\providecommand \EOS [0]{\spacefactor3000\relax}%
\providecommand \BibitemShut  [1]{\csname bibitem#1\endcsname}%
\let\auto@bib@innerbib\@empty
\bibitem [{\citenamefont {Christensen}(2018)}]{Christensen_2018}%
  \BibitemOpen
  \bibfield  {author} {\bibinfo {author} {\bibfnamefont {N.}~\bibnamefont
  {Christensen}},\ }\href {\doibase 10.1088/1361-6633/aae6b5} {\bibfield
  {journal} {\bibinfo  {journal} {Reports on Progress in Physics}\ }\textbf
  {\bibinfo {volume} {82}},\ \bibinfo {pages} {016903} (\bibinfo {year}
  {2018})}\BibitemShut {NoStop}%
\bibitem [{\citenamefont {{Schumann}}(1952{\natexlab{a}})}]{Schumann1}%
  \BibitemOpen
  \bibfield  {author} {\bibinfo {author} {\bibfnamefont {W.~O.}\ \bibnamefont
  {{Schumann}}},\ }\href {\doibase 10.1515/zna-1952-0202} {\bibfield  {journal}
  {\bibinfo  {journal} {Zeitschrift Naturforschung Teil A}\ }\textbf {\bibinfo
  {volume} {7}},\ \bibinfo {pages} {149} (\bibinfo {year}
  {1952}{\natexlab{a}})}\BibitemShut {NoStop}%
\bibitem [{\citenamefont {{Schumann}}(1952{\natexlab{b}})}]{Schumann2}%
  \BibitemOpen
  \bibfield  {author} {\bibinfo {author} {\bibfnamefont {W.~O.}\ \bibnamefont
  {{Schumann}}},\ }\href {\doibase 10.1515/zna-1952-3-404} {\bibfield
  {journal} {\bibinfo  {journal} {Zeitschrift Naturforschung Teil A}\ }\textbf
  {\bibinfo {volume} {7}},\ \bibinfo {pages} {250} (\bibinfo {year}
  {1952}{\natexlab{b}})}\BibitemShut {NoStop}%
\bibitem [{\citenamefont {Aasi}\ \emph {et~al.}(2015)\citenamefont {Aasi} \emph
  {et~al.}}]{2015}%
  \BibitemOpen
  \bibfield  {author} {\bibinfo {author} {\bibfnamefont {J.}~\bibnamefont
  {Aasi}} \emph {et~al.},\ }\href {\doibase 10.1088/0264-9381/32/7/074001}
  {\bibfield  {journal} {\bibinfo  {journal} {Classical and Quantum Gravity}\
  }\textbf {\bibinfo {volume} {32}},\ \bibinfo {pages} {074001} (\bibinfo
  {year} {2015})}\BibitemShut {NoStop}%
\bibitem [{\citenamefont {Acernese}\ \emph {et~al.}(2015)\citenamefont
  {Acernese} \emph {et~al.}}]{VIRGO:2014yos}%
  \BibitemOpen
  \bibfield  {author} {\bibinfo {author} {\bibfnamefont {F.}~\bibnamefont
  {Acernese}} \emph {et~al.} (\bibinfo {collaboration} {VIRGO}),\ }\href
  {\doibase 10.1088/0264-9381/32/2/024001} {\bibfield  {journal} {\bibinfo
  {journal} {Class. Quant. Grav.}\ }\textbf {\bibinfo {volume} {32}},\ \bibinfo
  {pages} {024001} (\bibinfo {year} {2015})},\ \Eprint
  {http://arxiv.org/abs/1408.3978} {arXiv:1408.3978 [gr-qc]} \BibitemShut
  {NoStop}%
\bibitem [{\citenamefont {Aso}\ \emph {et~al.}(2013)\citenamefont {Aso},
  \citenamefont {Michimura}, \citenamefont {Somiya}, \citenamefont {Ando},
  \citenamefont {Miyakawa}, \citenamefont {Sekiguchi}, \citenamefont
  {Tatsumi},\ and\ \citenamefont {Yamamoto}}]{PhysRevD.88.043007}%
  \BibitemOpen
  \bibfield  {author} {\bibinfo {author} {\bibfnamefont {Y.}~\bibnamefont
  {Aso}}, \bibinfo {author} {\bibfnamefont {Y.}~\bibnamefont {Michimura}},
  \bibinfo {author} {\bibfnamefont {K.}~\bibnamefont {Somiya}}, \bibinfo
  {author} {\bibfnamefont {M.}~\bibnamefont {Ando}}, \bibinfo {author}
  {\bibfnamefont {O.}~\bibnamefont {Miyakawa}}, \bibinfo {author}
  {\bibfnamefont {T.}~\bibnamefont {Sekiguchi}}, \bibinfo {author}
  {\bibfnamefont {D.}~\bibnamefont {Tatsumi}}, \ and\ \bibinfo {author}
  {\bibfnamefont {H.}~\bibnamefont {Yamamoto}} (\bibinfo {collaboration} {The
  KAGRA Collaboration}),\ }\href {\doibase 10.1103/PhysRevD.88.043007}
  {\bibfield  {journal} {\bibinfo  {journal} {Phys. Rev. D}\ }\textbf {\bibinfo
  {volume} {88}},\ \bibinfo {pages} {043007} (\bibinfo {year}
  {2013})}\BibitemShut {NoStop}%
\bibitem [{\citenamefont {Thrane}\ \emph {et~al.}(2013)\citenamefont {Thrane},
  \citenamefont {Christensen},\ and\ \citenamefont
  {Schofield}}]{Thrane:2013npa}%
  \BibitemOpen
  \bibfield  {author} {\bibinfo {author} {\bibfnamefont {E.}~\bibnamefont
  {Thrane}}, \bibinfo {author} {\bibfnamefont {N.}~\bibnamefont {Christensen}},
  \ and\ \bibinfo {author} {\bibfnamefont {R.~M.~S.}\ \bibnamefont
  {Schofield}},\ }\href {\doibase 10.1103/PhysRevD.87.123009} {\bibfield
  {journal} {\bibinfo  {journal} {Phys. Rev. D}\ }\textbf {\bibinfo {volume}
  {87}},\ \bibinfo {pages} {123009} (\bibinfo {year} {2013})},\ \Eprint
  {http://arxiv.org/abs/1303.2613} {arXiv:1303.2613 [astro-ph.IM]} \BibitemShut
  {NoStop}%
\bibitem [{\citenamefont {Thrane}\ \emph {et~al.}(2014)\citenamefont {Thrane},
  \citenamefont {Christensen}, \citenamefont {Schofield},\ and\ \citenamefont
  {Effler}}]{Thrane:2014yza}%
  \BibitemOpen
  \bibfield  {author} {\bibinfo {author} {\bibfnamefont {E.}~\bibnamefont
  {Thrane}}, \bibinfo {author} {\bibfnamefont {N.}~\bibnamefont {Christensen}},
  \bibinfo {author} {\bibfnamefont {R.~M.~S.}\ \bibnamefont {Schofield}}, \
  and\ \bibinfo {author} {\bibfnamefont {A.}~\bibnamefont {Effler}},\ }\href
  {\doibase 10.1103/PhysRevD.90.023013} {\bibfield  {journal} {\bibinfo
  {journal} {Phys. Rev. D}\ }\textbf {\bibinfo {volume} {90}},\ \bibinfo
  {pages} {023013} (\bibinfo {year} {2014})},\ \Eprint
  {http://arxiv.org/abs/1406.2367} {arXiv:1406.2367 [astro-ph.IM]} \BibitemShut
  {NoStop}%
\bibitem [{\citenamefont {Coughlin}\ \emph
  {et~al.}(2016{\natexlab{a}})\citenamefont {Coughlin} \emph
  {et~al.}}]{Coughlin:2016vor}%
  \BibitemOpen
  \bibfield  {author} {\bibinfo {author} {\bibfnamefont {M.~W.}\ \bibnamefont
  {Coughlin}} \emph {et~al.},\ }\href {\doibase 10.1088/0264-9381/33/22/224003}
  {\bibfield  {journal} {\bibinfo  {journal} {Class. Quant. Grav.}\ }\textbf
  {\bibinfo {volume} {33}},\ \bibinfo {pages} {224003} (\bibinfo {year}
  {2016}{\natexlab{a}})},\ \Eprint {http://arxiv.org/abs/1606.01011}
  {arXiv:1606.01011 [gr-qc]} \BibitemShut {NoStop}%
\bibitem [{\citenamefont {Himemoto}\ and\ \citenamefont
  {Taruya}(2017)}]{Himemoto:2017gnw}%
  \BibitemOpen
  \bibfield  {author} {\bibinfo {author} {\bibfnamefont {Y.}~\bibnamefont
  {Himemoto}}\ and\ \bibinfo {author} {\bibfnamefont {A.}~\bibnamefont
  {Taruya}},\ }\href {\doibase 10.1103/PhysRevD.96.022004} {\bibfield
  {journal} {\bibinfo  {journal} {Phys. Rev. D}\ }\textbf {\bibinfo {volume}
  {96}},\ \bibinfo {pages} {022004} (\bibinfo {year} {2017})},\ \Eprint
  {http://arxiv.org/abs/1704.07084} {arXiv:1704.07084 [astro-ph.IM]}
  \BibitemShut {NoStop}%
\bibitem [{\citenamefont {Coughlin}\ \emph
  {et~al.}(2018{\natexlab{a}})\citenamefont {Coughlin}, \citenamefont {Cirone},
  \citenamefont {Meyers}, \citenamefont {Atsuta}, \citenamefont {Boschi},
  \citenamefont {Chincarini}, \citenamefont {Christensen}, \citenamefont
  {De~Rosa}, \citenamefont {Effler}, \citenamefont {Fiori}, \citenamefont
  {Go\l{}kowski}, \citenamefont {Guidry}, \citenamefont {Harms}, \citenamefont
  {Hayama}, \citenamefont {Kataoka}, \citenamefont {Kubisz}, \citenamefont
  {Kulak}, \citenamefont {Laxen}, \citenamefont {Matas}, \citenamefont
  {Mlynarczyk}, \citenamefont {Ogawa}, \citenamefont {Paoletti}, \citenamefont
  {Salvador}, \citenamefont {Schofield}, \citenamefont {Somiya},\ and\
  \citenamefont {Thrane}}]{Coughlin:2018tjc}%
  \BibitemOpen
  \bibfield  {author} {\bibinfo {author} {\bibfnamefont {M.~W.}\ \bibnamefont
  {Coughlin}}, \bibinfo {author} {\bibfnamefont {A.}~\bibnamefont {Cirone}},
  \bibinfo {author} {\bibfnamefont {P.}~\bibnamefont {Meyers}}, \bibinfo
  {author} {\bibfnamefont {S.}~\bibnamefont {Atsuta}}, \bibinfo {author}
  {\bibfnamefont {V.}~\bibnamefont {Boschi}}, \bibinfo {author} {\bibfnamefont
  {A.}~\bibnamefont {Chincarini}}, \bibinfo {author} {\bibfnamefont {N.~L.}\
  \bibnamefont {Christensen}}, \bibinfo {author} {\bibfnamefont
  {R.}~\bibnamefont {De~Rosa}}, \bibinfo {author} {\bibfnamefont
  {A.}~\bibnamefont {Effler}}, \bibinfo {author} {\bibfnamefont
  {I.}~\bibnamefont {Fiori}}, \bibinfo {author} {\bibfnamefont
  {M.}~\bibnamefont {Go\l{}kowski}}, \bibinfo {author} {\bibfnamefont
  {M.}~\bibnamefont {Guidry}}, \bibinfo {author} {\bibfnamefont
  {J.}~\bibnamefont {Harms}}, \bibinfo {author} {\bibfnamefont
  {K.}~\bibnamefont {Hayama}}, \bibinfo {author} {\bibfnamefont
  {Y.}~\bibnamefont {Kataoka}}, \bibinfo {author} {\bibfnamefont
  {J.}~\bibnamefont {Kubisz}}, \bibinfo {author} {\bibfnamefont
  {A.}~\bibnamefont {Kulak}}, \bibinfo {author} {\bibfnamefont
  {M.}~\bibnamefont {Laxen}}, \bibinfo {author} {\bibfnamefont
  {A.}~\bibnamefont {Matas}}, \bibinfo {author} {\bibfnamefont
  {J.}~\bibnamefont {Mlynarczyk}}, \bibinfo {author} {\bibfnamefont
  {T.}~\bibnamefont {Ogawa}}, \bibinfo {author} {\bibfnamefont
  {F.}~\bibnamefont {Paoletti}}, \bibinfo {author} {\bibfnamefont
  {J.}~\bibnamefont {Salvador}}, \bibinfo {author} {\bibfnamefont
  {R.}~\bibnamefont {Schofield}}, \bibinfo {author} {\bibfnamefont
  {K.}~\bibnamefont {Somiya}}, \ and\ \bibinfo {author} {\bibfnamefont
  {E.}~\bibnamefont {Thrane}},\ }\href {\doibase 10.1103/PhysRevD.97.102007}
  {\bibfield  {journal} {\bibinfo  {journal} {Phys. Rev. D}\ }\textbf {\bibinfo
  {volume} {97}},\ \bibinfo {pages} {102007} (\bibinfo {year}
  {2018}{\natexlab{a}})}\BibitemShut {NoStop}%
\bibitem [{\citenamefont {Himemoto}\ and\ \citenamefont
  {Taruya}(2019)}]{Himemoto:2019iwd}%
  \BibitemOpen
  \bibfield  {author} {\bibinfo {author} {\bibfnamefont {Y.}~\bibnamefont
  {Himemoto}}\ and\ \bibinfo {author} {\bibfnamefont {A.}~\bibnamefont
  {Taruya}},\ }\href {\doibase 10.1103/PhysRevD.100.082001} {\bibfield
  {journal} {\bibinfo  {journal} {Phys. Rev. D}\ }\textbf {\bibinfo {volume}
  {100}},\ \bibinfo {pages} {082001} (\bibinfo {year} {2019})},\ \Eprint
  {http://arxiv.org/abs/1908.10635} {arXiv:1908.10635 [astro-ph.IM]}
  \BibitemShut {NoStop}%
\bibitem [{\citenamefont {Meyers}\ \emph {et~al.}(2020)\citenamefont {Meyers},
  \citenamefont {Martinovic}, \citenamefont {Christensen},\ and\ \citenamefont
  {Sakellariadou}}]{Meyers:2020qrb}%
  \BibitemOpen
  \bibfield  {author} {\bibinfo {author} {\bibfnamefont {P.~M.}\ \bibnamefont
  {Meyers}}, \bibinfo {author} {\bibfnamefont {K.}~\bibnamefont {Martinovic}},
  \bibinfo {author} {\bibfnamefont {N.}~\bibnamefont {Christensen}}, \ and\
  \bibinfo {author} {\bibfnamefont {M.}~\bibnamefont {Sakellariadou}},\ }\href
  {\doibase 10.1103/PhysRevD.102.102005} {\bibfield  {journal} {\bibinfo
  {journal} {Phys. Rev. D}\ }\textbf {\bibinfo {volume} {102}},\ \bibinfo
  {pages} {102005} (\bibinfo {year} {2020})},\ \Eprint
  {http://arxiv.org/abs/2008.00789} {arXiv:2008.00789 [gr-qc]} \BibitemShut
  {NoStop}%
\bibitem [{\citenamefont {Janssens}\ \emph {et~al.}(2023)\citenamefont
  {Janssens}, \citenamefont {Ball}, \citenamefont {Schofield}, \citenamefont
  {Christensen}, \citenamefont {Frey}, \citenamefont {van Remortel},
  \citenamefont {Banagiri}, \citenamefont {Coughlin}, \citenamefont {Effler},
  \citenamefont {Go\l{}kowski}, \citenamefont {Kubisz},\ and\ \citenamefont
  {Ostrowski}}]{PhysRevD.107.022004}%
  \BibitemOpen
  \bibfield  {author} {\bibinfo {author} {\bibfnamefont {K.}~\bibnamefont
  {Janssens}}, \bibinfo {author} {\bibfnamefont {M.}~\bibnamefont {Ball}},
  \bibinfo {author} {\bibfnamefont {R.~M.~S.}\ \bibnamefont {Schofield}},
  \bibinfo {author} {\bibfnamefont {N.}~\bibnamefont {Christensen}}, \bibinfo
  {author} {\bibfnamefont {R.}~\bibnamefont {Frey}}, \bibinfo {author}
  {\bibfnamefont {N.}~\bibnamefont {van Remortel}}, \bibinfo {author}
  {\bibfnamefont {S.}~\bibnamefont {Banagiri}}, \bibinfo {author}
  {\bibfnamefont {M.~W.}\ \bibnamefont {Coughlin}}, \bibinfo {author}
  {\bibfnamefont {A.}~\bibnamefont {Effler}}, \bibinfo {author} {\bibfnamefont
  {M.}~\bibnamefont {Go\l{}kowski}}, \bibinfo {author} {\bibfnamefont
  {J.}~\bibnamefont {Kubisz}}, \ and\ \bibinfo {author} {\bibfnamefont
  {M.}~\bibnamefont {Ostrowski}},\ }\href {\doibase
  10.1103/PhysRevD.107.022004} {\bibfield  {journal} {\bibinfo  {journal}
  {Phys. Rev. D}\ }\textbf {\bibinfo {volume} {107}},\ \bibinfo {pages}
  {022004} (\bibinfo {year} {2023})}\BibitemShut {NoStop}%
\bibitem [{\citenamefont {Kowalska-Leszczynska}\ \emph
  {et~al.}(2017)\citenamefont {Kowalska-Leszczynska} \emph
  {et~al.}}]{Kowalska-Leszczynska:2016low}%
  \BibitemOpen
  \bibfield  {author} {\bibinfo {author} {\bibfnamefont {I.}~\bibnamefont
  {Kowalska-Leszczynska}} \emph {et~al.},\ }\href {\doibase
  10.1088/1361-6382/aa60eb} {\bibfield  {journal} {\bibinfo  {journal} {Class.
  Quant. Grav.}\ }\textbf {\bibinfo {volume} {34}},\ \bibinfo {pages} {074002}
  (\bibinfo {year} {2017})},\ \Eprint {http://arxiv.org/abs/1612.01102}
  {arXiv:1612.01102 [astro-ph.IM]} \BibitemShut {NoStop}%
\bibitem [{\citenamefont {Janssens}\ \emph {et~al.}(2021)\citenamefont
  {Janssens}, \citenamefont {Martinovic}, \citenamefont {Christensen},
  \citenamefont {Meyers},\ and\ \citenamefont
  {Sakellariadou}}]{PhysRevD.104.122006}%
  \BibitemOpen
  \bibfield  {author} {\bibinfo {author} {\bibfnamefont {K.}~\bibnamefont
  {Janssens}}, \bibinfo {author} {\bibfnamefont {K.}~\bibnamefont
  {Martinovic}}, \bibinfo {author} {\bibfnamefont {N.}~\bibnamefont
  {Christensen}}, \bibinfo {author} {\bibfnamefont {P.~M.}\ \bibnamefont
  {Meyers}}, \ and\ \bibinfo {author} {\bibfnamefont {M.}~\bibnamefont
  {Sakellariadou}},\ }\href {\doibase 10.1103/PhysRevD.104.122006} {\bibfield
  {journal} {\bibinfo  {journal} {Phys. Rev. D}\ }\textbf {\bibinfo {volume}
  {104}},\ \bibinfo {pages} {122006} (\bibinfo {year} {2021})}\BibitemShut
  {NoStop}%
\bibitem [{\citenamefont {Punturo}\ \emph {et~al.}(2010)\citenamefont {Punturo}
  \emph {et~al.}}]{Punturo:2010zz}%
  \BibitemOpen
  \bibfield  {author} {\bibinfo {author} {\bibfnamefont {M.}~\bibnamefont
  {Punturo}} \emph {et~al.},\ }\href {\doibase 10.1088/0264-9381/27/19/194002}
  {\bibfield  {journal} {\bibinfo  {journal} {Class. Quant. Grav.}\ }\textbf
  {\bibinfo {volume} {27}},\ \bibinfo {pages} {194002} (\bibinfo {year}
  {2010})}\BibitemShut {NoStop}%
\bibitem [{\citenamefont {Hild}\ \emph {et~al.}(2011)\citenamefont {Hild} \emph
  {et~al.}}]{Hild:2010id}%
  \BibitemOpen
  \bibfield  {author} {\bibinfo {author} {\bibfnamefont {S.}~\bibnamefont
  {Hild}} \emph {et~al.},\ }\href {\doibase 10.1088/0264-9381/28/9/094013}
  {\bibfield  {journal} {\bibinfo  {journal} {Class. Quant. Grav.}\ }\textbf
  {\bibinfo {volume} {28}},\ \bibinfo {pages} {094013} (\bibinfo {year}
  {2011})},\ \Eprint {http://arxiv.org/abs/1012.0908} {arXiv:1012.0908 [gr-qc]}
  \BibitemShut {NoStop}%
\bibitem [{\citenamefont {Naticchioni}\ \emph {et~al.}(2014)\citenamefont
  {Naticchioni}, \citenamefont {Perciballi}, \citenamefont {Ricci},
  \citenamefont {Coccia}, \citenamefont {Malvezzi}, \citenamefont {Acernese},
  \citenamefont {Barone}, \citenamefont {Giordano}, \citenamefont {Romano},
  \citenamefont {Punturo}, \citenamefont {Rosa}, \citenamefont {Calia},\ and\
  \citenamefont {Loddo}}]{Naticchioni_2014}%
  \BibitemOpen
  \bibfield  {author} {\bibinfo {author} {\bibfnamefont {L.}~\bibnamefont
  {Naticchioni}}, \bibinfo {author} {\bibfnamefont {M.}~\bibnamefont
  {Perciballi}}, \bibinfo {author} {\bibfnamefont {F.}~\bibnamefont {Ricci}},
  \bibinfo {author} {\bibfnamefont {E.}~\bibnamefont {Coccia}}, \bibinfo
  {author} {\bibfnamefont {V.}~\bibnamefont {Malvezzi}}, \bibinfo {author}
  {\bibfnamefont {F.}~\bibnamefont {Acernese}}, \bibinfo {author}
  {\bibfnamefont {F.}~\bibnamefont {Barone}}, \bibinfo {author} {\bibfnamefont
  {G.}~\bibnamefont {Giordano}}, \bibinfo {author} {\bibfnamefont
  {R.}~\bibnamefont {Romano}}, \bibinfo {author} {\bibfnamefont
  {M.}~\bibnamefont {Punturo}}, \bibinfo {author} {\bibfnamefont {R.~D.}\
  \bibnamefont {Rosa}}, \bibinfo {author} {\bibfnamefont {P.}~\bibnamefont
  {Calia}}, \ and\ \bibinfo {author} {\bibfnamefont {G.}~\bibnamefont
  {Loddo}},\ }\href {\doibase 10.1088/0264-9381/31/10/105016} {\bibfield
  {journal} {\bibinfo  {journal} {Classical and Quantum Gravity}\ }\textbf
  {\bibinfo {volume} {31}},\ \bibinfo {pages} {105016} (\bibinfo {year}
  {2014})}\BibitemShut {NoStop}%
\bibitem [{\citenamefont {Naticchioni}\ \emph {et~al.}(2020)\citenamefont
  {Naticchioni}, \citenamefont {Boschi}, \citenamefont {Calloni}, \citenamefont
  {Capello}, \citenamefont {Cardini}, \citenamefont {Carpinelli}, \citenamefont
  {Cuccuru}, \citenamefont {D’Ambrosio}, \citenamefont {de~Rosa},
  \citenamefont {Giovanni}, \citenamefont {d’Urso}, \citenamefont {Fiori},
  \citenamefont {Gaviano}, \citenamefont {Giunchi}, \citenamefont {Majorana},
  \citenamefont {Migoni}, \citenamefont {Oggiano}, \citenamefont {Olivieri},
  \citenamefont {Paoletti}, \citenamefont {Paratore}, \citenamefont
  {Perciballi}, \citenamefont {Piccinini}, \citenamefont {Punturo},
  \citenamefont {Puppo}, \citenamefont {Rapagnani}, \citenamefont {Ricci},
  \citenamefont {Saccorotti}, \citenamefont {Sipala},\ and\ \citenamefont
  {Tringali}}]{Naticchioni_2020}%
  \BibitemOpen
  \bibfield  {author} {\bibinfo {author} {\bibfnamefont {L.}~\bibnamefont
  {Naticchioni}}, \bibinfo {author} {\bibfnamefont {V.}~\bibnamefont {Boschi}},
  \bibinfo {author} {\bibfnamefont {E.}~\bibnamefont {Calloni}}, \bibinfo
  {author} {\bibfnamefont {M.}~\bibnamefont {Capello}}, \bibinfo {author}
  {\bibfnamefont {A.}~\bibnamefont {Cardini}}, \bibinfo {author} {\bibfnamefont
  {M.}~\bibnamefont {Carpinelli}}, \bibinfo {author} {\bibfnamefont
  {S.}~\bibnamefont {Cuccuru}}, \bibinfo {author} {\bibfnamefont
  {M.}~\bibnamefont {D’Ambrosio}}, \bibinfo {author} {\bibfnamefont
  {R.}~\bibnamefont {de~Rosa}}, \bibinfo {author} {\bibfnamefont {M.~D.}\
  \bibnamefont {Giovanni}}, \bibinfo {author} {\bibfnamefont {D.}~\bibnamefont
  {d’Urso}}, \bibinfo {author} {\bibfnamefont {I.}~\bibnamefont {Fiori}},
  \bibinfo {author} {\bibfnamefont {S.}~\bibnamefont {Gaviano}}, \bibinfo
  {author} {\bibfnamefont {C.}~\bibnamefont {Giunchi}}, \bibinfo {author}
  {\bibfnamefont {E.}~\bibnamefont {Majorana}}, \bibinfo {author}
  {\bibfnamefont {C.}~\bibnamefont {Migoni}}, \bibinfo {author} {\bibfnamefont
  {G.}~\bibnamefont {Oggiano}}, \bibinfo {author} {\bibfnamefont
  {M.}~\bibnamefont {Olivieri}}, \bibinfo {author} {\bibfnamefont
  {F.}~\bibnamefont {Paoletti}}, \bibinfo {author} {\bibfnamefont
  {M.}~\bibnamefont {Paratore}}, \bibinfo {author} {\bibfnamefont
  {M.}~\bibnamefont {Perciballi}}, \bibinfo {author} {\bibfnamefont
  {D.}~\bibnamefont {Piccinini}}, \bibinfo {author} {\bibfnamefont
  {M.}~\bibnamefont {Punturo}}, \bibinfo {author} {\bibfnamefont
  {P.}~\bibnamefont {Puppo}}, \bibinfo {author} {\bibfnamefont
  {P.}~\bibnamefont {Rapagnani}}, \bibinfo {author} {\bibfnamefont
  {F.}~\bibnamefont {Ricci}}, \bibinfo {author} {\bibfnamefont
  {G.}~\bibnamefont {Saccorotti}}, \bibinfo {author} {\bibfnamefont
  {V.}~\bibnamefont {Sipala}}, \ and\ \bibinfo {author} {\bibfnamefont {M.~C.}\
  \bibnamefont {Tringali}},\ }\href {\doibase 10.1088/1742-6596/1468/1/012242}
  {\bibfield  {journal} {\bibinfo  {journal} {Journal of Physics: Conference
  Series}\ }\textbf {\bibinfo {volume} {1468}},\ \bibinfo {pages} {012242}
  (\bibinfo {year} {2020})}\BibitemShut {NoStop}%
\bibitem [{\citenamefont {Di~Giovanni}\ \emph {et~al.}(2020)\citenamefont
  {Di~Giovanni}, \citenamefont {Giunchi}, \citenamefont {Saccorotti},
  \citenamefont {Berbellini}, \citenamefont {Boschi}, \citenamefont {Olivieri},
  \citenamefont {De~Rosa}, \citenamefont {Naticchioni}, \citenamefont
  {Oggiano}, \citenamefont {Carpinelli}, \citenamefont {D’Urso},
  \citenamefont {Cuccuru}, \citenamefont {Sipala}, \citenamefont {Calloni},
  \citenamefont {Di~Fiore}, \citenamefont {Grado}, \citenamefont {Migoni},
  \citenamefont {Cardini}, \citenamefont {Paoletti}, \citenamefont {Fiori},
  \citenamefont {Harms}, \citenamefont {Majorana}, \citenamefont {Rapagnani},
  \citenamefont {Ricci},\ and\ \citenamefont {Punturo}}]{10.1785/0220200186}%
  \BibitemOpen
  \bibfield  {author} {\bibinfo {author} {\bibfnamefont {M.}~\bibnamefont
  {Di~Giovanni}}, \bibinfo {author} {\bibfnamefont {C.}~\bibnamefont
  {Giunchi}}, \bibinfo {author} {\bibfnamefont {G.}~\bibnamefont {Saccorotti}},
  \bibinfo {author} {\bibfnamefont {A.}~\bibnamefont {Berbellini}}, \bibinfo
  {author} {\bibfnamefont {L.}~\bibnamefont {Boschi}}, \bibinfo {author}
  {\bibfnamefont {M.}~\bibnamefont {Olivieri}}, \bibinfo {author}
  {\bibfnamefont {R.}~\bibnamefont {De~Rosa}}, \bibinfo {author} {\bibfnamefont
  {L.}~\bibnamefont {Naticchioni}}, \bibinfo {author} {\bibfnamefont
  {G.}~\bibnamefont {Oggiano}}, \bibinfo {author} {\bibfnamefont
  {M.}~\bibnamefont {Carpinelli}}, \bibinfo {author} {\bibfnamefont
  {D.}~\bibnamefont {D’Urso}}, \bibinfo {author} {\bibfnamefont
  {S.}~\bibnamefont {Cuccuru}}, \bibinfo {author} {\bibfnamefont
  {V.}~\bibnamefont {Sipala}}, \bibinfo {author} {\bibfnamefont
  {E.}~\bibnamefont {Calloni}}, \bibinfo {author} {\bibfnamefont
  {L.}~\bibnamefont {Di~Fiore}}, \bibinfo {author} {\bibfnamefont
  {A.}~\bibnamefont {Grado}}, \bibinfo {author} {\bibfnamefont
  {C.}~\bibnamefont {Migoni}}, \bibinfo {author} {\bibfnamefont
  {A.}~\bibnamefont {Cardini}}, \bibinfo {author} {\bibfnamefont
  {F.}~\bibnamefont {Paoletti}}, \bibinfo {author} {\bibfnamefont
  {I.}~\bibnamefont {Fiori}}, \bibinfo {author} {\bibfnamefont
  {J.}~\bibnamefont {Harms}}, \bibinfo {author} {\bibfnamefont
  {E.}~\bibnamefont {Majorana}}, \bibinfo {author} {\bibfnamefont
  {P.}~\bibnamefont {Rapagnani}}, \bibinfo {author} {\bibfnamefont
  {F.}~\bibnamefont {Ricci}}, \ and\ \bibinfo {author} {\bibfnamefont
  {M.}~\bibnamefont {Punturo}},\ }\href {\doibase 10.1785/0220200186}
  {\bibfield  {journal} {\bibinfo  {journal} {Seismological Research Letters}\
  }\textbf {\bibinfo {volume} {92}},\ \bibinfo {pages} {352} (\bibinfo {year}
  {2020})}\BibitemShut {NoStop}%
\bibitem [{\citenamefont {Allocca}\ \emph {et~al.}(2021)\citenamefont {Allocca}
  \emph {et~al.}}]{Allocca:2021opl}%
  \BibitemOpen
  \bibfield  {author} {\bibinfo {author} {\bibfnamefont {A.}~\bibnamefont
  {Allocca}} \emph {et~al.},\ }\href {\doibase 10.1140/epjp/s13360-021-01450-8}
  {\bibfield  {journal} {\bibinfo  {journal} {Eur. Phys. J. Plus}\ }\textbf
  {\bibinfo {volume} {136}},\ \bibinfo {pages} {511} (\bibinfo {year}
  {2021})},\ \bibinfo {note} {[Erratum: Eur.Phys.J.Plus 136, 607
  (2021)]}\BibitemShut {NoStop}%
\bibitem [{\citenamefont {Di~Giovanni}\ \emph {et~al.}(2023)\citenamefont
  {Di~Giovanni}, \citenamefont {Koley}, \citenamefont {Ensing}, \citenamefont
  {Andric}, \citenamefont {Harms}, \citenamefont {D'Urso}, \citenamefont
  {Naticchioni}, \citenamefont {De~Rosa}, \citenamefont {Giunchi},
  \citenamefont {Allocca}, \citenamefont {Cadoni}, \citenamefont {Calloni},
  \citenamefont {Cardini}, \citenamefont {Carpinelli}, \citenamefont {Contu},
  \citenamefont {Errico}, \citenamefont {Mangano}, \citenamefont {Olivieri},
  \citenamefont {Punturo}, \citenamefont {Rapagnani}, \citenamefont {Ricci},
  \citenamefont {Rozza}, \citenamefont {Saccorotti}, \citenamefont {Trozzo},
  \citenamefont {Dell'Aquila}, \citenamefont {Pesenti}, \citenamefont
  {Sipala},\ and\ \citenamefont {Tosta~e Melo}}]{digiovanni23}%
  \BibitemOpen
  \bibfield  {author} {\bibinfo {author} {\bibfnamefont {M.}~\bibnamefont
  {Di~Giovanni}}, \bibinfo {author} {\bibfnamefont {S.}~\bibnamefont {Koley}},
  \bibinfo {author} {\bibfnamefont {J.~X.}\ \bibnamefont {Ensing}}, \bibinfo
  {author} {\bibfnamefont {T.}~\bibnamefont {Andric}}, \bibinfo {author}
  {\bibfnamefont {J.}~\bibnamefont {Harms}}, \bibinfo {author} {\bibfnamefont
  {D.}~\bibnamefont {D'Urso}}, \bibinfo {author} {\bibfnamefont
  {L.}~\bibnamefont {Naticchioni}}, \bibinfo {author} {\bibfnamefont
  {R.}~\bibnamefont {De~Rosa}}, \bibinfo {author} {\bibfnamefont
  {C.}~\bibnamefont {Giunchi}}, \bibinfo {author} {\bibfnamefont
  {A.}~\bibnamefont {Allocca}}, \bibinfo {author} {\bibfnamefont
  {M.}~\bibnamefont {Cadoni}}, \bibinfo {author} {\bibfnamefont
  {E.}~\bibnamefont {Calloni}}, \bibinfo {author} {\bibfnamefont
  {A.}~\bibnamefont {Cardini}}, \bibinfo {author} {\bibfnamefont
  {M.}~\bibnamefont {Carpinelli}}, \bibinfo {author} {\bibfnamefont
  {A.}~\bibnamefont {Contu}}, \bibinfo {author} {\bibfnamefont
  {L.}~\bibnamefont {Errico}}, \bibinfo {author} {\bibfnamefont
  {V.}~\bibnamefont {Mangano}}, \bibinfo {author} {\bibfnamefont
  {M.}~\bibnamefont {Olivieri}}, \bibinfo {author} {\bibfnamefont
  {M.}~\bibnamefont {Punturo}}, \bibinfo {author} {\bibfnamefont
  {P.}~\bibnamefont {Rapagnani}}, \bibinfo {author} {\bibfnamefont
  {F.}~\bibnamefont {Ricci}}, \bibinfo {author} {\bibfnamefont
  {D.}~\bibnamefont {Rozza}}, \bibinfo {author} {\bibfnamefont
  {G.}~\bibnamefont {Saccorotti}}, \bibinfo {author} {\bibfnamefont
  {L.}~\bibnamefont {Trozzo}}, \bibinfo {author} {\bibfnamefont
  {D.}~\bibnamefont {Dell'Aquila}}, \bibinfo {author} {\bibfnamefont
  {L.}~\bibnamefont {Pesenti}}, \bibinfo {author} {\bibfnamefont
  {V.}~\bibnamefont {Sipala}}, \ and\ \bibinfo {author} {\bibfnamefont
  {I.}~\bibnamefont {Tosta~e Melo}},\ }\href {\doibase 10.1093/gji/ggad178}
  {\bibfield  {journal} {\bibinfo  {journal} {Geophysical Journal
  International}\ }\textbf {\bibinfo {volume} {234}},\ \bibinfo {pages} {1943}
  (\bibinfo {year} {2023})}\BibitemShut {NoStop}%
\bibitem [{\citenamefont {Saccorotti}\ \emph {et~al.}(2023)\citenamefont
  {Saccorotti}, \citenamefont {Giunchi}, \citenamefont {D'Ambrosio},
  \citenamefont {Gaviano}, \citenamefont {Naticchioni}, \citenamefont {D'Urso},
  \citenamefont {Rozza}, \citenamefont {Cardini}, \citenamefont {Contu},
  \citenamefont {Dordei}, \citenamefont {Cadeddu}, \citenamefont {Tuveri},
  \citenamefont {Migoni}, \citenamefont {Punturo}, \citenamefont {Allocca},
  \citenamefont {Calloni}, \citenamefont {Cardello}, \citenamefont {D'Onofrio},
  \citenamefont {Davari}, \citenamefont {Dell'Aquila}, \citenamefont {Rosa},
  \citenamefont {Carpinelli}, \citenamefont {Fiore}, \citenamefont
  {di~Giovanni}, \citenamefont {Errico}, \citenamefont {Fiori}, \citenamefont
  {Tringali}, \citenamefont {Harms}, \citenamefont {Koley}, \citenamefont
  {Longo}, \citenamefont {Majorana}, \citenamefont {Mangano}, \citenamefont
  {Olivieri}, \citenamefont {Paoletti}, \citenamefont {Pesenti}, \citenamefont
  {Puppo}, \citenamefont {Rapagnani}, \citenamefont {Razzano}, \citenamefont
  {Ricci}, \citenamefont {Sipala}, \citenamefont {e~Melo},\ and\ \citenamefont
  {Trozzo}}]{Saccorotti_2023}%
  \BibitemOpen
  \bibfield  {author} {\bibinfo {author} {\bibfnamefont {G.}~\bibnamefont
  {Saccorotti}}, \bibinfo {author} {\bibfnamefont {C.}~\bibnamefont {Giunchi}},
  \bibinfo {author} {\bibfnamefont {M.}~\bibnamefont {D'Ambrosio}}, \bibinfo
  {author} {\bibfnamefont {S.}~\bibnamefont {Gaviano}}, \bibinfo {author}
  {\bibfnamefont {L.}~\bibnamefont {Naticchioni}}, \bibinfo {author}
  {\bibfnamefont {D.}~\bibnamefont {D'Urso}}, \bibinfo {author} {\bibfnamefont
  {D.}~\bibnamefont {Rozza}}, \bibinfo {author} {\bibfnamefont
  {A.}~\bibnamefont {Cardini}}, \bibinfo {author} {\bibfnamefont
  {A.}~\bibnamefont {Contu}}, \bibinfo {author} {\bibfnamefont
  {F.}~\bibnamefont {Dordei}}, \bibinfo {author} {\bibfnamefont
  {M.}~\bibnamefont {Cadeddu}}, \bibinfo {author} {\bibfnamefont
  {M.}~\bibnamefont {Tuveri}}, \bibinfo {author} {\bibfnamefont
  {C.}~\bibnamefont {Migoni}}, \bibinfo {author} {\bibfnamefont
  {M.}~\bibnamefont {Punturo}}, \bibinfo {author} {\bibfnamefont
  {A.}~\bibnamefont {Allocca}}, \bibinfo {author} {\bibfnamefont
  {E.}~\bibnamefont {Calloni}}, \bibinfo {author} {\bibfnamefont {G.~L.}\
  \bibnamefont {Cardello}}, \bibinfo {author} {\bibfnamefont {L.}~\bibnamefont
  {D'Onofrio}}, \bibinfo {author} {\bibfnamefont {N.}~\bibnamefont {Davari}},
  \bibinfo {author} {\bibfnamefont {D.}~\bibnamefont {Dell'Aquila}}, \bibinfo
  {author} {\bibfnamefont {R.~D.}\ \bibnamefont {Rosa}}, \bibinfo {author}
  {\bibfnamefont {M.}~\bibnamefont {Carpinelli}}, \bibinfo {author}
  {\bibfnamefont {L.~D.}\ \bibnamefont {Fiore}}, \bibinfo {author}
  {\bibfnamefont {M.}~\bibnamefont {di~Giovanni}}, \bibinfo {author}
  {\bibfnamefont {L.}~\bibnamefont {Errico}}, \bibinfo {author} {\bibfnamefont
  {I.}~\bibnamefont {Fiori}}, \bibinfo {author} {\bibfnamefont {M.~C.}\
  \bibnamefont {Tringali}}, \bibinfo {author} {\bibfnamefont {J.}~\bibnamefont
  {Harms}}, \bibinfo {author} {\bibfnamefont {S.}~\bibnamefont {Koley}},
  \bibinfo {author} {\bibfnamefont {V.}~\bibnamefont {Longo}}, \bibinfo
  {author} {\bibfnamefont {E.}~\bibnamefont {Majorana}}, \bibinfo {author}
  {\bibfnamefont {V.}~\bibnamefont {Mangano}}, \bibinfo {author} {\bibfnamefont
  {M.}~\bibnamefont {Olivieri}}, \bibinfo {author} {\bibfnamefont
  {F.}~\bibnamefont {Paoletti}}, \bibinfo {author} {\bibfnamefont
  {L.}~\bibnamefont {Pesenti}}, \bibinfo {author} {\bibfnamefont
  {P.}~\bibnamefont {Puppo}}, \bibinfo {author} {\bibfnamefont
  {P.}~\bibnamefont {Rapagnani}}, \bibinfo {author} {\bibfnamefont
  {M.}~\bibnamefont {Razzano}}, \bibinfo {author} {\bibfnamefont
  {F.}~\bibnamefont {Ricci}}, \bibinfo {author} {\bibfnamefont
  {V.}~\bibnamefont {Sipala}}, \bibinfo {author} {\bibfnamefont {I.~T.}\
  \bibnamefont {e~Melo}}, \ and\ \bibinfo {author} {\bibfnamefont
  {L.}~\bibnamefont {Trozzo}},\ }\href {\doibase
  10.1140/epjp/s13360-023-04395-2} {\bibfield  {journal} {\bibinfo  {journal}
  {The European Physical Journal Plus}\ }\textbf {\bibinfo {volume} {138}},\
  \bibinfo {pages} {793} (\bibinfo {year} {2023})}\BibitemShut {NoStop}%
\bibitem [{\citenamefont {Amann}\ \emph {et~al.}(2020)\citenamefont {Amann}
  \emph {et~al.}}]{Amann:2020jgo}%
  \BibitemOpen
  \bibfield  {author} {\bibinfo {author} {\bibfnamefont {F.}~\bibnamefont
  {Amann}} \emph {et~al.},\ }\href {\doibase 10.1063/5.0018414} {\bibfield
  {journal} {\bibinfo  {journal} {Rev. Sci. Instrum.}\ }\textbf {\bibinfo
  {volume} {91}},\ \bibinfo {pages} {9} (\bibinfo {year} {2020})},\ \Eprint
  {http://arxiv.org/abs/2003.03434} {arXiv:2003.03434 [physics.ins-det]}
  \BibitemShut {NoStop}%
\bibitem [{\citenamefont {Koley}\ \emph {et~al.}(2022)\citenamefont {Koley},
  \citenamefont {Bader}, \citenamefont {van~den Brand}, \citenamefont
  {Campman}, \citenamefont {Bulten}, \citenamefont {Linde},\ and\ \citenamefont
  {Vink}}]{Koley_2022}%
  \BibitemOpen
  \bibfield  {author} {\bibinfo {author} {\bibfnamefont {S.}~\bibnamefont
  {Koley}}, \bibinfo {author} {\bibfnamefont {M.}~\bibnamefont {Bader}},
  \bibinfo {author} {\bibfnamefont {J.}~\bibnamefont {van~den Brand}}, \bibinfo
  {author} {\bibfnamefont {X.}~\bibnamefont {Campman}}, \bibinfo {author}
  {\bibfnamefont {H.~J.}\ \bibnamefont {Bulten}}, \bibinfo {author}
  {\bibfnamefont {F.}~\bibnamefont {Linde}}, \ and\ \bibinfo {author}
  {\bibfnamefont {B.}~\bibnamefont {Vink}},\ }\href {\doibase
  10.1088/1361-6382/ac2b08} {\bibfield  {journal} {\bibinfo  {journal}
  {Classical and Quantum Gravity}\ }\textbf {\bibinfo {volume} {39}},\ \bibinfo
  {pages} {025008} (\bibinfo {year} {2022})}\BibitemShut {NoStop}%
\bibitem [{\citenamefont {Bader}\ \emph {et~al.}(2022)\citenamefont {Bader},
  \citenamefont {Koley}, \citenamefont {van~den Brand}, \citenamefont
  {Campman}, \citenamefont {Bulten}, \citenamefont {Linde},\ and\ \citenamefont
  {Vink}}]{Bader_2022}%
  \BibitemOpen
  \bibfield  {author} {\bibinfo {author} {\bibfnamefont {M.}~\bibnamefont
  {Bader}}, \bibinfo {author} {\bibfnamefont {S.}~\bibnamefont {Koley}},
  \bibinfo {author} {\bibfnamefont {J.}~\bibnamefont {van~den Brand}}, \bibinfo
  {author} {\bibfnamefont {X.}~\bibnamefont {Campman}}, \bibinfo {author}
  {\bibfnamefont {H.~J.}\ \bibnamefont {Bulten}}, \bibinfo {author}
  {\bibfnamefont {F.}~\bibnamefont {Linde}}, \ and\ \bibinfo {author}
  {\bibfnamefont {B.}~\bibnamefont {Vink}},\ }\href {\doibase
  10.1088/1361-6382/ac1be4} {\bibfield  {journal} {\bibinfo  {journal}
  {Classical and Quantum Gravity}\ }\textbf {\bibinfo {volume} {39}},\ \bibinfo
  {pages} {025009} (\bibinfo {year} {2022})}\BibitemShut {NoStop}%
\bibitem [{\citenamefont {Reitze}\ \emph {et~al.}(2019)\citenamefont {Reitze},
  \citenamefont {Adhikari}, \citenamefont {Ballmer}, \citenamefont {Barish},
  \citenamefont {Barsotti}, \citenamefont {Billingsley}, \citenamefont {Brown},
  \citenamefont {Chen}, \citenamefont {Coyne}, \citenamefont {Eisenstein},
  \citenamefont {Evans}, \citenamefont {Fritschel}, \citenamefont {Hall},
  \citenamefont {Lazzarini}, \citenamefont {Lovelace}, \citenamefont {Read},
  \citenamefont {Sathyaprakash}, \citenamefont {Shoemaker}, \citenamefont
  {Smith}, \citenamefont {Torrie}, \citenamefont {Vitale}, \citenamefont
  {Weiss}, \citenamefont {Wipf},\ and\ \citenamefont
  {Zucker}}]{Reitze2019Cosmic}%
  \BibitemOpen
  \bibfield  {author} {\bibinfo {author} {\bibfnamefont {D.}~\bibnamefont
  {Reitze}}, \bibinfo {author} {\bibfnamefont {R.~X.}\ \bibnamefont
  {Adhikari}}, \bibinfo {author} {\bibfnamefont {S.}~\bibnamefont {Ballmer}},
  \bibinfo {author} {\bibfnamefont {B.}~\bibnamefont {Barish}}, \bibinfo
  {author} {\bibfnamefont {L.}~\bibnamefont {Barsotti}}, \bibinfo {author}
  {\bibfnamefont {G.}~\bibnamefont {Billingsley}}, \bibinfo {author}
  {\bibfnamefont {D.~A.}\ \bibnamefont {Brown}}, \bibinfo {author}
  {\bibfnamefont {Y.}~\bibnamefont {Chen}}, \bibinfo {author} {\bibfnamefont
  {D.}~\bibnamefont {Coyne}}, \bibinfo {author} {\bibfnamefont
  {R.}~\bibnamefont {Eisenstein}}, \bibinfo {author} {\bibfnamefont
  {M.}~\bibnamefont {Evans}}, \bibinfo {author} {\bibfnamefont
  {P.}~\bibnamefont {Fritschel}}, \bibinfo {author} {\bibfnamefont {E.~D.}\
  \bibnamefont {Hall}}, \bibinfo {author} {\bibfnamefont {A.}~\bibnamefont
  {Lazzarini}}, \bibinfo {author} {\bibfnamefont {G.}~\bibnamefont {Lovelace}},
  \bibinfo {author} {\bibfnamefont {J.}~\bibnamefont {Read}}, \bibinfo {author}
  {\bibfnamefont {B.~S.}\ \bibnamefont {Sathyaprakash}}, \bibinfo {author}
  {\bibfnamefont {D.}~\bibnamefont {Shoemaker}}, \bibinfo {author}
  {\bibfnamefont {J.}~\bibnamefont {Smith}}, \bibinfo {author} {\bibfnamefont
  {C.}~\bibnamefont {Torrie}}, \bibinfo {author} {\bibfnamefont
  {S.}~\bibnamefont {Vitale}}, \bibinfo {author} {\bibfnamefont
  {R.}~\bibnamefont {Weiss}}, \bibinfo {author} {\bibfnamefont
  {C.}~\bibnamefont {Wipf}}, \ and\ \bibinfo {author} {\bibfnamefont
  {M.}~\bibnamefont {Zucker}},\ }\href {https://baas.aas.org/pub/2020n7i035}
  {\bibfield  {journal} {\bibinfo  {journal} {Bulletin of the AAS}\ }\textbf
  {\bibinfo {volume} {51}} (\bibinfo {year} {2019})},\ \bibinfo {note}
  {https://baas.aas.org/pub/2020n7i035}\BibitemShut {NoStop}%
\bibitem [{\citenamefont {Mandic}\ \emph {et~al.}(2018)\citenamefont {Mandic},
  \citenamefont {Tsai}, \citenamefont {Pavlis}, \citenamefont {Prestegard},
  \citenamefont {Bowden}, \citenamefont {Meyers},\ and\ \citenamefont
  {Caton}}]{10.1785/0220170228}%
  \BibitemOpen
  \bibfield  {author} {\bibinfo {author} {\bibfnamefont {V.}~\bibnamefont
  {Mandic}}, \bibinfo {author} {\bibfnamefont {V.~C.}\ \bibnamefont {Tsai}},
  \bibinfo {author} {\bibfnamefont {G.~L.}\ \bibnamefont {Pavlis}}, \bibinfo
  {author} {\bibfnamefont {T.}~\bibnamefont {Prestegard}}, \bibinfo {author}
  {\bibfnamefont {D.~C.}\ \bibnamefont {Bowden}}, \bibinfo {author}
  {\bibfnamefont {P.}~\bibnamefont {Meyers}}, \ and\ \bibinfo {author}
  {\bibfnamefont {R.}~\bibnamefont {Caton}},\ }\href {\doibase
  10.1785/0220170228} {\bibfield  {journal} {\bibinfo  {journal} {Seismological
  Research Letters}\ }\textbf {\bibinfo {volume} {89}},\ \bibinfo {pages}
  {2420} (\bibinfo {year} {2018})}\BibitemShut {NoStop}%
\bibitem [{\citenamefont {{TUBS, Wikimedia Commons}}(2013)}]{mapUSA}%
  \BibitemOpen
  \bibfield  {author} {\bibinfo {author} {\bibnamefont {{TUBS, Wikimedia
  Commons}}},\ }\href@noop {} {\enquote {\bibinfo {title} {File:{USA} edcp (+hi
  +ak) relief location map.png},}\ }\bibinfo {howpublished}
  {\url{https://commons.wikimedia.org/wiki/File:Usa\_edcp\_\%28\%2BHI\_\%2BAK\%29\_relief\_location\_map.png}}
  (\bibinfo {year} {2013})\BibitemShut {NoStop}%
\bibitem [{\citenamefont {{TUBS, Wikimedia Commons}}(2011)}]{mapEU}%
  \BibitemOpen
  \bibfield  {author} {\bibinfo {author} {\bibnamefont {{TUBS, Wikimedia
  Commons}}},\ }\href@noop {} {\enquote {\bibinfo {title} {File:blank in
  "europe" (relief) (-mini map).svg},}\ }\bibinfo {howpublished}
  {\url{https://commons.wikimedia.org/wiki/File:BLANK\_in\_Europe\_\%28relief\%29\_\%28-mini\_map\%29.svg}}
  (\bibinfo {year} {2011})\BibitemShut {NoStop}%
\bibitem [{\citenamefont {Janssens}\ \emph {et~al.}(2022)\citenamefont
  {Janssens}, \citenamefont {Boileau}, \citenamefont {Christensen},
  \citenamefont {Badaracco},\ and\ \citenamefont {van
  Remortel}}]{PhysRevD.106.042008}%
  \BibitemOpen
  \bibfield  {author} {\bibinfo {author} {\bibfnamefont {K.}~\bibnamefont
  {Janssens}}, \bibinfo {author} {\bibfnamefont {G.}~\bibnamefont {Boileau}},
  \bibinfo {author} {\bibfnamefont {N.}~\bibnamefont {Christensen}}, \bibinfo
  {author} {\bibfnamefont {F.}~\bibnamefont {Badaracco}}, \ and\ \bibinfo
  {author} {\bibfnamefont {N.}~\bibnamefont {van Remortel}},\ }\href {\doibase
  10.1103/PhysRevD.106.042008} {\bibfield  {journal} {\bibinfo  {journal}
  {Phys. Rev. D}\ }\textbf {\bibinfo {volume} {106}},\ \bibinfo {pages}
  {042008} (\bibinfo {year} {2022})}\BibitemShut {NoStop}%
\bibitem [{\citenamefont {Coughlin}\ \emph {et~al.}(2014)\citenamefont
  {Coughlin}, \citenamefont {Harms}, \citenamefont {Christensen}, \citenamefont
  {Dergachev}, \citenamefont {DeSalvo}, \citenamefont {Kandhasamy},\ and\
  \citenamefont {Mandic}}]{Coughlin_2014_seismic}%
  \BibitemOpen
  \bibfield  {author} {\bibinfo {author} {\bibfnamefont {M.}~\bibnamefont
  {Coughlin}}, \bibinfo {author} {\bibfnamefont {J.}~\bibnamefont {Harms}},
  \bibinfo {author} {\bibfnamefont {N.}~\bibnamefont {Christensen}}, \bibinfo
  {author} {\bibfnamefont {V.}~\bibnamefont {Dergachev}}, \bibinfo {author}
  {\bibfnamefont {R.}~\bibnamefont {DeSalvo}}, \bibinfo {author} {\bibfnamefont
  {S.}~\bibnamefont {Kandhasamy}}, \ and\ \bibinfo {author} {\bibfnamefont
  {V.}~\bibnamefont {Mandic}},\ }\href {\doibase
  10.1088/0264-9381/31/21/215003} {\bibfield  {journal} {\bibinfo  {journal}
  {Classical and Quantum Gravity}\ }\textbf {\bibinfo {volume} {31}},\ \bibinfo
  {pages} {215003} (\bibinfo {year} {2014})}\BibitemShut {NoStop}%
\bibitem [{\citenamefont {Coughlin}\ \emph {et~al.}(2019)\citenamefont
  {Coughlin}, \citenamefont {Harms}, \citenamefont {Bowden}, \citenamefont
  {Meyers}, \citenamefont {Tsai}, \citenamefont {Mandic}, \citenamefont
  {Pavlis},\ and\ \citenamefont {Prestegard}}]{Coughlin2019_seismic}%
  \BibitemOpen
  \bibfield  {author} {\bibinfo {author} {\bibfnamefont {M.}~\bibnamefont
  {Coughlin}}, \bibinfo {author} {\bibfnamefont {J.}~\bibnamefont {Harms}},
  \bibinfo {author} {\bibfnamefont {D.~C.}\ \bibnamefont {Bowden}}, \bibinfo
  {author} {\bibfnamefont {P.}~\bibnamefont {Meyers}}, \bibinfo {author}
  {\bibfnamefont {V.~C.}\ \bibnamefont {Tsai}}, \bibinfo {author}
  {\bibfnamefont {V.}~\bibnamefont {Mandic}}, \bibinfo {author} {\bibfnamefont
  {G.}~\bibnamefont {Pavlis}}, \ and\ \bibinfo {author} {\bibfnamefont
  {T.}~\bibnamefont {Prestegard}},\ }\href {\doibase 10.1029/2018jb016608}
  {\bibfield  {journal} {\bibinfo  {journal} {Journal of Geophysical Research:
  Solid Earth}\ }\textbf {\bibinfo {volume} {124}},\ \bibinfo {pages}
  {2941–2956} (\bibinfo {year} {2019})}\BibitemShut {NoStop}%
\bibitem [{\citenamefont {Harms}(2019)}]{Harms2019}%
  \BibitemOpen
  \bibfield  {author} {\bibinfo {author} {\bibfnamefont {J.}~\bibnamefont
  {Harms}},\ }\href {\doibase 10.1007/s41114-019-0022-2} {\bibfield  {journal}
  {\bibinfo  {journal} {Living Reviews in Relativity}\ }\textbf {\bibinfo
  {volume} {22}} (\bibinfo {year} {2019}),\
  10.1007/s41114-019-0022-2}\BibitemShut {NoStop}%
\bibitem [{\citenamefont {Saulson}(1984)}]{PhysRevD.30.732}%
  \BibitemOpen
  \bibfield  {author} {\bibinfo {author} {\bibfnamefont {P.~R.}\ \bibnamefont
  {Saulson}},\ }\href {\doibase 10.1103/PhysRevD.30.732} {\bibfield  {journal}
  {\bibinfo  {journal} {Phys. Rev. D}\ }\textbf {\bibinfo {volume} {30}},\
  \bibinfo {pages} {732} (\bibinfo {year} {1984})}\BibitemShut {NoStop}%
\bibitem [{\citenamefont {Hughes}\ and\ \citenamefont
  {Thorne}(1998)}]{PhysRevD.58.122002}%
  \BibitemOpen
  \bibfield  {author} {\bibinfo {author} {\bibfnamefont {S.~A.}\ \bibnamefont
  {Hughes}}\ and\ \bibinfo {author} {\bibfnamefont {K.~S.}\ \bibnamefont
  {Thorne}},\ }\href {\doibase 10.1103/PhysRevD.58.122002} {\bibfield
  {journal} {\bibinfo  {journal} {Phys. Rev. D}\ }\textbf {\bibinfo {volume}
  {58}},\ \bibinfo {pages} {122002} (\bibinfo {year} {1998})}\BibitemShut
  {NoStop}%
\bibitem [{\citenamefont {Branchesi}\ \emph {et~al.}(2023)\citenamefont
  {Branchesi} \emph {et~al.}}]{Branchesi:2023mws}%
  \BibitemOpen
  \bibfield  {author} {\bibinfo {author} {\bibfnamefont {M.}~\bibnamefont
  {Branchesi}} \emph {et~al.},\ }\href@noop {} {\  (\bibinfo {year} {2023})},\
  \Eprint {http://arxiv.org/abs/2303.15923} {arXiv:2303.15923 [gr-qc]}
  \BibitemShut {NoStop}%
\bibitem [{\citenamefont {Canuel}\ \emph {et~al.}(2018)\citenamefont {Canuel},
  \citenamefont {Bertoldi}, \citenamefont {Amand}, \citenamefont {Pozzo~di
  Borgo}, \citenamefont {Chantrait}, \citenamefont {Danquigny}, \citenamefont
  {Dovale~Álvarez}, \citenamefont {Fang}, \citenamefont {Freise},
  \citenamefont {Geiger}, \citenamefont {Gillot}, \citenamefont {Henry},
  \citenamefont {Hinderer}, \citenamefont {Holleville}, \citenamefont {Junca},
  \citenamefont {Lefèvre}, \citenamefont {Merzougui}, \citenamefont {Mielec},
  \citenamefont {Monfret}, \citenamefont {Pelisson}, \citenamefont
  {Prevedelli}, \citenamefont {Reynaud}, \citenamefont {Riou}, \citenamefont
  {Rogister}, \citenamefont {Rosat}, \citenamefont {Cormier}, \citenamefont
  {Landragin}, \citenamefont {Chaibi}, \citenamefont {Gaffet},\ and\
  \citenamefont {Bouyer}}]{Canuel_2018}%
  \BibitemOpen
  \bibfield  {author} {\bibinfo {author} {\bibfnamefont {B.}~\bibnamefont
  {Canuel}}, \bibinfo {author} {\bibfnamefont {A.}~\bibnamefont {Bertoldi}},
  \bibinfo {author} {\bibfnamefont {L.}~\bibnamefont {Amand}}, \bibinfo
  {author} {\bibfnamefont {E.}~\bibnamefont {Pozzo~di Borgo}}, \bibinfo
  {author} {\bibfnamefont {T.}~\bibnamefont {Chantrait}}, \bibinfo {author}
  {\bibfnamefont {C.}~\bibnamefont {Danquigny}}, \bibinfo {author}
  {\bibfnamefont {M.}~\bibnamefont {Dovale~Álvarez}}, \bibinfo {author}
  {\bibfnamefont {B.}~\bibnamefont {Fang}}, \bibinfo {author} {\bibfnamefont
  {A.}~\bibnamefont {Freise}}, \bibinfo {author} {\bibfnamefont
  {R.}~\bibnamefont {Geiger}}, \bibinfo {author} {\bibfnamefont
  {J.}~\bibnamefont {Gillot}}, \bibinfo {author} {\bibfnamefont
  {S.}~\bibnamefont {Henry}}, \bibinfo {author} {\bibfnamefont
  {J.}~\bibnamefont {Hinderer}}, \bibinfo {author} {\bibfnamefont
  {D.}~\bibnamefont {Holleville}}, \bibinfo {author} {\bibfnamefont
  {J.}~\bibnamefont {Junca}}, \bibinfo {author} {\bibfnamefont
  {G.}~\bibnamefont {Lefèvre}}, \bibinfo {author} {\bibfnamefont
  {M.}~\bibnamefont {Merzougui}}, \bibinfo {author} {\bibfnamefont
  {N.}~\bibnamefont {Mielec}}, \bibinfo {author} {\bibfnamefont
  {T.}~\bibnamefont {Monfret}}, \bibinfo {author} {\bibfnamefont
  {S.}~\bibnamefont {Pelisson}}, \bibinfo {author} {\bibfnamefont
  {M.}~\bibnamefont {Prevedelli}}, \bibinfo {author} {\bibfnamefont
  {S.}~\bibnamefont {Reynaud}}, \bibinfo {author} {\bibfnamefont
  {I.}~\bibnamefont {Riou}}, \bibinfo {author} {\bibfnamefont {Y.}~\bibnamefont
  {Rogister}}, \bibinfo {author} {\bibfnamefont {S.}~\bibnamefont {Rosat}},
  \bibinfo {author} {\bibfnamefont {E.}~\bibnamefont {Cormier}}, \bibinfo
  {author} {\bibfnamefont {A.}~\bibnamefont {Landragin}}, \bibinfo {author}
  {\bibfnamefont {W.}~\bibnamefont {Chaibi}}, \bibinfo {author} {\bibfnamefont
  {S.}~\bibnamefont {Gaffet}}, \ and\ \bibinfo {author} {\bibfnamefont
  {P.}~\bibnamefont {Bouyer}},\ }\href {\doibase 10.1038/s41598-018-32165-z}
  {\bibfield  {journal} {\bibinfo  {journal} {Scientific Reports}\ }\textbf
  {\bibinfo {volume} {8}} (\bibinfo {year} {2018}),\
  10.1038/s41598-018-32165-z}\BibitemShut {NoStop}%
\bibitem [{\citenamefont {Canuel}\ \emph {et~al.}(2020)\citenamefont {Canuel}
  \emph {et~al.}}]{Canuel:2019abg}%
  \BibitemOpen
  \bibfield  {author} {\bibinfo {author} {\bibfnamefont {B.}~\bibnamefont
  {Canuel}} \emph {et~al.},\ }\href {\doibase 10.1088/1361-6382/aba80e}
  {\bibfield  {journal} {\bibinfo  {journal} {Class. Quant. Grav.}\ }\textbf
  {\bibinfo {volume} {37}},\ \bibinfo {pages} {225017} (\bibinfo {year}
  {2020})},\ \Eprint {http://arxiv.org/abs/1911.03701} {arXiv:1911.03701
  [physics.atom-ph]} \BibitemShut {NoStop}%
\bibitem [{\citenamefont {Chaibi}\ \emph {et~al.}(2016)\citenamefont {Chaibi},
  \citenamefont {Geiger}, \citenamefont {Canuel}, \citenamefont {Bertoldi},
  \citenamefont {Landragin},\ and\ \citenamefont {Bouyer}}]{Chaibi2016}%
  \BibitemOpen
  \bibfield  {author} {\bibinfo {author} {\bibfnamefont {W.}~\bibnamefont
  {Chaibi}}, \bibinfo {author} {\bibfnamefont {R.}~\bibnamefont {Geiger}},
  \bibinfo {author} {\bibfnamefont {B.}~\bibnamefont {Canuel}}, \bibinfo
  {author} {\bibfnamefont {A.}~\bibnamefont {Bertoldi}}, \bibinfo {author}
  {\bibfnamefont {A.}~\bibnamefont {Landragin}}, \ and\ \bibinfo {author}
  {\bibfnamefont {P.}~\bibnamefont {Bouyer}},\ }\href
  {http://dx.doi.org/10.1103/PhysRevD.93.021101} {\bibfield  {journal}
  {\bibinfo  {journal} {Phys. Rev. D}\ }\textbf {\bibinfo {volume} {93}},\
  \bibinfo {pages} {021101} (\bibinfo {year} {2016})}\BibitemShut {NoStop}%
\bibitem [{\citenamefont {Junca}\ \emph {et~al.}(2019)\citenamefont {Junca},
  \citenamefont {Bertoldi}, \citenamefont {Sabulsky}, \citenamefont {Lefèvre},
  \citenamefont {Zou}, \citenamefont {Decitre}, \citenamefont {Geiger},
  \citenamefont {Landragin}, \citenamefont {Gaffet}, \citenamefont {Bouyer},\
  and\ \citenamefont {Canuel}}]{Junca_2019}%
  \BibitemOpen
  \bibfield  {author} {\bibinfo {author} {\bibfnamefont {J.}~\bibnamefont
  {Junca}}, \bibinfo {author} {\bibfnamefont {A.}~\bibnamefont {Bertoldi}},
  \bibinfo {author} {\bibfnamefont {D.}~\bibnamefont {Sabulsky}}, \bibinfo
  {author} {\bibfnamefont {G.}~\bibnamefont {Lefèvre}}, \bibinfo {author}
  {\bibfnamefont {X.}~\bibnamefont {Zou}}, \bibinfo {author} {\bibfnamefont
  {J.-B.}\ \bibnamefont {Decitre}}, \bibinfo {author} {\bibfnamefont
  {R.}~\bibnamefont {Geiger}}, \bibinfo {author} {\bibfnamefont
  {A.}~\bibnamefont {Landragin}}, \bibinfo {author} {\bibfnamefont
  {S.}~\bibnamefont {Gaffet}}, \bibinfo {author} {\bibfnamefont
  {P.}~\bibnamefont {Bouyer}}, \ and\ \bibinfo {author} {\bibfnamefont
  {B.}~\bibnamefont {Canuel}},\ }\href {\doibase 10.1103/physrevd.99.104026}
  {\bibfield  {journal} {\bibinfo  {journal} {Physical Review D}\ }\textbf
  {\bibinfo {volume} {99}} (\bibinfo {year} {2019}),\
  10.1103/physrevd.99.104026}\BibitemShut {NoStop}%
\bibitem [{\citenamefont {Driggers}\ \emph {et~al.}(2012)\citenamefont
  {Driggers}, \citenamefont {Harms},\ and\ \citenamefont
  {Adhikari}}]{PhysRevD.86.102001}%
  \BibitemOpen
  \bibfield  {author} {\bibinfo {author} {\bibfnamefont {J.~C.}\ \bibnamefont
  {Driggers}}, \bibinfo {author} {\bibfnamefont {J.}~\bibnamefont {Harms}}, \
  and\ \bibinfo {author} {\bibfnamefont {R.~X.}\ \bibnamefont {Adhikari}},\
  }\href {\doibase 10.1103/PhysRevD.86.102001} {\bibfield  {journal} {\bibinfo
  {journal} {Phys. Rev. D}\ }\textbf {\bibinfo {volume} {86}},\ \bibinfo
  {pages} {102001} (\bibinfo {year} {2012})}\BibitemShut {NoStop}%
\bibitem [{\citenamefont {Harms}\ and\ \citenamefont
  {Paik}(2015)}]{PhysRevD.92.022001}%
  \BibitemOpen
  \bibfield  {author} {\bibinfo {author} {\bibfnamefont {J.}~\bibnamefont
  {Harms}}\ and\ \bibinfo {author} {\bibfnamefont {H.~J.}\ \bibnamefont
  {Paik}},\ }\href {\doibase 10.1103/PhysRevD.92.022001} {\bibfield  {journal}
  {\bibinfo  {journal} {Phys. Rev. D}\ }\textbf {\bibinfo {volume} {92}},\
  \bibinfo {pages} {022001} (\bibinfo {year} {2015})}\BibitemShut {NoStop}%
\bibitem [{\citenamefont {Coughlin}\ \emph
  {et~al.}(2016{\natexlab{b}})\citenamefont {Coughlin}, \citenamefont {Mukund},
  \citenamefont {Harms}, \citenamefont {Driggers}, \citenamefont {Adhikari},\
  and\ \citenamefont {Mitra}}]{Coughlin_2016}%
  \BibitemOpen
  \bibfield  {author} {\bibinfo {author} {\bibfnamefont {M.}~\bibnamefont
  {Coughlin}}, \bibinfo {author} {\bibfnamefont {N.}~\bibnamefont {Mukund}},
  \bibinfo {author} {\bibfnamefont {J.}~\bibnamefont {Harms}}, \bibinfo
  {author} {\bibfnamefont {J.}~\bibnamefont {Driggers}}, \bibinfo {author}
  {\bibfnamefont {R.}~\bibnamefont {Adhikari}}, \ and\ \bibinfo {author}
  {\bibfnamefont {S.}~\bibnamefont {Mitra}},\ }\href {\doibase
  10.1088/0264-9381/33/24/244001} {\bibfield  {journal} {\bibinfo  {journal}
  {Classical and Quantum Gravity}\ }\textbf {\bibinfo {volume} {33}},\ \bibinfo
  {pages} {244001} (\bibinfo {year} {2016}{\natexlab{b}})}\BibitemShut
  {NoStop}%
\bibitem [{\citenamefont {Coughlin}\ \emph
  {et~al.}(2018{\natexlab{b}})\citenamefont {Coughlin}, \citenamefont {Harms},
  \citenamefont {Driggers}, \citenamefont {McManus}, \citenamefont {Mukund},
  \citenamefont {Ross}, \citenamefont {Slagmolen},\ and\ \citenamefont
  {Venkateswara}}]{PhysRevLett.121.221104}%
  \BibitemOpen
  \bibfield  {author} {\bibinfo {author} {\bibfnamefont {M.~W.}\ \bibnamefont
  {Coughlin}}, \bibinfo {author} {\bibfnamefont {J.}~\bibnamefont {Harms}},
  \bibinfo {author} {\bibfnamefont {J.}~\bibnamefont {Driggers}}, \bibinfo
  {author} {\bibfnamefont {D.~J.}\ \bibnamefont {McManus}}, \bibinfo {author}
  {\bibfnamefont {N.}~\bibnamefont {Mukund}}, \bibinfo {author} {\bibfnamefont
  {M.~P.}\ \bibnamefont {Ross}}, \bibinfo {author} {\bibfnamefont {B.~J.~J.}\
  \bibnamefont {Slagmolen}}, \ and\ \bibinfo {author} {\bibfnamefont
  {K.}~\bibnamefont {Venkateswara}},\ }\href {\doibase
  10.1103/PhysRevLett.121.221104} {\bibfield  {journal} {\bibinfo  {journal}
  {Phys. Rev. Lett.}\ }\textbf {\bibinfo {volume} {121}},\ \bibinfo {pages}
  {221104} (\bibinfo {year} {2018}{\natexlab{b}})}\BibitemShut {NoStop}%
\bibitem [{\citenamefont {Tringali}\ \emph {et~al.}(2019)\citenamefont
  {Tringali}, \citenamefont {Bulik}, \citenamefont {Harms}, \citenamefont
  {Fiori}, \citenamefont {Paoletti}, \citenamefont {Singh}, \citenamefont
  {Idzkowski}, \citenamefont {Kutynia}, \citenamefont {Nikliborc},
  \citenamefont {Suchi{\'{n}}ski}, \citenamefont {Bertolini},\ and\
  \citenamefont {Koley}}]{Tringali_2019}%
  \BibitemOpen
  \bibfield  {author} {\bibinfo {author} {\bibfnamefont {M.~C.}\ \bibnamefont
  {Tringali}}, \bibinfo {author} {\bibfnamefont {T.}~\bibnamefont {Bulik}},
  \bibinfo {author} {\bibfnamefont {J.}~\bibnamefont {Harms}}, \bibinfo
  {author} {\bibfnamefont {I.}~\bibnamefont {Fiori}}, \bibinfo {author}
  {\bibfnamefont {F.}~\bibnamefont {Paoletti}}, \bibinfo {author}
  {\bibfnamefont {N.}~\bibnamefont {Singh}}, \bibinfo {author} {\bibfnamefont
  {B.}~\bibnamefont {Idzkowski}}, \bibinfo {author} {\bibfnamefont
  {A.}~\bibnamefont {Kutynia}}, \bibinfo {author} {\bibfnamefont
  {K.}~\bibnamefont {Nikliborc}}, \bibinfo {author} {\bibfnamefont
  {M.}~\bibnamefont {Suchi{\'{n}}ski}}, \bibinfo {author} {\bibfnamefont
  {A.}~\bibnamefont {Bertolini}}, \ and\ \bibinfo {author} {\bibfnamefont
  {S.}~\bibnamefont {Koley}},\ }\href {\doibase 10.1088/1361-6382/ab5c43}
  {\bibfield  {journal} {\bibinfo  {journal} {Classical and Quantum Gravity}\
  }\textbf {\bibinfo {volume} {37}},\ \bibinfo {pages} {025005} (\bibinfo
  {year} {2019})}\BibitemShut {NoStop}%
\bibitem [{\citenamefont {Badaracco}\ \emph {et~al.}(2020)\citenamefont
  {Badaracco}, \citenamefont {Harms}, \citenamefont {Bertolini}, \citenamefont
  {Bulik}, \citenamefont {Fiori}, \citenamefont {Idzkowski}, \citenamefont
  {Kutynia}, \citenamefont {Nikliborc}, \citenamefont {Paoletti}, \citenamefont
  {Paoli}, \citenamefont {Rei},\ and\ \citenamefont
  {Suchinski}}]{Badaracco_2020}%
  \BibitemOpen
  \bibfield  {author} {\bibinfo {author} {\bibfnamefont {F.}~\bibnamefont
  {Badaracco}}, \bibinfo {author} {\bibfnamefont {J.}~\bibnamefont {Harms}},
  \bibinfo {author} {\bibfnamefont {A.}~\bibnamefont {Bertolini}}, \bibinfo
  {author} {\bibfnamefont {T.}~\bibnamefont {Bulik}}, \bibinfo {author}
  {\bibfnamefont {I.}~\bibnamefont {Fiori}}, \bibinfo {author} {\bibfnamefont
  {B.}~\bibnamefont {Idzkowski}}, \bibinfo {author} {\bibfnamefont
  {A.}~\bibnamefont {Kutynia}}, \bibinfo {author} {\bibfnamefont
  {K.}~\bibnamefont {Nikliborc}}, \bibinfo {author} {\bibfnamefont
  {F.}~\bibnamefont {Paoletti}}, \bibinfo {author} {\bibfnamefont
  {A.}~\bibnamefont {Paoli}}, \bibinfo {author} {\bibfnamefont
  {L.}~\bibnamefont {Rei}}, \ and\ \bibinfo {author} {\bibfnamefont
  {M.}~\bibnamefont {Suchinski}},\ }\href {\doibase 10.1088/1361-6382/abab64}
  {\bibfield  {journal} {\bibinfo  {journal} {Classical and Quantum Gravity}\
  }\textbf {\bibinfo {volume} {37}},\ \bibinfo {pages} {195016} (\bibinfo
  {year} {2020})}\BibitemShut {NoStop}%
\bibitem [{\citenamefont {Badaracco}\ and\ \citenamefont
  {Harms}(2019)}]{Badaracco_2019}%
  \BibitemOpen
  \bibfield  {author} {\bibinfo {author} {\bibfnamefont {F.}~\bibnamefont
  {Badaracco}}\ and\ \bibinfo {author} {\bibfnamefont {J.}~\bibnamefont
  {Harms}},\ }\href {\doibase 10.1088/1361-6382/ab28c1} {\bibfield  {journal}
  {\bibinfo  {journal} {Classical and Quantum Gravity}\ }\textbf {\bibinfo
  {volume} {36}},\ \bibinfo {pages} {145006} (\bibinfo {year}
  {2019})}\BibitemShut {NoStop}%
\bibitem [{\citenamefont {Andric}\ and\ \citenamefont
  {Harms}(2020)}]{NN_Sardinia2020}%
  \BibitemOpen
  \bibfield  {author} {\bibinfo {author} {\bibfnamefont {T.}~\bibnamefont
  {Andric}}\ and\ \bibinfo {author} {\bibfnamefont {J.}~\bibnamefont {Harms}},\
  }\href {\doibase https://doi.org/10.1029/2020JB020401} {\bibfield  {journal}
  {\bibinfo  {journal} {Journal of Geophysical Research: Solid Earth}\ }\textbf
  {\bibinfo {volume} {125}},\ \bibinfo {pages} {e2020JB020401} (\bibinfo {year}
  {2020})},\ \bibinfo {note} {e2020JB020401 10.1029/2020JB020401}\BibitemShut
  {NoStop}%
\bibitem [{\citenamefont {van Beveren}\ \emph {et~al.}(2023)\citenamefont {van
  Beveren}, \citenamefont {Bader}, \citenamefont {van~den Brand}, \citenamefont
  {Bulten}, \citenamefont {Campman}, \citenamefont {Koley},\ and\ \citenamefont
  {Linde}}]{vanBeveren_2023}%
  \BibitemOpen
  \bibfield  {author} {\bibinfo {author} {\bibfnamefont {V.}~\bibnamefont {van
  Beveren}}, \bibinfo {author} {\bibfnamefont {M.}~\bibnamefont {Bader}},
  \bibinfo {author} {\bibfnamefont {J.}~\bibnamefont {van~den Brand}}, \bibinfo
  {author} {\bibfnamefont {H.~J.}\ \bibnamefont {Bulten}}, \bibinfo {author}
  {\bibfnamefont {X.}~\bibnamefont {Campman}}, \bibinfo {author} {\bibfnamefont
  {S.}~\bibnamefont {Koley}}, \ and\ \bibinfo {author} {\bibfnamefont
  {F.}~\bibnamefont {Linde}},\ }\href {\doibase 10.1088/1361-6382/acf3c8}
  {\bibfield  {journal} {\bibinfo  {journal} {Classical and Quantum Gravity}\
  }\textbf {\bibinfo {volume} {40}},\ \bibinfo {pages} {205008} (\bibinfo
  {year} {2023})}\BibitemShut {NoStop}%
\bibitem [{\citenamefont {Peterson}(1993)}]{Pet1993}%
  \BibitemOpen
  \bibfield  {author} {\bibinfo {author} {\bibfnamefont {J.}~\bibnamefont
  {Peterson}},\ }\href@noop {} {\bibfield  {journal} {\bibinfo  {journal}
  {Open-file report}\ }\textbf {\bibinfo {volume} {93-322}} (\bibinfo {year}
  {1993})}\BibitemShut {NoStop}%
\bibitem [{\citenamefont {Stutzmann}\ \emph {et~al.}(2009)\citenamefont
  {Stutzmann}, \citenamefont {Schimmel}, \citenamefont {Patau},\ and\
  \citenamefont {Maggi}}]{Stutzmann2009}%
  \BibitemOpen
  \bibfield  {author} {\bibinfo {author} {\bibfnamefont {E.}~\bibnamefont
  {Stutzmann}}, \bibinfo {author} {\bibfnamefont {M.}~\bibnamefont {Schimmel}},
  \bibinfo {author} {\bibfnamefont {G.}~\bibnamefont {Patau}}, \ and\ \bibinfo
  {author} {\bibfnamefont {A.}~\bibnamefont {Maggi}},\ }\href {\doibase
  https://doi.org/10.1029/2009GC002619} {\bibfield  {journal} {\bibinfo
  {journal} {Geochemistry, Geophysics, Geosystems}\ }\textbf {\bibinfo {volume}
  {10}} (\bibinfo {year} {2009}),\ https://doi.org/10.1029/2009GC002619},\
  \Eprint
  {http://arxiv.org/abs/https://agupubs.onlinelibrary.wiley.com/doi/pdf/10.1029/2009GC002619}
  {https://agupubs.onlinelibrary.wiley.com/doi/pdf/10.1029/2009GC002619}
  \BibitemShut {NoStop}%
\bibitem [{\citenamefont {Misek}\ \emph {et~al.}(2023)\citenamefont {Misek},
  \citenamefont {Jakus}, \citenamefont {Hamza~Sladicekova}, \citenamefont
  {Zastko}, \citenamefont {Veternik}, \citenamefont {Jakusova},\ and\
  \citenamefont {Belyaev}}]{10.3389/fphy.2023.1094921}%
  \BibitemOpen
  \bibfield  {author} {\bibinfo {author} {\bibfnamefont {J.}~\bibnamefont
  {Misek}}, \bibinfo {author} {\bibfnamefont {J.}~\bibnamefont {Jakus}},
  \bibinfo {author} {\bibfnamefont {K.}~\bibnamefont {Hamza~Sladicekova}},
  \bibinfo {author} {\bibfnamefont {L.}~\bibnamefont {Zastko}}, \bibinfo
  {author} {\bibfnamefont {M.}~\bibnamefont {Veternik}}, \bibinfo {author}
  {\bibfnamefont {V.}~\bibnamefont {Jakusova}}, \ and\ \bibinfo {author}
  {\bibfnamefont {I.}~\bibnamefont {Belyaev}},\ }\href {\doibase
  10.3389/fphy.2023.1094921} {\bibfield  {journal} {\bibinfo  {journal}
  {Frontiers in Physics}\ }\textbf {\bibinfo {volume} {11}} (\bibinfo {year}
  {2023}),\ 10.3389/fphy.2023.1094921}\BibitemShut {NoStop}%
\bibitem [{\citenamefont {Weaver}(1982)}]{f24a9b6cf61740668d35dac9c3c317ab}%
  \BibitemOpen
  \bibfield  {author} {\bibinfo {author} {\bibfnamefont {R.}~\bibnamefont
  {Weaver}},\ }\href {\doibase 10.1121/1.387816} {\bibfield  {journal}
  {\bibinfo  {journal} {Journal of the Acoustical Society of America}\ }\textbf
  {\bibinfo {volume} {71}},\ \bibinfo {pages} {1608} (\bibinfo {year}
  {1982})}\BibitemShut {NoStop}%
\bibitem [{\citenamefont {Badaracco}\ \emph {et~al.}(2024)\citenamefont
  {Badaracco}, \citenamefont {Harms},\ and\ \citenamefont
  {Rei}}]{Badaracco_2024}%
  \BibitemOpen
  \bibfield  {author} {\bibinfo {author} {\bibfnamefont {F.}~\bibnamefont
  {Badaracco}}, \bibinfo {author} {\bibfnamefont {J.}~\bibnamefont {Harms}}, \
  and\ \bibinfo {author} {\bibfnamefont {L.}~\bibnamefont {Rei}},\ }\href
  {\doibase 10.1088/1361-6382/ad1715} {\bibfield  {journal} {\bibinfo
  {journal} {Classical and Quantum Gravity}\ }\textbf {\bibinfo {volume}
  {41}},\ \bibinfo {pages} {025013} (\bibinfo {year} {2024})}\BibitemShut
  {NoStop}%
\bibitem [{\citenamefont {Christensen}(1992)}]{PhysRevD.46.5250}%
  \BibitemOpen
  \bibfield  {author} {\bibinfo {author} {\bibfnamefont {N.}~\bibnamefont
  {Christensen}},\ }\href {\doibase 10.1103/PhysRevD.46.5250} {\bibfield
  {journal} {\bibinfo  {journal} {Phys. Rev. D}\ }\textbf {\bibinfo {volume}
  {46}},\ \bibinfo {pages} {5250} (\bibinfo {year} {1992})}\BibitemShut
  {NoStop}%
\bibitem [{\citenamefont {Allen}\ and\ \citenamefont
  {Romano}(1999)}]{PhysRevD.59.102001}%
  \BibitemOpen
  \bibfield  {author} {\bibinfo {author} {\bibfnamefont {B.}~\bibnamefont
  {Allen}}\ and\ \bibinfo {author} {\bibfnamefont {J.~D.}\ \bibnamefont
  {Romano}},\ }\href {\doibase 10.1103/PhysRevD.59.102001} {\bibfield
  {journal} {\bibinfo  {journal} {Phys. Rev. D}\ }\textbf {\bibinfo {volume}
  {59}},\ \bibinfo {pages} {102001} (\bibinfo {year} {1999})}\BibitemShut
  {NoStop}%
\bibitem [{\citenamefont {Romano}\ and\ \citenamefont
  {Cornish}(2017{\natexlab{a}})}]{LivingRevRelativ20}%
  \BibitemOpen
  \bibfield  {author} {\bibinfo {author} {\bibfnamefont {J.~D.}\ \bibnamefont
  {Romano}}\ and\ \bibinfo {author} {\bibfnamefont {N.~J.}\ \bibnamefont
  {Cornish}},\ }\href {\doibase 10.1007/s41114-017-0004-1} {\bibfield
  {journal} {\bibinfo  {journal} {Living Rev. Relativ.}\ }\textbf {\bibinfo
  {volume} {20}},\ \bibinfo {pages} {2} (\bibinfo {year}
  {2017}{\natexlab{a}})}\BibitemShut {NoStop}%
\bibitem [{\citenamefont {Ade}\ \emph {et~al.}(2016)\citenamefont {Ade} \emph
  {et~al.}}]{Planck:2015fie}%
  \BibitemOpen
  \bibfield  {author} {\bibinfo {author} {\bibfnamefont {P.~A.~R.}\
  \bibnamefont {Ade}} \emph {et~al.} (\bibinfo {collaboration} {Planck}),\
  }\href {\doibase 10.1051/0004-6361/201525830} {\bibfield  {journal} {\bibinfo
   {journal} {Astron. Astrophys.}\ }\textbf {\bibinfo {volume} {594}},\
  \bibinfo {pages} {A13} (\bibinfo {year} {2016})},\ \Eprint
  {http://arxiv.org/abs/1502.01589} {arXiv:1502.01589 [astro-ph.CO]}
  \BibitemShut {NoStop}%
\bibitem [{\citenamefont {Romano}\ and\ \citenamefont
  {Cornish}(2017{\natexlab{b}})}]{Romano:2016dpx}%
  \BibitemOpen
  \bibfield  {author} {\bibinfo {author} {\bibfnamefont {J.~D.}\ \bibnamefont
  {Romano}}\ and\ \bibinfo {author} {\bibfnamefont {N.~J.}\ \bibnamefont
  {Cornish}},\ }\href {\doibase 10.1007/s41114-017-0004-1} {\bibfield
  {journal} {\bibinfo  {journal} {Living Rev. Rel.}\ }\textbf {\bibinfo
  {volume} {20}},\ \bibinfo {pages} {2} (\bibinfo {year}
  {2017}{\natexlab{b}})},\ \Eprint {http://arxiv.org/abs/1608.06889}
  {arXiv:1608.06889 [gr-qc]} \BibitemShut {NoStop}%
\bibitem [{\citenamefont {Thrane}\ and\ \citenamefont
  {Romano}(2013)}]{Thrane:2013oya}%
  \BibitemOpen
  \bibfield  {author} {\bibinfo {author} {\bibfnamefont {E.}~\bibnamefont
  {Thrane}}\ and\ \bibinfo {author} {\bibfnamefont {J.~D.}\ \bibnamefont
  {Romano}},\ }\href {\doibase 10.1103/PhysRevD.88.124032} {\bibfield
  {journal} {\bibinfo  {journal} {Phys. Rev. D}\ }\textbf {\bibinfo {volume}
  {88}},\ \bibinfo {pages} {124032} (\bibinfo {year} {2013})},\ \Eprint
  {http://arxiv.org/abs/1310.5300} {arXiv:1310.5300 [astro-ph.IM]} \BibitemShut
  {NoStop}%
\bibitem [{\citenamefont {{LIGO Scientific Collaboration, Virgo Collaboration
  and KAGRA Collaboration}}()}]{O3IsotropicDataset}%
  \BibitemOpen
  \bibfield  {author} {\bibinfo {author} {\bibnamefont {{LIGO Scientific
  Collaboration, Virgo Collaboration and KAGRA Collaboration}}},\ }\href
  {https://dcc.ligo.org/G2001287/public} {\enquote {\bibinfo {title} {Data for
  upper limits on the isotropic gravitational-wave background from advanced
  {LIGO}'s and advanced {Virgo}'s third observing run},}\ }\BibitemShut
  {NoStop}%
\bibitem [{\citenamefont {Abbott}\ \emph
  {et~al.}(2021{\natexlab{a}})\citenamefont {Abbott} \emph
  {et~al.}}]{KAGRA:2021kbb}%
  \BibitemOpen
  \bibfield  {author} {\bibinfo {author} {\bibfnamefont {R.}~\bibnamefont
  {Abbott}} \emph {et~al.} (\bibinfo {collaboration} {KAGRA, Virgo, LIGO
  Scientific}),\ }\href {\doibase 10.1103/PhysRevD.104.022004} {\bibfield
  {journal} {\bibinfo  {journal} {Phys. Rev. D}\ }\textbf {\bibinfo {volume}
  {104}},\ \bibinfo {pages} {022004} (\bibinfo {year} {2021}{\natexlab{a}})},\
  \Eprint {http://arxiv.org/abs/2101.12130} {arXiv:2101.12130 [gr-qc]}
  \BibitemShut {NoStop}%
\bibitem [{\citenamefont {Maggiore}(2008)}]{MM_PartI}%
  \BibitemOpen
  \bibfield  {author} {\bibinfo {author} {\bibfnamefont {M.}~\bibnamefont
  {Maggiore}},\ }\href@noop {} {\emph {\bibinfo {title} {Gravitational waves -
  Volume 1: Theory and experiments}}}\ (\bibinfo  {publisher} {Oxford
  university press},\ \bibinfo {year} {2008})\BibitemShut {NoStop}%
\bibitem [{\citenamefont {Abbott}\ \emph {et~al.}(2016)\citenamefont {Abbott},
  \citenamefont {Abbott}, \citenamefont {Abbott} \emph
  {et~al.}}]{PhysRevLett.116.131102}%
  \BibitemOpen
  \bibfield  {author} {\bibinfo {author} {\bibfnamefont {B.~P.}\ \bibnamefont
  {Abbott}}, \bibinfo {author} {\bibfnamefont {R.}~\bibnamefont {Abbott}},
  \bibinfo {author} {\bibfnamefont {T.~D.}\ \bibnamefont {Abbott}},  \emph
  {et~al.} (\bibinfo {collaboration} {LIGO Scientific Collaboration and Virgo
  Collaboration}),\ }\href {\doibase 10.1103/PhysRevLett.116.131102} {\bibfield
   {journal} {\bibinfo  {journal} {Phys. Rev. Lett.}\ }\textbf {\bibinfo
  {volume} {116}},\ \bibinfo {pages} {131102} (\bibinfo {year}
  {2016})}\BibitemShut {NoStop}%
\bibitem [{\citenamefont {Regimbau}(2011)}]{Regimbau_2011}%
  \BibitemOpen
  \bibfield  {author} {\bibinfo {author} {\bibfnamefont {T.}~\bibnamefont
  {Regimbau}},\ }\href {\doibase 10.1088/1674-4527/11/4/001} {\bibfield
  {journal} {\bibinfo  {journal} {Research in Astronomy and Astrophysics}\
  }\textbf {\bibinfo {volume} {11}},\ \bibinfo {pages} {369} (\bibinfo {year}
  {2011})}\BibitemShut {NoStop}%
\bibitem [{\citenamefont {Ferrari}\ \emph {et~al.}(1999)\citenamefont
  {Ferrari}, \citenamefont {Matarrese},\ and\ \citenamefont
  {Schneider}}]{10.1046/j.1365-8711.1999.02194.x}%
  \BibitemOpen
  \bibfield  {author} {\bibinfo {author} {\bibfnamefont {V.}~\bibnamefont
  {Ferrari}}, \bibinfo {author} {\bibfnamefont {S.}~\bibnamefont {Matarrese}},
  \ and\ \bibinfo {author} {\bibfnamefont {R.}~\bibnamefont {Schneider}},\
  }\href {\doibase 10.1046/j.1365-8711.1999.02194.x} {\bibfield  {journal}
  {\bibinfo  {journal} {Monthly Notices of the Royal Astronomical Society}\
  }\textbf {\bibinfo {volume} {303}},\ \bibinfo {pages} {247} (\bibinfo {year}
  {1999})},\ \Eprint
  {http://arxiv.org/abs/https://academic.oup.com/mnras/article-pdf/303/2/247/18630520/303-2-247.pdf}
  {https://academic.oup.com/mnras/article-pdf/303/2/247/18630520/303-2-247.pdf}
  \BibitemShut {NoStop}%
\bibitem [{\citenamefont {Abbott}\ \emph
  {et~al.}(2021{\natexlab{b}})\citenamefont {Abbott} \emph
  {et~al.}}]{LIGOScientific:2021nrg}%
  \BibitemOpen
  \bibfield  {author} {\bibinfo {author} {\bibfnamefont {R.}~\bibnamefont
  {Abbott}} \emph {et~al.} (\bibinfo {collaboration} {{LIGO Scientific, Virgo,
  KAGRA collaborations}}),\ }\href {\doibase 10.1103/PhysRevLett.126.241102}
  {\bibfield  {journal} {\bibinfo  {journal} {Phys. Rev. Lett.}\ }\textbf
  {\bibinfo {volume} {126}},\ \bibinfo {pages} {241102} (\bibinfo {year}
  {2021}{\natexlab{b}})},\ \Eprint {http://arxiv.org/abs/2101.12248}
  {arXiv:2101.12248 [gr-qc]} \BibitemShut {NoStop}%
\bibitem [{\citenamefont {{ET Steering Committee Editorial
  Team}}(2020)}]{ETdesignRep}%
  \BibitemOpen
  \bibfield  {author} {\bibinfo {author} {\bibnamefont {{ET Steering Committee
  Editorial Team}}},\ }\href {https://apps.et-gw.eu/tds/?content=3&r=17245} {\
  (\bibinfo {year} {2020})},\ \Eprint {http://arxiv.org/abs/ET-0007B-20}
  {ET-0007B-20} \BibitemShut {NoStop}%
\bibitem [{\citenamefont {Davis}\ \emph {et~al.}(2021)\citenamefont {Davis},
  \citenamefont {Areeda} \emph {et~al.}}]{Davis_2021}%
  \BibitemOpen
  \bibfield  {author} {\bibinfo {author} {\bibfnamefont {D.}~\bibnamefont
  {Davis}}, \bibinfo {author} {\bibfnamefont {J.~S.}\ \bibnamefont {Areeda}},
  \emph {et~al.},\ }\href {\doibase 10.1088/1361-6382/abfd85} {\bibfield
  {journal} {\bibinfo  {journal} {Classical and Quantum Gravity}\ }\textbf
  {\bibinfo {volume} {38}},\ \bibinfo {pages} {135014} (\bibinfo {year}
  {2021})}\BibitemShut {NoStop}%
\end{thebibliography}%

\section*{Appendix: August results}
\label{appendix:SiteComp-Aug}

In this Appendix we provide the results for the different sites during the month of August. We will however not further discuss these as they are very similar from the January results presented in the main text. The main difference is the lower amplitude of the seismic correlations by about an order magnitude around the micro-seism peaks.

\begin{figure*}
    \centering
    \includegraphics[width=0.49\textwidth]{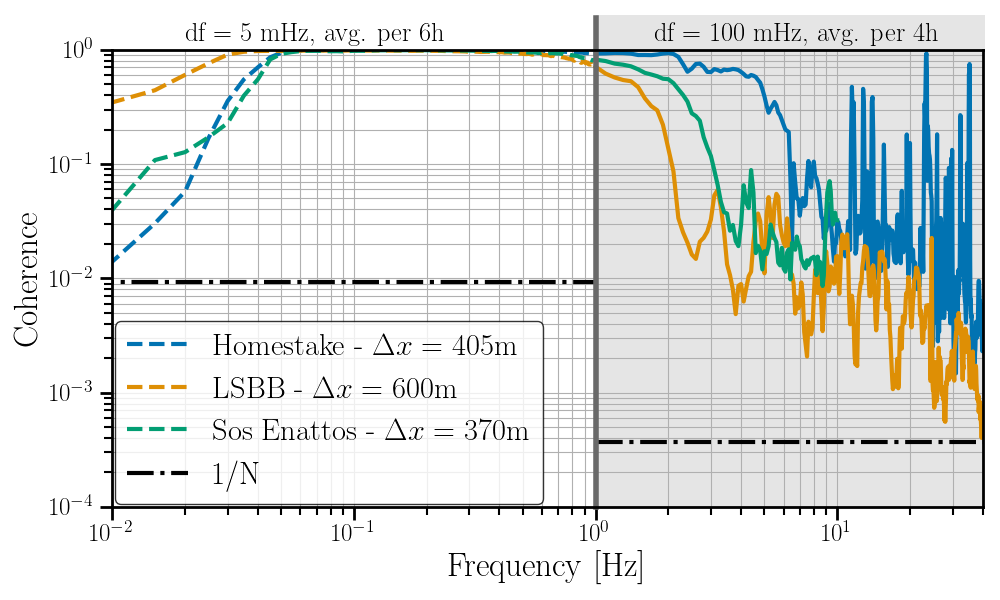}
    \includegraphics[width=0.49\textwidth]{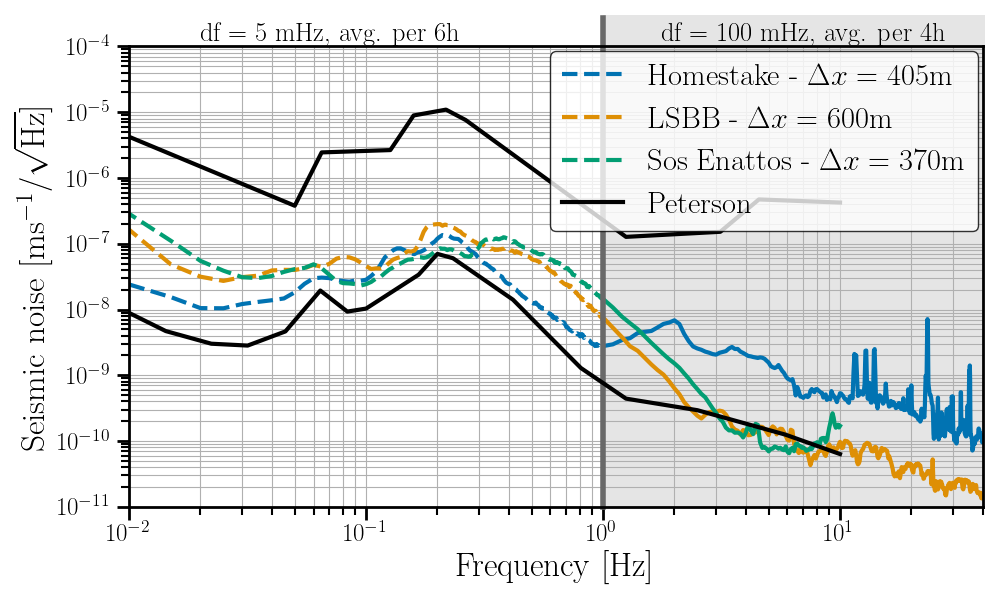}
    \caption{The median coherence (left panel) and CSD (right panel) of the underground seismometers for the different geographical locations studied in this paper for the month of August. 
    For more details on the sensor's specifications, see Tab. \ref{tab:sensors}.
    Note: the data is not from the same year. The data $<$1Hz (dashed curves) are analysed using 200 second long segments which are averaged per 6\,h-window. Above 1Hz the data (full curves) are analysed using 10 second long segments which are averaged per 4\,h-window. The black dot-dashed line represents the level of coherence expected from Gaussian data. The Peterson low and high noise models are shown in black (right panel).}
    \label{fig:Comparison_Aug_distance_CohCSD_5mHz}
\end{figure*}

\begin{figure}
    \centering
    \includegraphics[width=\linewidth]{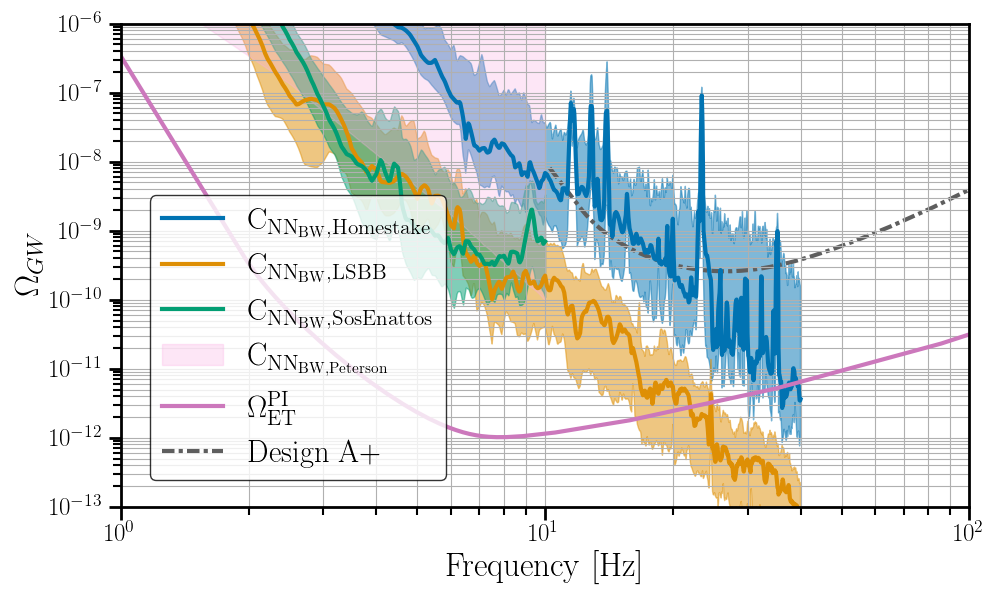}
    \caption{The projected impact from correlated NN from body-waves, as calculated in this section, for the seismic data from the month of January for the different locations, see the text for details on the used sensors, their distances and depths. 
    As a comparison we make the same projection using the Peterson low noise and high noise models. For the broadband ($\Omega^{\rm PI}_{{\rm ET}}$) sensitivity to a GWB we assumed one year of observation time (100\% duty cycle). The one year PI curve of the A+ design for the LIGO Hanford LIGO Livingston and Virgo detectors is represented by the dot-dashed curve. This curve was obtained using the open data provided by the LVK collaborations \cite{O3IsotropicDataset} and was first presented in \cite{KAGRA:2021kbb}. Please note: in this paper we present the 1$\sigma$ PI-curve, whereas in \cite{KAGRA:2021kbb} the 2$\sigma$ PI-curve is shown.}
    \label{fig:StochBudget_Aug_NN}
\end{figure}

\end{document}